\documentclass[%
 reprint,
 amsmath,amssymb,
 aps,
]{revtex4-2}

\usepackage{graphicx}
\usepackage{dcolumn}
\usepackage{bm}
\usepackage{siunitx}
\DeclareSIUnit\barn{b}
\usepackage{multirow}

\begin{document}

\preprint{APS/123-QED}

\title{Post Long Shutdown 2 CERN Proton Synchrotron\\ Transverse Impedance Model: Description and Beam-Based Validation}

\author{Sébastien Joly}
 \email{sebastien.joly@helmholtz-berlin.de}
 \altaffiliation[Also at ]{CERN, Geneva, Switzerland.}
\author{Mauro Migliorati}%
 \altaffiliation[Also at ]{Istituto Nazionale di Fisica Nucleare, Roma, Italy.}
\affiliation{%
 University of Rome ‘La Sapienza’, Roma, Italy
}%


\author{Nicolas Mounet}
\author{Benoît Salvant}
\affiliation{CERN, Geneva, Switzerland}


\date{\today}

\begin{abstract}
In the framework of the LHC Injectors Upgrade project, the CERN Proton Synchrotron underwent hardware changes during the Long Shutdown 2. These changes required an update of the CERN Proton Synchrotron impedance model and an assessment of their impact on beam quality. In the following, the updated CERN Proton Synchrotron impedance model is presented and then validated with beam-based measurements, comparing impedance-induced tune shifts and instability growth rates with simulations performed with a macroparticle tracking code, PyHEADTAIL, which includes the effects of beam-coupling impedance, betatron motion, non-linear synchrotron motion, and space charge.

The interplay between first and second-order chromaticity, space-charge forces, and their effect on beam observables is thus studied in detail. Their respective impacts have been assessed and benchmarked with measurements to validate the PS transverse impedance model.
\end{abstract}

\maketitle


\section{Introduction}


The pursuit of high-brightness beams in many facilities across the world has intensified the impact of collective effects, particularly those driven by impedance and space charge. The consequences of these effects range from beam quality degradation (e.g. emittance blow-up) to beam intensity losses. For example, the interplay between space-charge-induced tune spread and resonances can strongly limit the choice of the working point in SIS100~\cite{PhysRevAccelBeams.25.054402}. In addition to this, transverse instabilities arising from the combined effect of impedance and space charge prevents increasing the maximum beam intensity by causing severe beam losses as observed in the Fermilab Recycler~\cite{mohsen:hb2023-thafp04}, ISIS RCS~\cite{williamson:hb2023-tha1i2, williamson2023a} and JPARC RCS~\cite{PhysRevAccelBeams.21.024203, 10.1093/ptep/ptw169}. In addition, space charge effects were recently predicted to be also relevant in fourth generation light sources~\cite{Antipov:2024ppx}.

One of the most demanding projects to push beam performance further is the High-Luminosity LHC (HL-LHC) project~\cite{BejarAlonso:2069130}, which aims at increasing the integrated luminosity, i.e.\ the ratio of the total number of events over a period divided by the event cross section, by one order of magnitude. It would allow to surpass the initial LHC goal of approximately~\SI{300}{\per \femto \barn} to reach~\SI{3000}{\per \femto \barn} over the 12 years following the upgrade~\cite{BejarAlonso:2069130}.

In preparation for HL-LHC, its injectors were upgraded during the LHC Injectors Upgrade (LIU)~\cite{Damerau:1976692} to enable a twofold increase of the injected beam intensity with a slightly smaller transverse emittance required by the higher luminosity. The PSB, PS, and SPS underwent significant hardware refurbishment and upgrade during the Long Shutdown 2 (LS2, 2019 - 2021). The PS has a history of transverse instabilities~\cite{Gareyte:1974vf, Metral:2020gny, Migliorati:2018}, and the hardware modifications introduced during LS2 raised concerns about its beam stability with the new LIU parameter range.
To address these concerns, the PS impedance model was updated and validated with beam-based measurements.
The updated impedance model proved to be critical in the mitigation of recent instabilities~\cite{Joly:IPAC23-WEPL148, Joly:2024acd}. It is also expected to provide key insights into PS beam stability in future scenarios, such as those envisioned in the Physics Beyond Collider (PBC) project~\cite{Beacham:2019nyx}.

In this paper, we first describe the updated transverse impedance model of the PS. Then, beam-based reference measurement campaigns, carried out to validate the transverse impedance model, are reviewed. Both tune shifts and instability growth rates were acquired to study respectively the imaginary and real part of the PS transverse impedance. 
Several mechanisms prone to impacting the impedance-induced beam observables, namely the first and second-order chromaticity and direct space charge force, are then studied through simulations with PyHEADTAIL~\cite{Rumolo:702717, Oeftiger:2672381, pyheadtail}, a macroparticle tracking code, and a detailed assessment of the impact of each mechanism on the impedance-induced beam observables is performed. The beam observables are finally assessed against PyHEADTAIL simulations to validate the impedance model.

\section{Transverse impedance model of the CERN PS}

The role of an impedance model~\cite{Salvant:IPAC19} is first to predict instability thresholds and ensure that the nominal parameters of the accelerator can be achieved. Second, it allows identifying the dominant impedance sources and quantifying their impact on observed instabilities. Thirdly, it provides the necessary input for macroparticle tracking simulations that include the interplay of beam-coupling impedance with other effects (e.g. betatron coupling, space charge, potential well distortion, beam-beam, feedback system, etc). 

The use of impedance models is well established in the study of collective effects in both particle colliders~\cite{Ishibashi:2024, Mounet:2012} and light sources~\cite{PhysRevAccelBeams.24.104801, Wang:2022tnt, Smaluk:2022tnt, Carver:2023xyy}.
Once an impedance model is deemed accurate, its Fourier transform -- the wake function -- can be used as input in macroparticle tracking codes (e.g. PyHEADTAIL, Xsuite~\cite{xsuite}, elegant~\cite{elegant}, mbtrack2~\cite{Gamelin:IPAC21-MOPAB070}) to simulate the turn-by-turn motion.
These simulations can reproduce impedance-induced beam observables such as the tune shift~\cite{Joly:IPAC23-WEPL149, Smaluk:2022tnt, Carver:IPAC21-MOPAB117}, closed-orbit distortion~\cite{Smaluk:2017, Marti:IPAC19-MOPGW066}, or instability growth rate~\cite{Joly:IPAC23-WEPL149, Zannini:IPAC15-MOPJE049, delaFuente:2024oro}. The validity of the impedance model is determined through its agreement with the beam-based measurements.

\subsection{PS transverse impedance model workflow and overview}

The PS transverse impedance model is generated through the following workflow. First, the Twiss parameters are computed at each relevant ring element using MADX~\cite{madx}. These parameters are used to calculate the beta weighting—defined as the ratio between the local beta function at an element and the average beta function around the ring, $\beta_{\mathrm{x,y}}^{\mathrm{element}} / \langle \beta_{\mathrm{x,y}} \rangle$—which is then passed to the impedance model handled by the XWakes code (formerly PyWIT)~\cite{pywit}.

The model includes the impedance contributions of the elements that have the greatest impact on the overall impedance. These contributions are obtained either from electromagnetic simulations (e.g., using CST Studio Suite~\cite{cst}), analytical formulas (e.g., IW2D~\cite{Mounet:2012}), or direct impedance measurements performed on a bench. Each contribution is scaled by its corresponding beta weighting and summed to obtain the total transverse impedance of the accelerator.

The resulting impedance model is then used to generate the total PS wake function from the computation of the Fourier transform of the impedance using the NEFFINT~\cite{neffint} algorithm. Finally, the wake function is used as an input in macroparticle tracking simulations to simulate the impedance-induced tune shift and instability growth rate, which will be discussed in a later section.

\begin{figure}[!htbp]
\centering
   \includegraphics[width=0.95\linewidth]{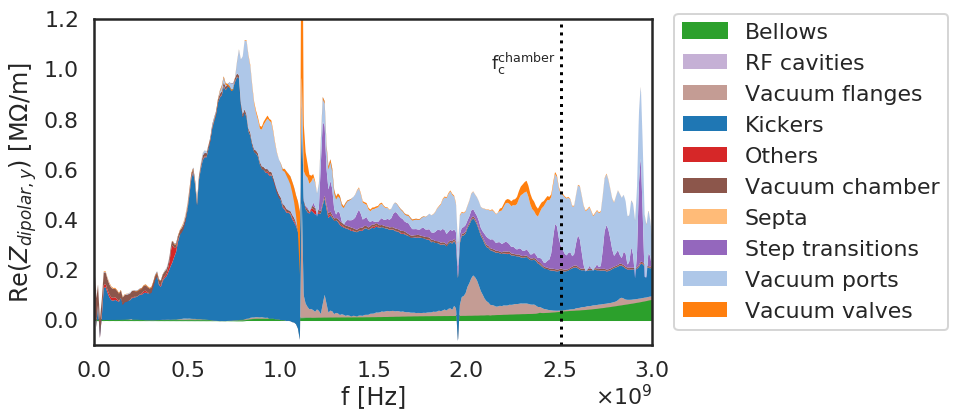}
 \hfill
   \includegraphics[width=0.95\linewidth]{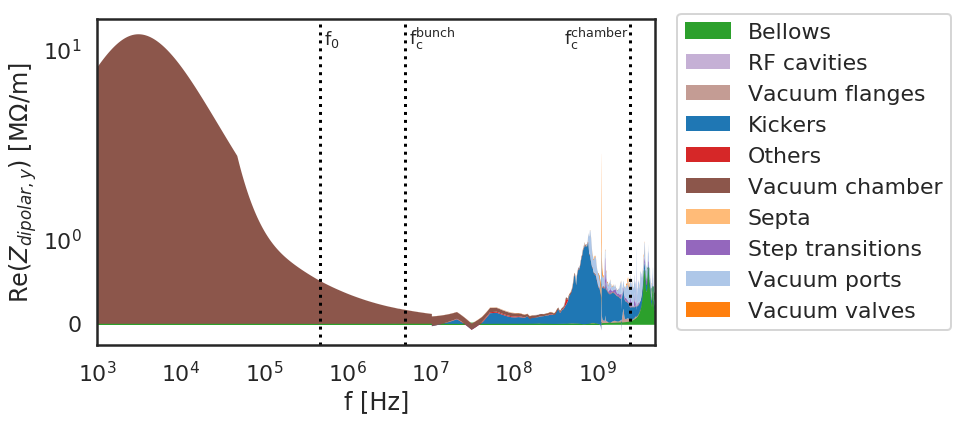}
\caption{Representation of the stacked real part of the vertical dipolar impedance contributions accounted for in the PS impedance model, at injection energy, in linear and logarithmic scales. The revolution frequency $f_0$, the frequency corresponding to the bunch length at injection $f_\mathrm{c}^\mathrm{bunch}$, and the vacuum chamber cutoff frequency $f_\mathrm{c}^\mathrm{chamber}$ are displayed as dashed black lines.}
\label{fig:zydip}
\end{figure}

The current PS transverse impedance model is the result of the work of several people over the years: L.~Ventura~\cite{Ventura:1625120}, S.~Persichelli~\cite{Persichelli:2015}, N.~Biancacci~\cite{Impedance_model}, B.~Popovic~\cite{popovic}.
In recent years, it was upgraded to its Post Long Shutdown 2 version, which is described in detail in~\cite{Joly:2024}. The PS impedance model is composed of a vacuum chamber made of Inconel alloy (30\%) and stainless steel (70\%) sections, kicker magnets for injection and extraction with their external circuits, accelerating cavities operating at 10, 40, 80, and~\SI{200}{\MHz}, a Finemet cavity for mitigating longitudinal instabilities, an electrostatic septum for extraction, vacuum instrumentation such as vacuum ports, bellows, valves, and metallic flanges, a stripline pick-up for tune acquisition, and two beam dumps. The individual contributions of these elements to the impedance model are illustrated in Fig.~\ref{fig:zydip}.

A complete layout of the machine and the specific elements taken into account in the impedance model can be found in~\cite{Impedance_website}. All the impedance components' stacked plots are shown in Appendix~\ref{app:A}.

According to Sacherer's formalism~\cite{Blewett:118362}, the effective impedance $Z_\mathrm{t}^\mathrm{eff}$ characterizes the interaction between a particle bunch and the impedance spectrum. This quantity is defined as the sum of the impedance overlapping with the mode power spectrum (offsetted by the chromatic frequency) along betatron lines:

\begin{equation}
    Z_\mathrm{t}^\mathrm{eff} = \frac{\sum_{p=-\infty}^{\infty} Z_\mathrm{t}(\omega')h_\mathrm{l}(\omega' - \omega_{\xi})}{\sum_{p=-\infty}^{\infty} h_\mathrm{l}(\omega' - \omega_{\xi})},
    \label{eq:Zeff}
\end{equation}
where $\omega$ is the angular frequency, $Z_\mathrm{t}(\omega)$ the transverse impedance spectrum, $h_\mathrm{l}$ the mode power spectrum of the azimuthal mode $l$, and $\omega_{\xi}$ the chromatic frequency. The betatron lines are sampled according to the following relation:
\begin{equation}
    \omega' = [(p + Q_0) + l Q_\mathrm{s}]\omega_0,
    \label{eq:betatron_lines}
\end{equation}
where $p$ is the spectral line number, $Q_0$ the unperturbed tune, $Q_\mathrm{s}$ the synchrotron tune, and $\omega_0$ is the angular revolution frequency.

Thus, $f_0$ indicates the density of the spectral lines in the bunch spectrum and $f_\mathrm{c}^\mathrm{bunch}$ the qualitative extent of the mode power spectrum ($l = 0$). The convention used for the frequency corresponding to the bunch length is $f_\mathrm{c}^\mathrm{bunch} = \frac{1}{\tau_\mathrm{b}} =$~\SI{5}{\MHz} assuming a Gaussian distribution of full bunch length at injection energy $\tau_\mathrm{b} =$~\SI{200}{\ns} (four times the standard deviation - this is the typical bunch length for LHC beams). Far above this value, the interaction of the bunch spectrum with the impedance spectrum is typically negligible.

Above the chamber cutoff frequency $f_\mathrm{c}^\mathrm{chamber}$, most of the EM fields generated by the bunch are no longer trapped by accelerator components and instead propagate into the surrounding vacuum chamber. Approximating the chamber as a circular waveguide with a radius equal to the smallest semi-axis of the vacuum chamber $r = \SI{35}{\mm}$, the cutoff frequency for the dominant TE$_{11}$ mode is $f_\mathrm{c}^\mathrm{chamber} = \frac{j'_{1,1} c}{2 \pi r} \approx \SI{2.5}{\GHz}$, where $j'_{1,1} \approx 1.84$ is the first zero of the derivative of the Bessel function of order one.

In the next section, we will study how the hardware modifications implemented during Long Shutdown 2 affected the PS transverse impedance model, with particular focus on the most significant change: the increased injection energy and its impact on the vacuum chamber impedance.

\subsection{Hardware Changes Introduced during the Long Shutdown 2}

One of the limitations in the PS before the LS2 and the LIU upgrade was the dominant effect of space charge at injection energy~\cite{Damerau:1976692}. Its impact translates into an incoherent tune shift depending on the particle's position in the bunch. In the presence of a strong space charge force, a significant emittance blow-up~\cite{Wasef:IPAC13-WEPEA070} caused by resonance crossing~\cite{Asvesta:2020qtd} was observed during the injection plateau.

Besides, the space charge tune spread~\cite{Lee} scales with the beam intensity and the inverse of the transverse emittance. Thus, doubling the beam intensity while preserving the transverse emittance could not be considered in this configuration. Yet, the space charge tune spread also scales inversely with the beam energy in $\beta^2 \gamma^3$. Consequently, increasing the injection energy can considerably reduce the detrimental impact of space charge~\cite{Damerau:1976692}. This solution was considered for the PS and the injection kinetic energy has been raised from 1.4 to~\SI{2}{\GeV}. The injection region of the accelerator was renovated during LS2 to accommodate this new energy range.

In addition, raising the injection energy also reduces the indirect space charge contribution to the vacuum chamber impedance. The indirect space charge impedance is defined as the contribution that would be there if the chamber boundaries were perfectly conducting~\cite{Salvant:2010dda}, it only depends on the vacuum chamber geometry and the beam energy. The indirect space charge contribution scales inversely in $\beta \gamma^2$ and thus vanishes for an ultra-relativistic beam. At injection energy, this contribution is dominant in the hundreds of kHz to the GHz range, and the new injection energy results in a 25\% reduction~\cite{Joly:IPAC21-WEPAB224, Joly:2024} of the impedance in this frequency range. The real part of the impedance is, however, left almost untouched with only a slight impedance increase~\cite{Joly:IPAC21-WEPAB224, Joly:2024} between~\SI {1}{\kHz} and~\SI{100}{\kHz}.

\begin{figure}[!htbp]
    \centering
    \includegraphics[width=0.9\linewidth]{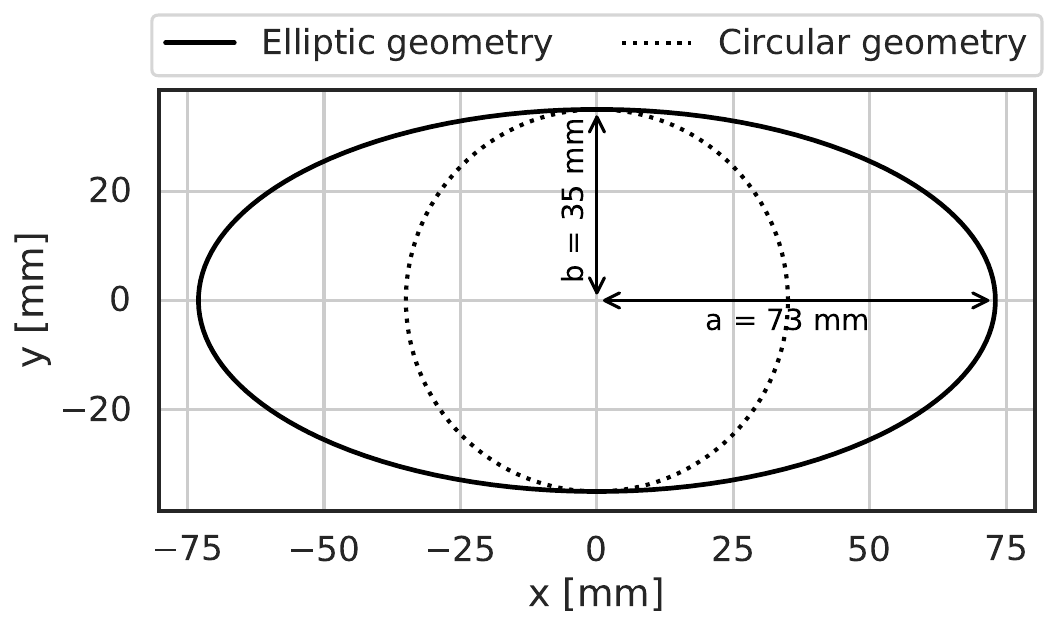}
    \caption{Comparison of the circular and elliptic vacuum chamber geometries used in impedance calculations.}
    \label{fig:PS_geometry}
\end{figure}

While the energy dependence of the vacuum chamber impedance is well understood, calculating this impedance for the PS elliptic vacuum chamber in the non-ultrarelativistic regime ($\beta \approx 0.95$) proves challenging. In this regime, the Yokoya factors~\cite{yokoya} do not apply and a different formalism~\cite{Migliorati:2705426, Joly:2024} relying on the Mathieu functions must be employed. Until then, the PS vacuum chamber was approximated as a circular chamber whose radius matches the smallest ellipse semi-axis. Now, it is accurately calculated for its elliptic cross-section as illustrated in Fig.~\ref{fig:PS_geometry}.

While the vertical dipolar impedance remains unchanged between both geometries, the horizontal dipolar impedance is reduced by approximately a factor of two in the elliptic geometry. The calculations for the elliptic geometry also properly account for the quadrupolar impedance, which was negligible in the circular geometry. At high frequencies where the Yokoya factors' validity range is exceeded~\cite{Joly:IPAC21-WEPAB224}, the impedance shows a steeper drop than predicted by the Yokoya factors. The frequency behavior and energy dependence of the real part of the impedance remain otherwise consistent between both geometries.

In addition to the vacuum chamber impedance, LS2 introduced several other changes to the PS impedance model, including the removal of the Continuous Transfer (CT) equipment~\cite{Koukovini-Platia:2640483, Joly:IPAC21-WEPAB224}, the installation of new beam dumps, RF equipment, and beam instrumentation. Further details can be found in~\cite{Joly:2024}.


Based on the updated PS impedance model, macroparticle tracking simulations can be performed to predict impedance-induced beam observables, in particular the impedance-induced tune shift and instability growth rate.

\section{Beam-based validation of the PS transverse impedance model}

Two single-bunch beam observables of interest, caused by the presence of transverse impedance, are the instability growth rate and the impedance-induced tune shift, both affecting the centroid motion in the following way:

\begin{equation*}
  \begin{split}
    x(t) & \propto e^{j 2 \pi f_0 (Q_{0} + \Delta Q)t} \\
    & \propto e^{j 2 \pi f_0 \left[Q_{0} + \Re{(\Delta Q)}\right] t} e^{- 2 \pi f_0 \Im{(\Delta Q)} t}
  \end{split}
\label{eq:effect_dQ}
\end{equation*}

Where $\Delta Q$ is the complex tune shift, $Q_0$ is the unperturbed tune, and $f_0$ is the revolution frequency. The real part of the impedance induces the instability growth rate ($2 \pi f_0 \Im{(\Delta Q)}$), which leads to an exponential increase of the bunch centroid turn after turn. The imaginary part generates a tune shift, resulting in an intensity-dependent shift of the coherent tune.

To study these effects, macroparticle tracking codes can be used, such as PyHEADTAIL, Xsuite, elegant, mbtrack2, etc. Here, PyHEADTAIL is being used as it enables the study of the interplay between wakefields, chromaticity, and space charge forces, which will be examined in the following sections.

All the parameters used in the PyHEADTAIL simulations were measured during the reference measurement campaigns and are listed in Appendix~\ref{app:B}. The injection energy cycle was based on a Neutron Time-of-Flight (nTOF) cycle, whereas the extraction energy one was based on an East Area experimental facility (EAST) cycle.

While simulations provide nearly ideal signals, beam-based measurements face several real-world limitations. These include limited equipment resolution, statistical and electronic noise, and various sources of non-linearities that cause tune spread. Consequently, methods to enhance the signal quality are essential to fully exploit the measured data. The following methods were used with all the beam-based measurements and the simulations presented in this paper.

\subsection{Tune Acquisition Method}

The measurement of impedance-induced tune shift requires a sufficient accuracy to allow the observation of its variation with beam intensity in the presence of tune spread. To achieve this precision, an external excitation is employed to enhance the frequency spectrum of coherent beam oscillations. The optimal approach uses a chirp signal, a sinusoidal excitation whose frequency varies linearly from $f_{min}$ to $f_{max}$ throughout the signal acquisition period. The excitation is produced by the PS feedback system.

In addition to exciting the beam, the frequency spectrum is calculated by using the Numerical Analysis of Fundamental Frequencies (NAFF) method~\cite{naff}, as implemented in PyNAFF~\cite{pynaff}. While an FFT spectrum remains useful to qualitatively observe the spectral content of a signal, its accuracy can be outperformed by NAFF algorithms, which allow us to determine the tune with the highest accuracy while requiring smaller sample sizes than an FFT.

\subsection{Instability Growth Rate Acquisition Method}

When an instability develops, the transverse beam amplitude exhibits exponential growth with respect to the number of turns. To quantify this growth, we first extract the oscillation envelope using a Hilbert transform of the signal. The (exponential) growth rate is then determined by applying a modified Least-Squares fitting technique~\cite{LS_exp}, which optimizes the parameters of an exponential function to minimize the sum of squared residuals between the fitted curve and the measured envelope.

Different approaches to measure and analyze the tune and instability growth rates are described in~\cite{Joly:2024}.

\section{Effect of chromaticity (linear and non-linear)}

The chromaticity is a crucial mechanism in the study of the impact of transverse impedance on beam dynamics as it couples the transverse and longitudinal planes. It gives rise to a tune shift experienced by off-momentum particles, which can be written as:

\begin{equation*}
    \Delta Q_\xi = Q' \delta + Q'' \delta^2,
\end{equation*}

where $\delta$ is the relative particle momentum deviation with respect to the synchronous particle, $Q'$ is the first-order chromaticity, and $Q''$ is the second-order chromaticity. The first-order chromaticity is responsible for the head-tail phase shift $\chi$~\cite{sands1969head, Cocq:370134}, which may allow the amplification of the impact of wakefields on a bunch after each synchrotron period and eventually lead to head-tail instabilities. The head-tail phase shift is usually expressed through the chromatic frequency $f_\xi$, defined as: 

\begin{equation}
    f_{\xi} = \chi / \tau_\mathrm{b} = \frac{f_0 Q'}{|\eta|},
    \label{eq:fxi}
\end{equation}

where $\tau_\mathrm{b}$ is the full bunch length and $\eta$ the slippage factor.
The chromatic frequency intervenes in Eq.~\eqref{eq:Zeff} under the form of $\omega_{\xi} = 2 \ pi f_{\xi}$ and causes a shift of the mode spectrum over the impedance spectrum. As a result, scanning the chromatic shift frequency by modifying the first-order chromaticity can be leveraged to probe the accelerator's effective impedance over different frequency ranges.

In the PS, the working point of particle beams with kinetic energy below~\SI{3.5}{\GeV} is primarily controlled using combined function magnets and additional quadrupoles distributed along the ring~\cite{Huschauer:2194332}. Above this energy, these elements become insufficient due to their limited number and gradient. Instead, Pole Face Windings (PFW) and Figure of Eight Loops (F8L) are used for working point control, though this comes with the drawback of introducing non-linear chromaticity~\cite{Huschauer_WP}.
Still, the chromaticity can be controlled to some extent using a measured response matrix that relates the currents in the PFW/F8L (5 degrees of freedom) to the tunes $Q_\mathrm{x,y}$, first-order chromaticities $Q'_\mathrm{x,y}$, and either horizontal or vertical second-order chromaticity $Q''_\mathrm{x}$ or $Q''_\mathrm{y}$. However, since the chromaticity control knobs was primarily designed for correction rather than wide-range scanning, the achievable chromaticity range during measurements is sometimes limited.

To understand these operational boundaries, we measured the achievable chromaticity ranges at both injection ($E_\mathrm{kin} =$~\SI{2.0}{\GeV}, under transition crossing) and extraction ($E_\mathrm{kin} =$~\SI{25.4}{\GeV}, above transition crossing) energies. At injection, we could vary first-order chromaticity between $Q'$ = 0 and -9, allowing to explore the impedance spectrum from 0 to~\SI{80}{\MHz}. While extraction energy measurements had a narrower chromaticity range ($Q'$ = 0 to 1.5), the smaller slippage factor still enabled the exploration of the impedance spectrum up to~\SI{60}{\MHz}. In addition, the bunch spectrum at extraction energy spans a wider frequency range than at injection energy due to the shorter bunch length, resulting from non-adiabatic damping during acceleration.

As it can be noted from the stacked plots shown in Appendix~\ref{app:A}, the 0 to~\SI{100}{\MHz} frequency range of the PS impedance is dominated by the vacuum chamber impedance. Moreover, the imaginary part of the impedance remains almost constant with frequency in this frequency range, except for the peak at low frequencies. Thus, independently of the chromatic shift, the tune shift is anticipated to be mainly affected by the proximity to the low frequency peak introduced by the vacuum chamber impedance. Consequently, $Q'$ is expected to have a minimal impact on the tune shifts once the bunch spectrum is sufficiently far from the peak.

\begin{figure}[!htbp]
\centering
  \includegraphics[width=\linewidth]{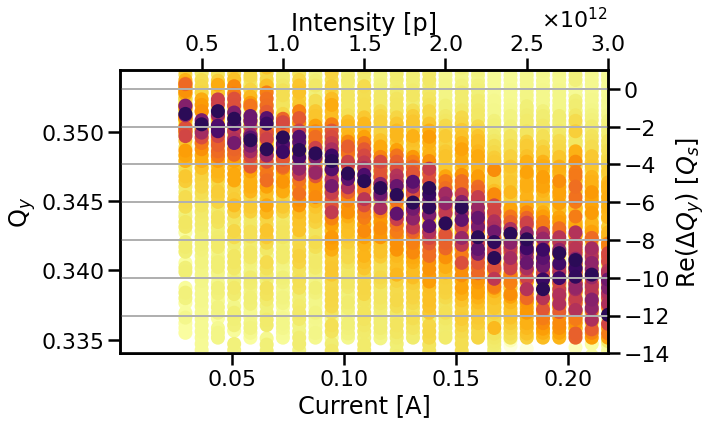}
  \includegraphics[width=\linewidth]{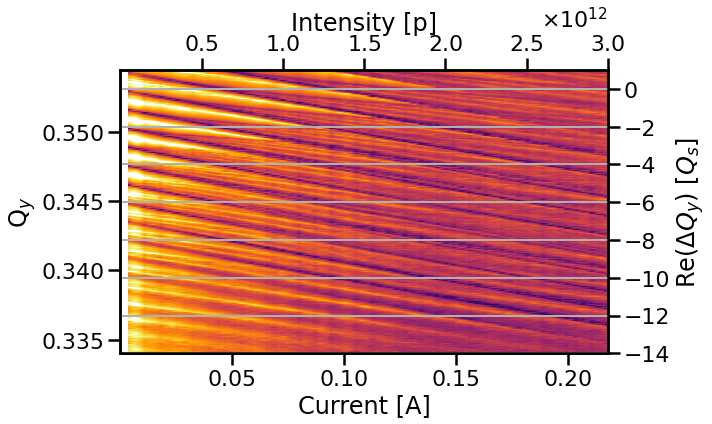}
\caption{Vertical tune shift measurements performed in 2022 (top) and simulations (bottom) at $E_{\mathrm{kinetic}} =$~\SI{2}{\GeV} for $Q'_\mathrm{y} \approx -7$. The color scale represents the tune magnitude, from high (purple) to low (yellow).}
\label{fig:dQy_FB_measurements_sim}
\end{figure}

The observation of the main mode tune shift has been proven challenging in simulations, especially for the injection energy case. All the azimuthal and radial modes exhibit a comparable magnitude, hence hiding the main mode, as visible in Fig.~\ref{fig:dQy_FB_measurements_sim}, for simulations including a negative first-order chromaticity, without non-linearities or space charge effects. In comparison, the main mode is well noticeable in measurements and its magnitude surpasses by orders of magnitude the one of the other modes.

Despite this effect, the main mode remains present in the simulations and can be observed for low chromaticities (i.e. $Q'_\mathrm{y} \geq -1$). As the focus is put on the impedance-induced tune shift, a workaround can be employed to increase the main mode magnitude. To do so, a positive $Q'$ is used, as a positive chromaticity below transition results in an unstable mode 0, thus further exciting the main mode peak in the tune shift spectrum. Then, it is known that the imaginary part of the transverse impedance exhibits even parity as a function of frequency~\cite{chao1993physics}. As a result, the overlap of the imaginary part of the impedance with the bunch power spectrum frequency (shifted by $f_\xi$) is independent of the chromaticity sign.

\begin{figure}[!htbp]
\centering
  \includegraphics[width=\linewidth]{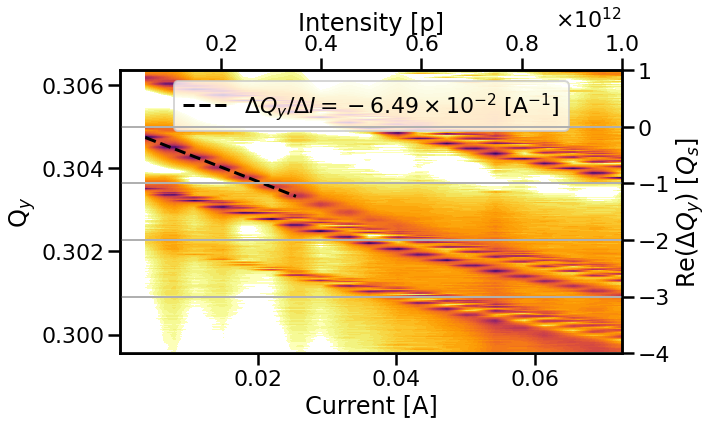}
  \includegraphics[width=\linewidth]{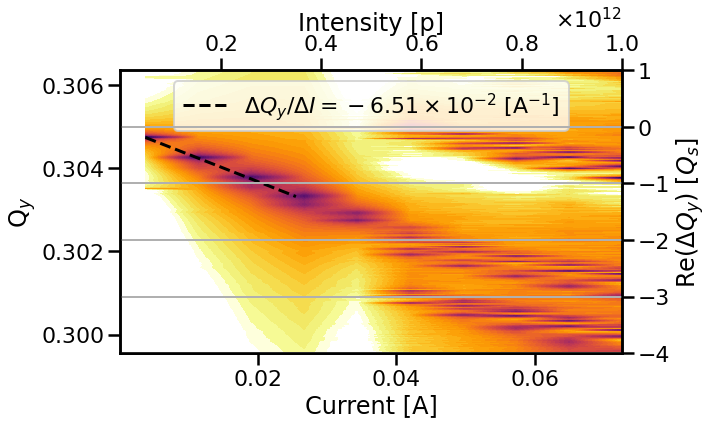}
\caption{Vertical tune shift for $Q'_\mathrm{y} = -1$ (top) and $Q'_\mathrm{y} = 1$ (bottom), from simulations at injection energy. The color scale represents the tune magnitude, from high (purple) to low (yellow).}
\label{fig:dQ_Qp_FB_artifice}
\end{figure}

The method's validity was assessed by using $Q'_\mathrm{y} = -1$ and 1 at injection energy, where the main mode could be observed in both cases, as depicted in Fig.~\ref{fig:dQ_Qp_FB_artifice}. By comparing both chromaticities, comparable tune shift slopes were found, as well as a significant increase of the main mode magnitude for intensities below~\SI{5e11}{p} with a positive chromaticity, compared to a negative one.
Based on this result, the method is deemed valid and has been used in simulations to estimate the tune shift slopes for the various chromaticities probed during the measurement campaigns. 

To fully understand the role of chromaticity on the impedance-induced tune shift, we analyze its effects systematically. We first examine how $Q'$ influences the measured and simulated tune shift slopes. Then, we investigate the impact of $Q''$, which proved crucial to interpret a discrepancy between measurements and simulations.

\subsection{First-order chromaticity}

\begin{figure}[!htbp]
\centering
  \includegraphics[width=0.85\linewidth]{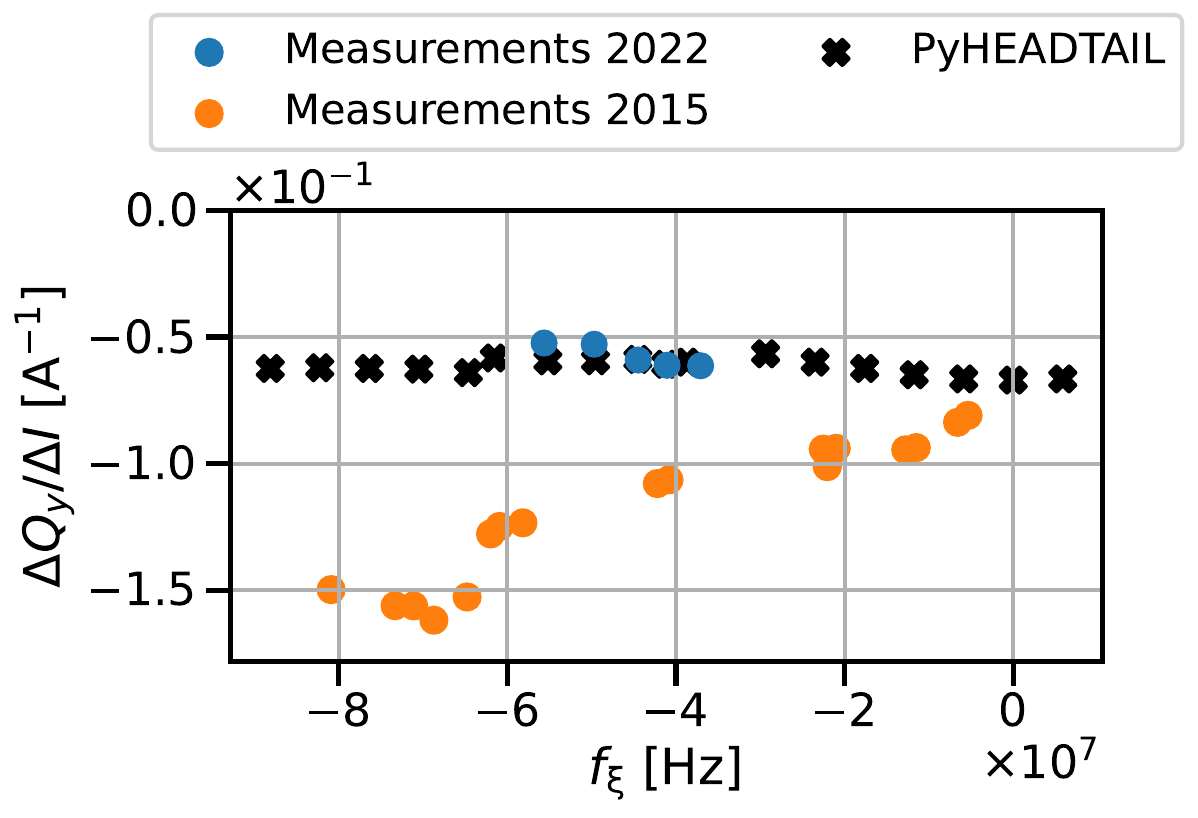}
  \includegraphics[width=0.85\linewidth]{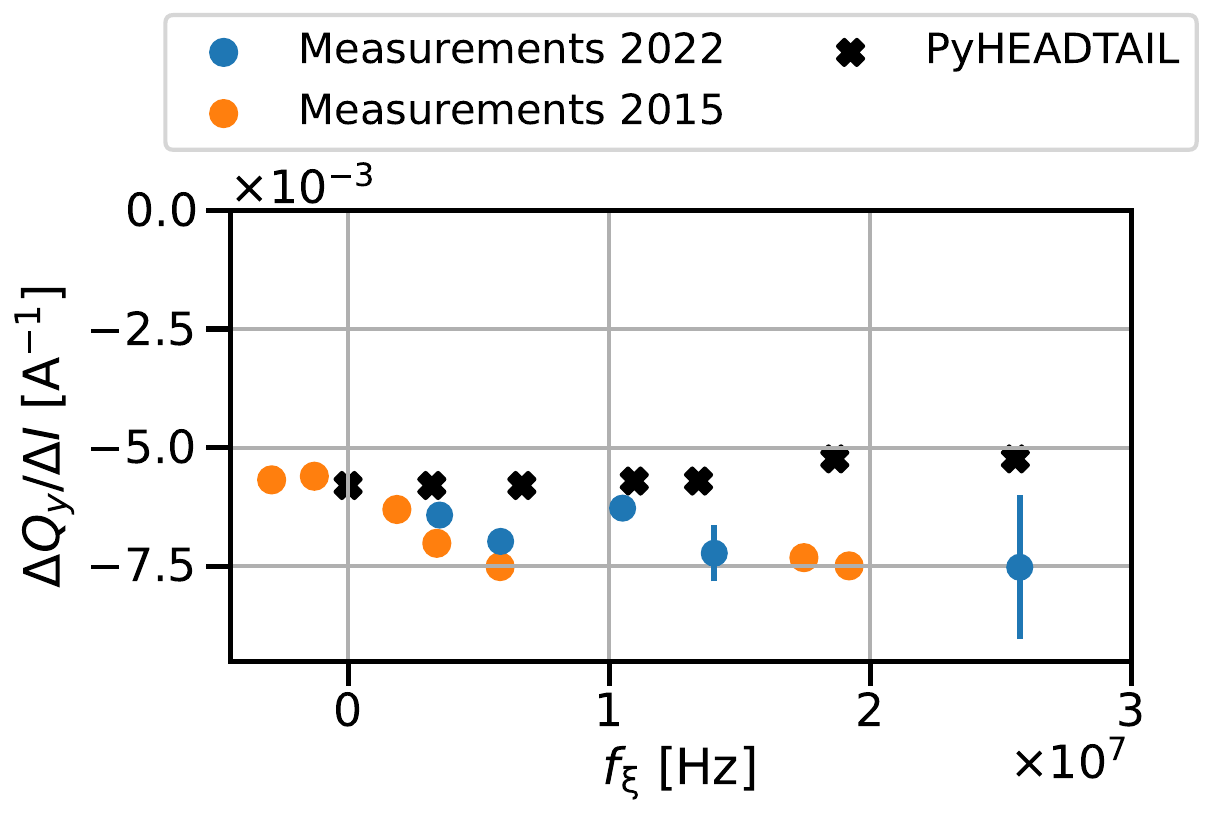}
\caption{Comparisons between measured and simulated vertical tune shift slopes against chromatic frequency $f_\xi$ at injection (top) and extraction (bottom) energies.}
\label{fig:dQ_Qp_FT}
\end{figure}

Making use of the method described above to observe the simulated main mode tune shift for $|Q'_\mathrm{y}| \geq 1$, we now compare the measured and simulated tune shift slopes across different $Q'$ values. Fig.~\ref{fig:dQ_Qp_FT} shows the vertical tune shift slopes obtained at both injection and extraction energies for various chromaticities. The simulation parameters are detailed in Appendix~\ref{app:B}, and all tune shift slopes are plotted against the chromatic frequency defined in Eq.~\eqref{eq:fxi}. The figure shows results until $f_\xi = \SI{30}{\MHz}$, as simulations above this frequency return the tune of a higher-order mode (mode with a -1 azimuthal mode number) instead of the main mode.

First, the tune shift slopes show a clear beam energy dependence: at injection energy, they are approximately one order of magnitude larger than at extraction energy, regardless of chromaticity. This difference originates from the reduced indirect space charge impedance contribution at higher beam energy.

Then, it can be observed that the 2015 and 2022 measurements exhibit different behaviors with $f_\xi$ at injection energy. The 2015 data shows significant variation in tune shift slopes with increasing chromaticity, while the 2022 measurements remain more stable. Additionally, while simulations agree well with 2022 measurements at injection energy, they underestimate the tune shift slopes at extraction energy for both 2015 and 2022 data. A similar discrepancy was observed in~\cite{abp_injectors_day} for non-zero chromaticity values.

Since the PS transverse impedance remained largely unchanged between 2015 and 2022, these observations suggest a missing mechanism in our simulations. A plausible candidate to explain these discrepancies is the second-order chromaticity $Q''$.

\subsection{Second-order chromaticity}

\begin{figure}[!htbp]
\centering
  \includegraphics[width=\linewidth]{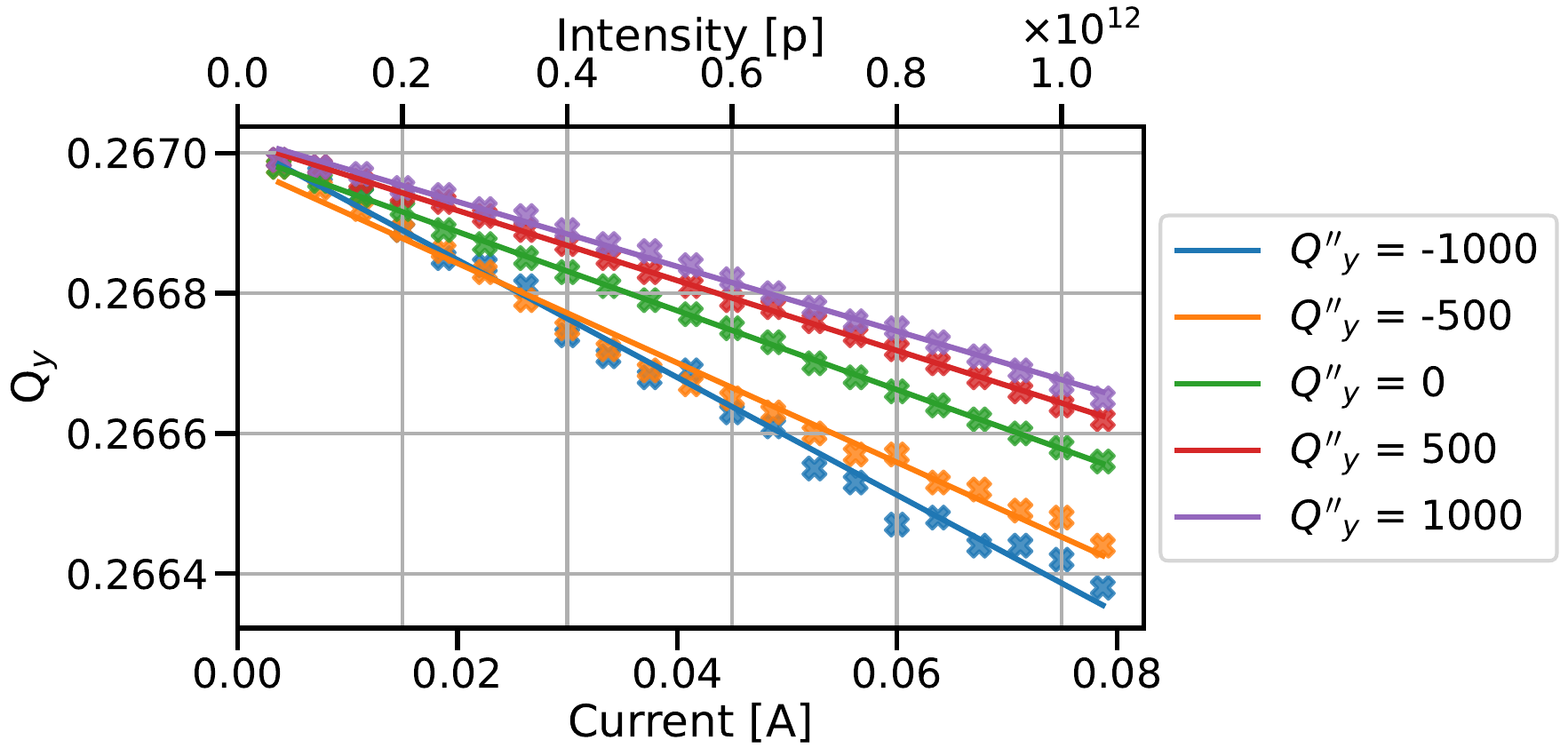}
\caption{Impact of second-order chromaticity on simulated vertical tune for various $Q''$ at extraction energy. Crosses represent the tune values used for the fit by a single tune shift slope (shown in solid lines).}
\label{fig:impact_Qpp_dQ_slopes}
\end{figure}

To investigate whether second-order chromaticity could explain the observed discrepancies, we performed additional PyHEADTAIL simulations with the extraction energy scenario, where the largest discrepancy between measurements and simulations was observed. A parameter scan with $Q' = 0$ while varying $Q''$ between $\pm 1000$ - a range achievable in the PS - was performed. The resulting simulated tune shifts are shown in Fig.~\ref{fig:impact_Qpp_dQ_slopes}.

\begin{figure}[!htbp]
\centering
  \includegraphics[width=\linewidth]{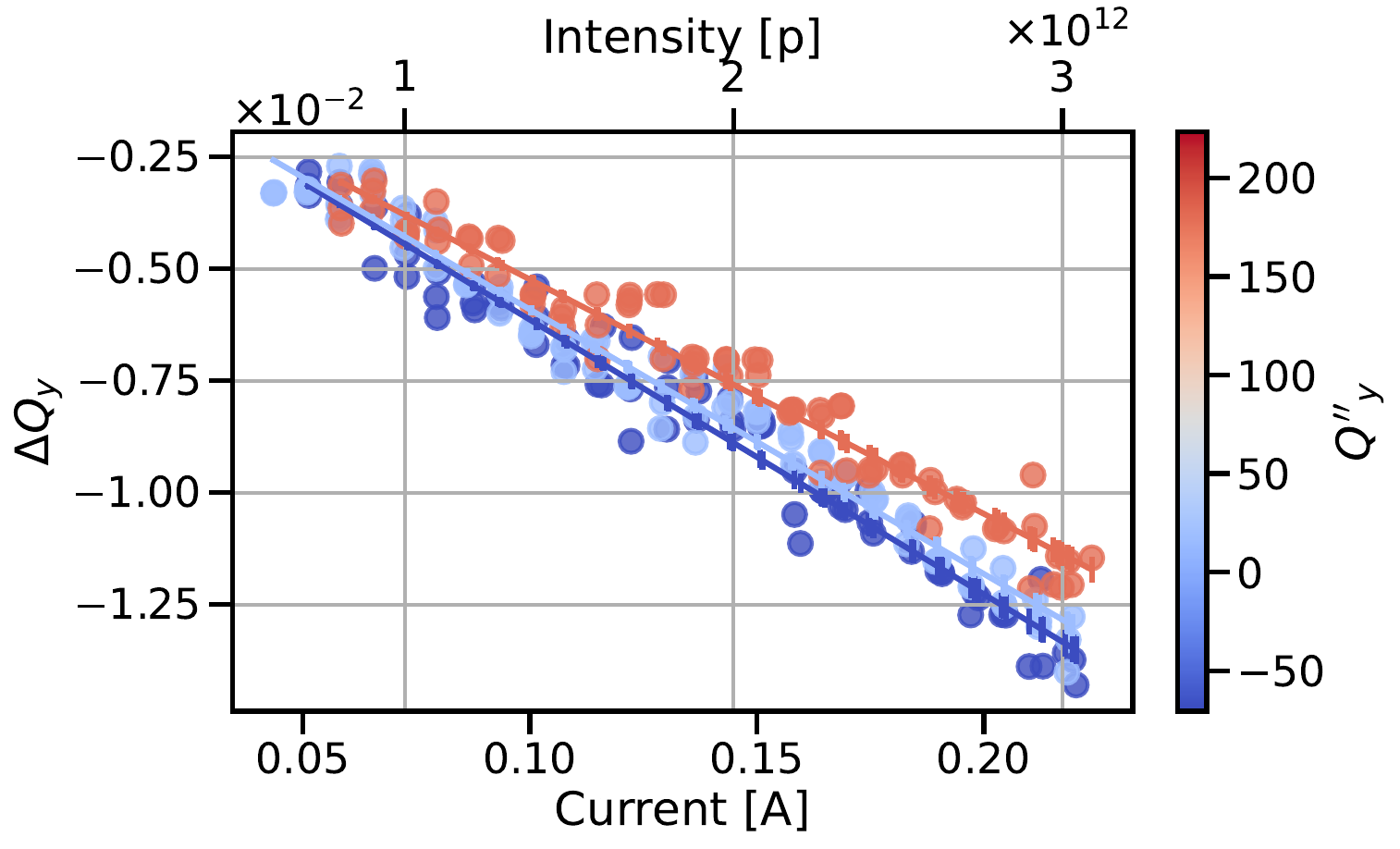}
  \includegraphics[width=\linewidth]{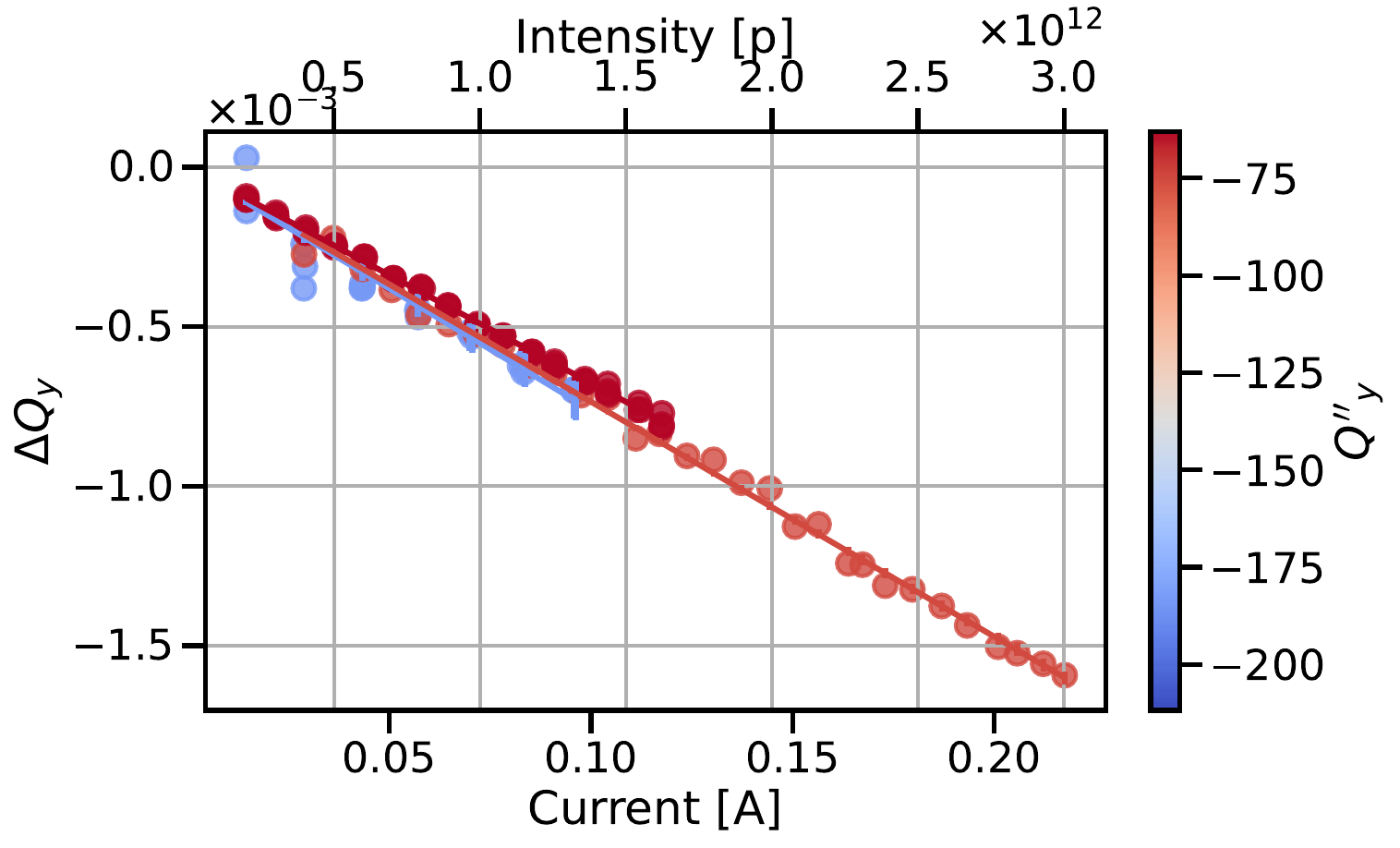}
\caption{Vertical tune shift slope fit with measured tunes at injection (top) and extraction energy (bottom) versus second-order chromaticity.}
\label{fig:dQy_Qpp_meas_slopes}
\end{figure}

A significant dependency of the main mode tune shift with $Q''$ is found in both planes. It can be observed that a negative $Q''$ increases the tune shift slope and a positive one decreases it. The same observation was made in the LHC, where they studied the generation of transverse Landau damping through the introduction of $Q''$~\cite{Schenk:2018pae}.
A chromaticity value of $Q''_\mathrm{y} = -500$ can be responsible for an increase of the tune shift slope by 30\%. Although initially observed in simulations at extraction energy, this correlation between $Q''$ and tune shift slopes extends to the injection energy case as well. Additionally, measurements at both injection and extraction energies, shown in Fig.~\ref{fig:dQy_Qpp_meas_slopes}, also confirm this behavior.

\begin{figure}[!htbp]
\centering
  \includegraphics[width=\linewidth]{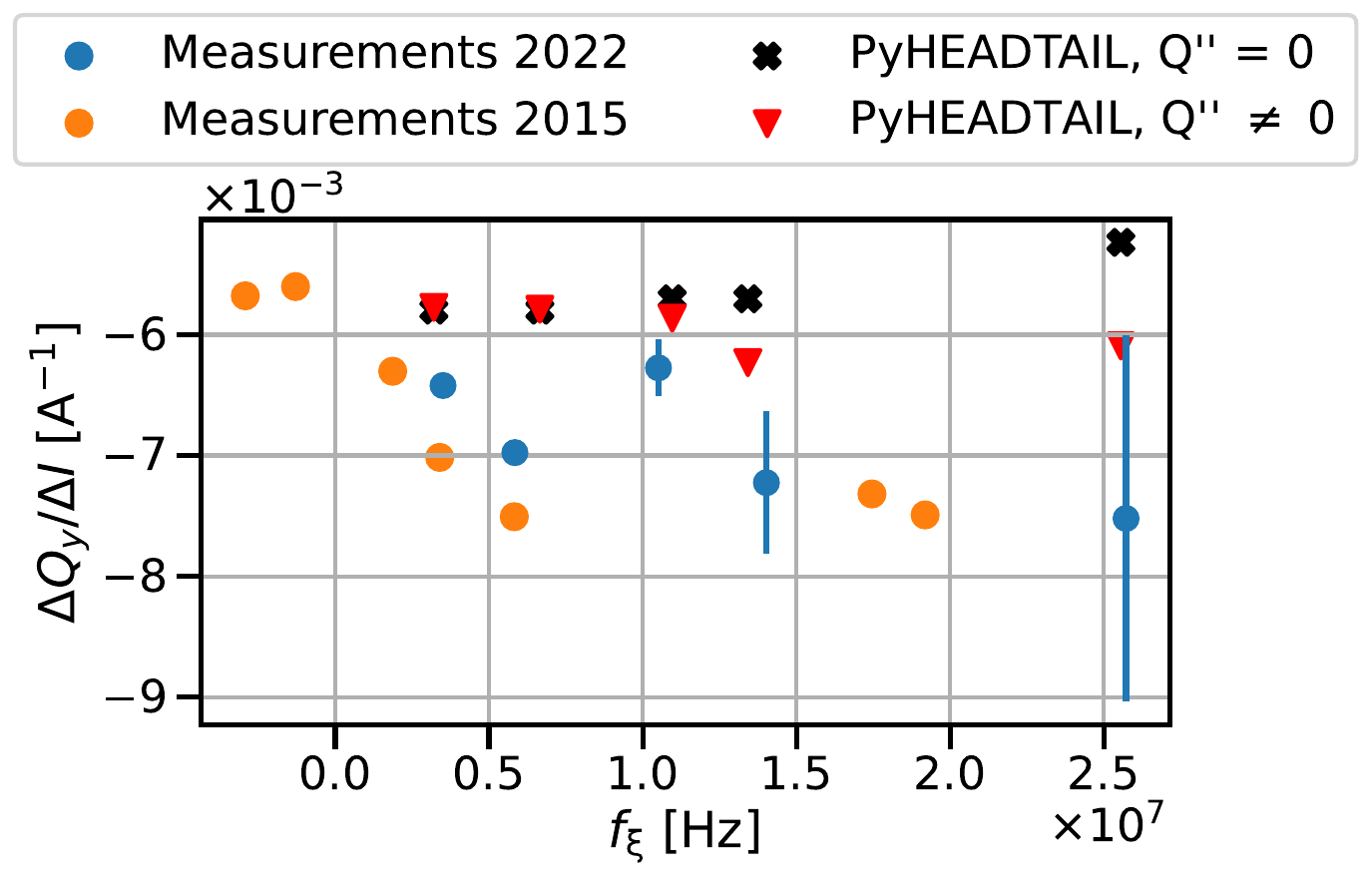}
\caption{Comparisons between measured and simulated vertical tune shift slopes against chromatic frequency $f_\xi$, including the effect of $Q''$.}
\label{fig:dQ_Qp_FT2}
\end{figure}

By including the second-order chromaticity measured during measurements in the simulations, we can refine the comparison with measurements at extraction energy, shown in Fig.~\ref{fig:dQ_Qp_FT}. The updated comparison is presented in Fig.~\ref{fig:dQ_Qp_FT2}, where both the simulated tune shift slopes with and without $Q''$ are plotted.

During these measurements, a correlation has been observed between $Q'$ and $Q''$. In other words, an acquisition exhibiting a low $Q'$ also shows a low $Q''$. Therefore, the simulated tune shift slope for low chromatic frequency values is similar whether $Q''$ is accounted for or not. In any case, the discrepancy between measurements and simulations has been reduced for large chromatic frequencies. Nonetheless, $Q''$ is not sufficient to fully explain it. To address this, we now focus on two mechanisms still missing from the simulations -- the octupolar order non-linearities and the space charge force.

\section{Effect of octupolar order non-linearities and direct space charge}

While second-order chromaticity explains some of the observed discrepancies, other mechanisms also have an important interplay with impedance effects. As highlighted in Fig.~\ref{fig:dQy_FB_measurements_sim}, the simulated main mode cannot be observed similarly to the measured one over a wide range of beam intensities. It remains visible for intensities lower than~\SI{5e11}{p} before being masked by higher-order modes. The simulations differ from the measurements by the absence of Landau damping from octupolar order non-linearities and direct space charge. 

The strength of the octupolar order non-linearities is quantified by the anharmonicity (or detuning) coefficients~\cite{Berg:318826} $a_\mathrm{xx}, a_\mathrm{xy}, a_\mathrm{yy}$ which result in an incoherent tune shift $Q^\mathrm{AD}_\mathrm{x,y}$ depending on the particle's transverse actions $J_\mathrm{x,y}$ as:

\begin{equation*}
  \begin{split}
    \Delta Q^\mathrm{AD}_\mathrm{x} = a_{\mathrm{xx}} . J_\mathrm{x} + a_{\mathrm{xy}} . J_\mathrm{y}, \\
    \Delta Q^\mathrm{AD}_\mathrm{y} = a_{\mathrm{yy}} . J_\mathrm{y} + a_{\mathrm{xy}} . J_\mathrm{x}.
  \end{split}
\end{equation*}

The anharmonicity coefficient values are obtained from MADX results.
At injection energy, the anharmonicity coefficients are $a_\mathrm{xx} = 6.11$, $a_\mathrm{xy} = -53.7$, $a_\mathrm{yy} = 32.6$. They give rise to a tune spread of $\Delta Q^\mathrm{AD}_\mathrm{x} \sim \SI{1e-4}{}$ and $\Delta Q^\mathrm{AD}_\mathrm{y} \sim \SI{1e-3}{}$ for a beam intensity of $\SI{1e12}{p}$. It can be noted that a tune spread of a similar order of magnitude can be reached with second-order chromaticity with values of $Q''_\mathrm{x} = 2.5$  and $Q''_\mathrm{y} = 25$.
The anharmonicity coefficients for the extraction energy cycles used during simulations can be found in Appendix~\ref{app:B}.

The direct space charge force is introduced in the simulations by using a 3D Particle In Cell (PIC) solver~\cite{Oeftiger:318826} to compute the interactions between particles. Its principle lies in the overlap of the bunch distribution with a 3D grid in which the Poisson equation is solved cell by cell, and the resulting change of momentum is propagated to all particles of a cell.

Both effects are closely tied to the beta functions and dispersion variations along the ring. Thus, the simulations require the accelerator to be divided into a sufficient number of segments to realistically model these interactions. It is handled by PyHEADTAIL by using the Floquet matrix defined in Eq.~\eqref{eq:floquet} for the \emph{i}th macroparticle in the \emph{n}th segment.

\begin{equation}
    \begin{aligned}
    \begin{pmatrix}
    u_{i,n+1} \\
    u'_{i,n+1}
    \end{pmatrix}
    &=
    \begin{pmatrix}
    \cos(2\pi \mu_{i,n}) & \beta \sin(2\pi \mu_{i,n}) \\
    - \frac{1}{\beta} \sin(2\pi \mu_{i,n}) & \cos(2\pi \mu_{i,n})
    \end{pmatrix}
    \begin{pmatrix}
    u_{i,n} \\
    u'_{i,n}
    \end{pmatrix} \\
    &+
    \begin{pmatrix}
        \eta \left(1 - \cos(2\pi \mu_{i,n}) \right) \\
        \frac{\eta}{\beta} \sin(2\pi \mu_{i,n})
    \end{pmatrix}
    \end{aligned}
\label{eq:floquet}
\end{equation}
Here $u$ and $u'$ represent the horizontal or vertical position and angle, $\beta$ is the beta function, $\mu_{i,n}$ is the betatron phase advance and $\eta$ is the dispersion. The impact of the chromaticity and anharmonicity coefficients is accounted for through $\mu_{i,n}$.
The convergence tests on the required number of segments and PIC solver parameters can be found in~\cite{Joly:2024}. The parameters used by PyHEADTAIL simulations, including space charge, are summarized in Table~\ref{tab:PIC_parameters}.

\begin{table}[!htbp]
\begin{ruledtabular}
    \caption{Parameters for the PyHEADTAIL tune shift simulations with space charge}
    \begin{tabular}{ccc}
         $n_\mathrm{macroparticles}$ & $n^\mathrm{cell}_\perp$ & Transverse grid span  \\
         \hline
         \SI{3e6}{} & 128 & $\pm 5 \sigma_\mathrm{x,y}$ \\
         \hline
         $n_\mathrm{segments}$ & $n^\mathrm{cell}_\mathrm{\parallel}$ & Longitudinal grid span \\
         \hline
         240 & 32 & $\pm 4 \sigma_\mathrm{z}$ \\
    \end{tabular}
    \label{tab:PIC_parameters}
\end{ruledtabular}
\end{table}

Compared to previous simulations, the Landau damping will change for each intensity due to the intensity-dependent transverse emittance in the PS and the scaling of the space charge force with intensity. The intensity dependency of the transverse emittance follows an affine function and is referenced in Appendix~\ref{app:B}. As intensity increases, so does the emittance, thus the Landau damping from the octupolar order non-linearities increases ($\propto \varepsilon$) while the one from the direct space charge slightly increases ($\propto N / \varepsilon$), as the brightness asymptotically increases towards its high-intensity value. The overall Landau damping thus increases with the beam intensity, especially for headtail modes. The main mode remains unaffected by direct space charge as its effect is invariant from the bunch position in the vacuum chamber~\cite{Burov:318826}, but still gets more stable with higher octupolar order non-linearities at high intensities.

\begin{figure}[!htbp]
\centering
  \includegraphics[width=\linewidth]{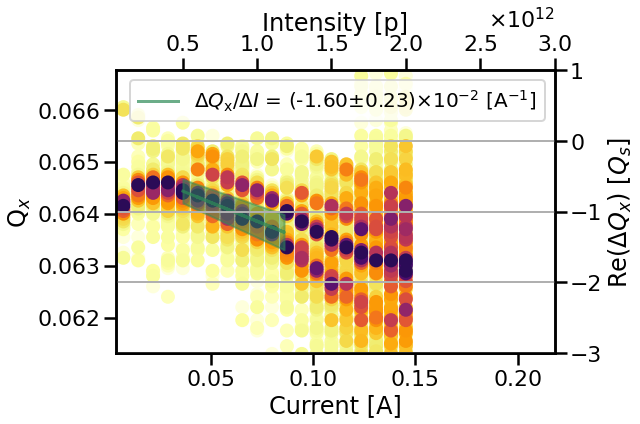}
  \includegraphics[width=\linewidth]{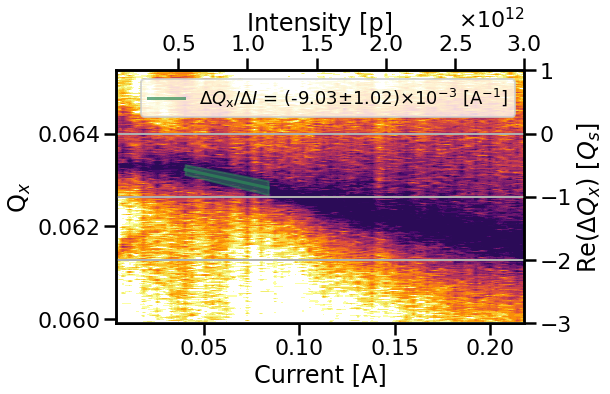}
  \includegraphics[width=\linewidth]{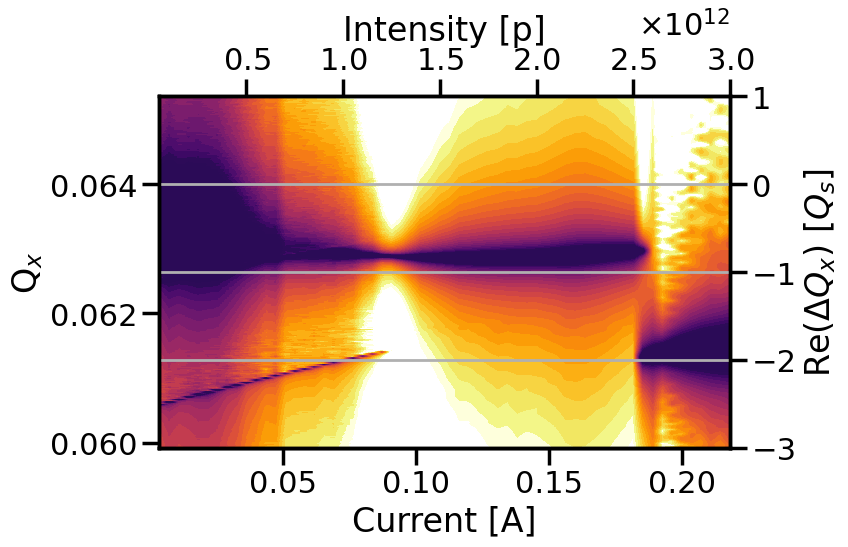}
\caption{Horizontal tune shift against beam intensity at injection energy: measurements (top), simulations with space charge (middle) and simulations without space charge (bottom). All simulations include first and second-order chromaticity and octupolar order non-linearities. The color scale represents the tune magnitude, from high (purple) to low (yellow). The fitted tune shift slope and its confidence interval are represented by a green line and envelope.}
\label{fig:FB_dQx_SC}
\end{figure}

\begin{figure}[!htbp]
\centering
  \includegraphics[width=\linewidth]{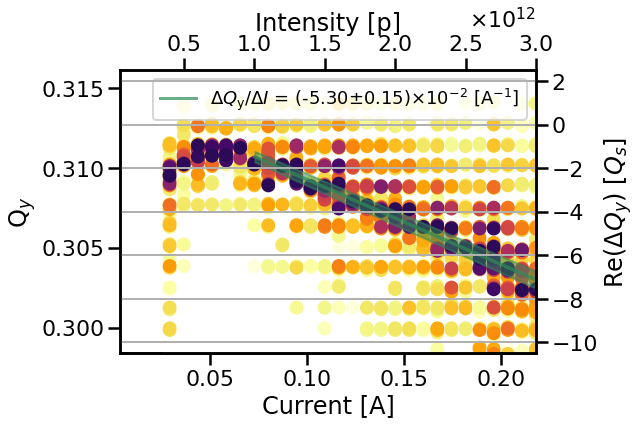}
  \includegraphics[width=\linewidth]{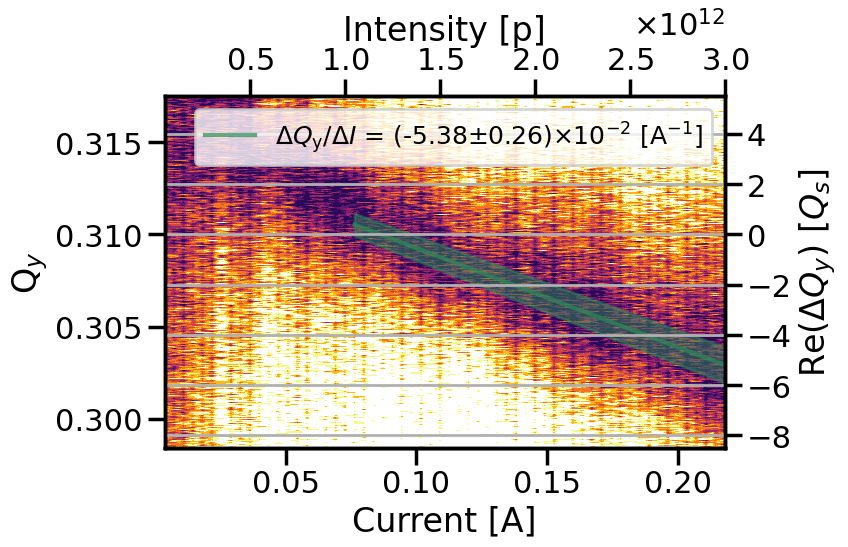}
  \includegraphics[width=\linewidth]{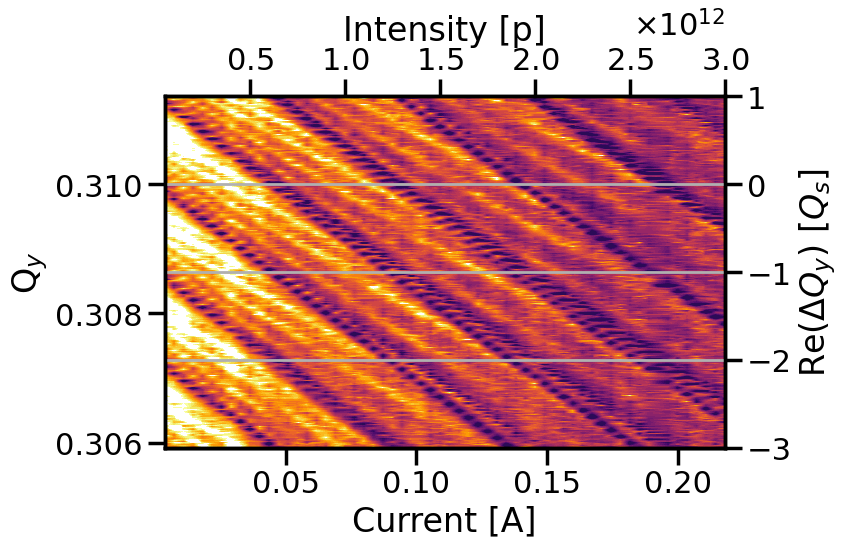}
  \caption{Vertical tune shift against beam intensity at injection energy: measurements (top), simulations with space charge (middle) and simulations without space charge (bottom). All simulations include first and second-order chromaticity and octupolar order non-linearities. The color scale represents the tune magnitude, from high (purple) to low (yellow). The fitted tune shift slope and its confidence interval are represented by a green line and envelope.}
\label{fig:FB_dQy_SC}
\end{figure}

To study this, PyHEADTAIL simulations were performed with all the relevant mechanisms: space charge, octupolar order non-linearities, first and second-order chromaticity, and non-linear synchrotron motion. The simulation parameters originate from the measurements to which they are compared. Two different acquisitions with different PFW settings were used for each plane. The measurements exhibit a horizontal chromaticity of $Q'_\mathrm{x} = 0$, $Q''_\mathrm{x} = -236$ and a vertical chromaticity of $Q'_\mathrm{y} = -9.4$, $Q''_\mathrm{x} = 178$.
Both sets of measurements are compared with simulations in Figs.~\ref{fig:FB_dQx_SC} and~\ref{fig:FB_dQy_SC}. The synchrotron sidebands are displayed on an additional axis, where the zeroth sideband is placed at the intercept of the tune shift slope with the vertical axis.

Firstly, it can be noted in Figs.~\ref{fig:FB_dQx_SC} and~\ref{fig:FB_dQy_SC} that the measured unperturbed tune differs from the simulated one. While simulations use the working point ($Q_\mathrm{x}, Q_\mathrm{y} = 6.064, 6.317$) obtained from the main mode tune measured at the lowest intensity, the measured main mode exhibits non-linear behavior at low intensities, leading to a different unperturbed tune.
Secondly, the inclusion of space charge in simulations is essential to reproduce the measurements, particularly the visibility of the main mode over other modes. However, neighboring modes restrict the fitting range of the tune shift slope at high intensities, especially in the horizontal plane, for both measurements and simulations.
In the vertical plane, measurements and simulations show good agreement on the tune shift slope. The horizontal plane exhibits a 50\% underestimation in simulations, where only one mode is observed compared to two parallel modes in measurements. The lower line, assumed to be the main mode, was used for the tune shift slope fitting.
Lastly, at low intensities in both planes, the main mode tune shift exhibits a bend with intensity. This behavior resembles the results from~\cite{Kornilov:2010zz}, where space charge significantly affects the mode tune shift when $\Delta Q_\mathrm{SC} \approx Q_\mathrm{s}$. In this intensity range, the transition from the low to intermediate space charge regime may influence both the mode tune shift and Landau damping. As a consequence, the previously dominant mode can become stabilized with a slight increase in intensity, allowing a different mode to become the new dominant one.

The 2022 measurement campaign achieved a major milestone: the clear observation of the main mode tune shift at both injection and extraction energies, with relative uncertainties as low as 5\%. This breakthrough was enabled by refining the chirp signal used to excite the beam~\cite{Joly:2024}. The PS impedance model successfully reproduces these measurements when including space charge effects, particularly in the vertical plane, where simulated and measured tune shift slopes closely match.

\section{Validation of the transverse impedance model}

The PS transverse impedance model should be able to accurately reproduce the impact of the impedance on beam dynamics, in particular the impedance-induced tune shift.
Similarly to the previous section, the space charge, octupolar order non-linearities, first and second-order chromaticity, and non-linear synchrotron motion are included in the simulations. Measurements from 2021 and 2022 with similar $Q'$ are compared with simulations based on 2022 measurement parameters. The different chromaticities for both sets of measurements at injection and extraction energies can be found in Table~\ref{tab:params_validate_model}.

\begin{table}[!htbp]
\begin{ruledtabular}
    \centering
    \begin{tabular}{cccccc}
        \hline
        Scenario & Year & $Q'_\mathrm{x}$ & $Q'_\mathrm{y}$ & $Q''_\mathrm{x}$ & $Q''_\mathrm{y}$ \\
        \hline
        \multirow{2}{*}{Injection} & 2021 & -5.4 & -9.6 & -206 & 215 \\
         & 2022 & -5.0 & -9.4 & 89 & 178 \\
        \hline
        \multirow{2}{*}{Extraction} & 2021 & 2.3 & 1.7 & 365 & -407 \\
         & 2022 & 0.5 & 1.4 & -46 & -212 \\
        \hline
    \end{tabular}
    \caption{Chromaticities measured at injection and extraction energies during 2021 and 2022 campaigns.}
    \label{tab:params_validate_model}
\end{ruledtabular}
\end{table}

\begin{figure}[!htbp]
\centering
  \includegraphics[width=0.9\linewidth]{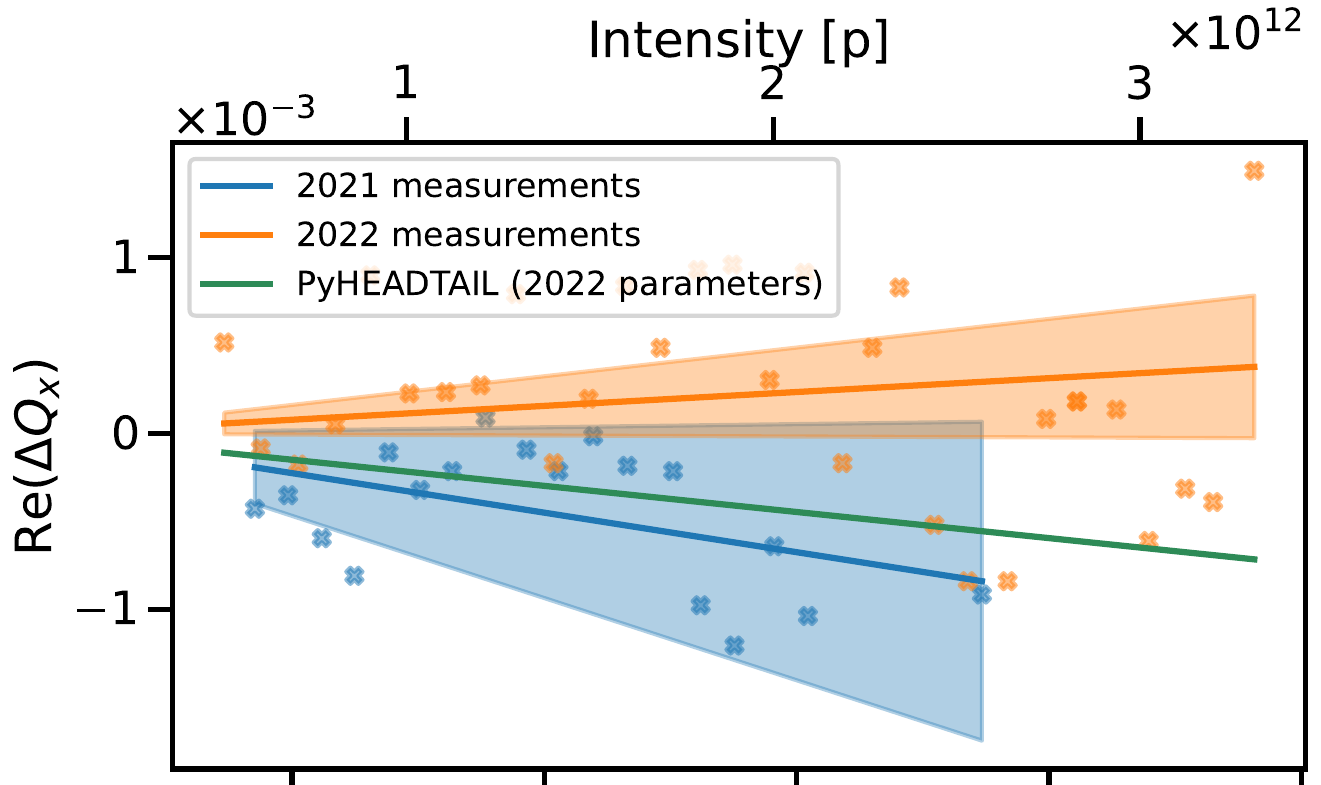}
  \hfill
  \includegraphics[width=0.9\linewidth]{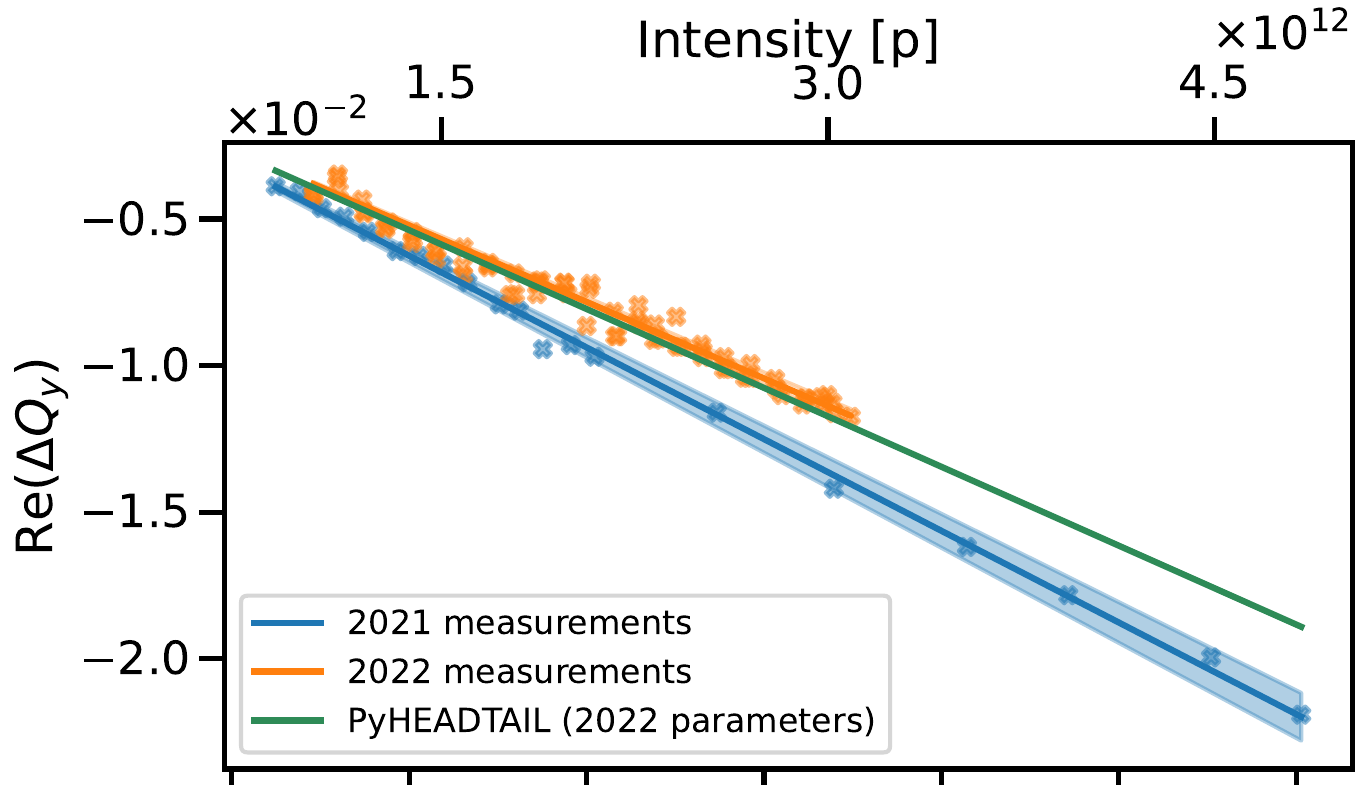}
\caption{Comparison between 2021 and 2022 measurements with PyHEADTAIL simulations, at injection energy, in horizontal (top) and vertical (bottom).}
\label{fig:dQ_validate_model_FB}
\end{figure}

\begin{figure}[!htbp]
\centering
  \includegraphics[width=0.9\linewidth]{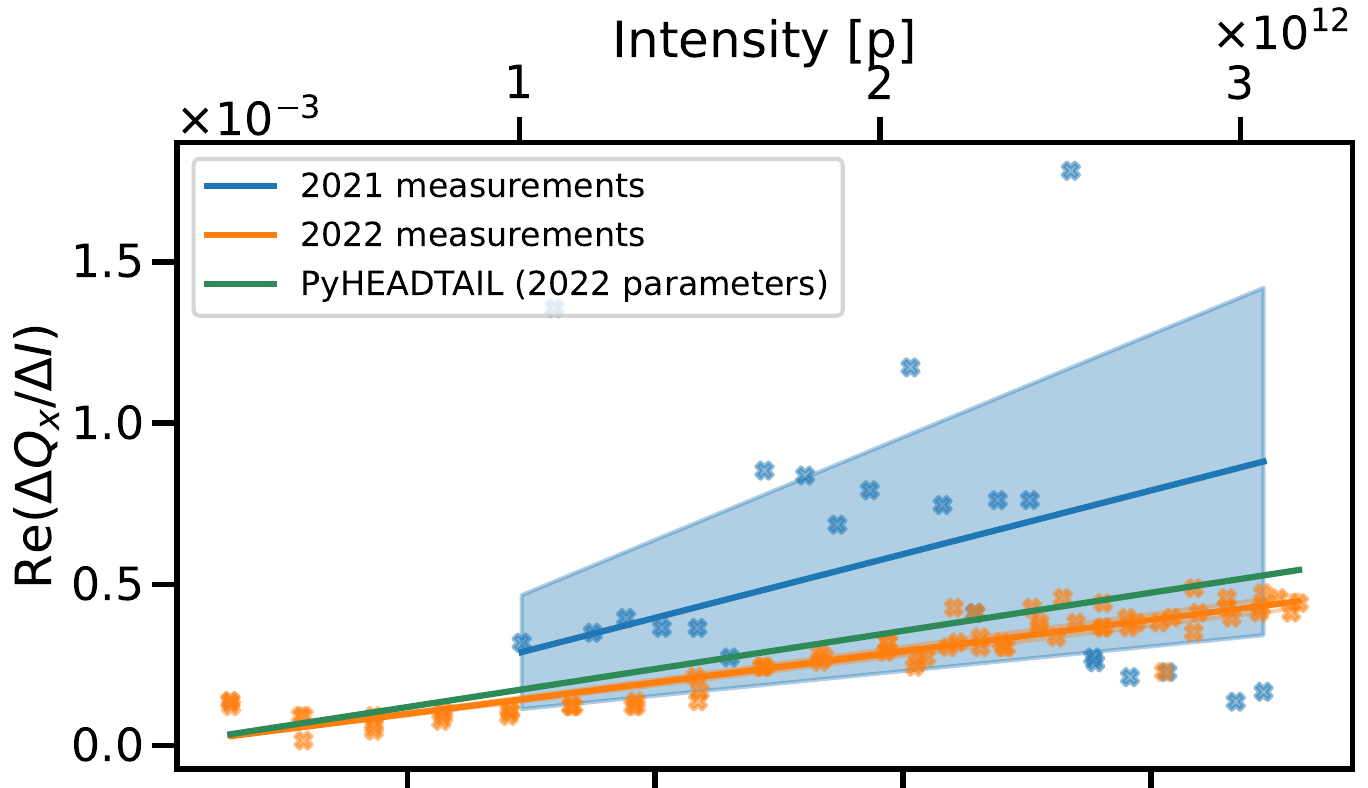}
  \hfill
  \includegraphics[width=0.9\linewidth]{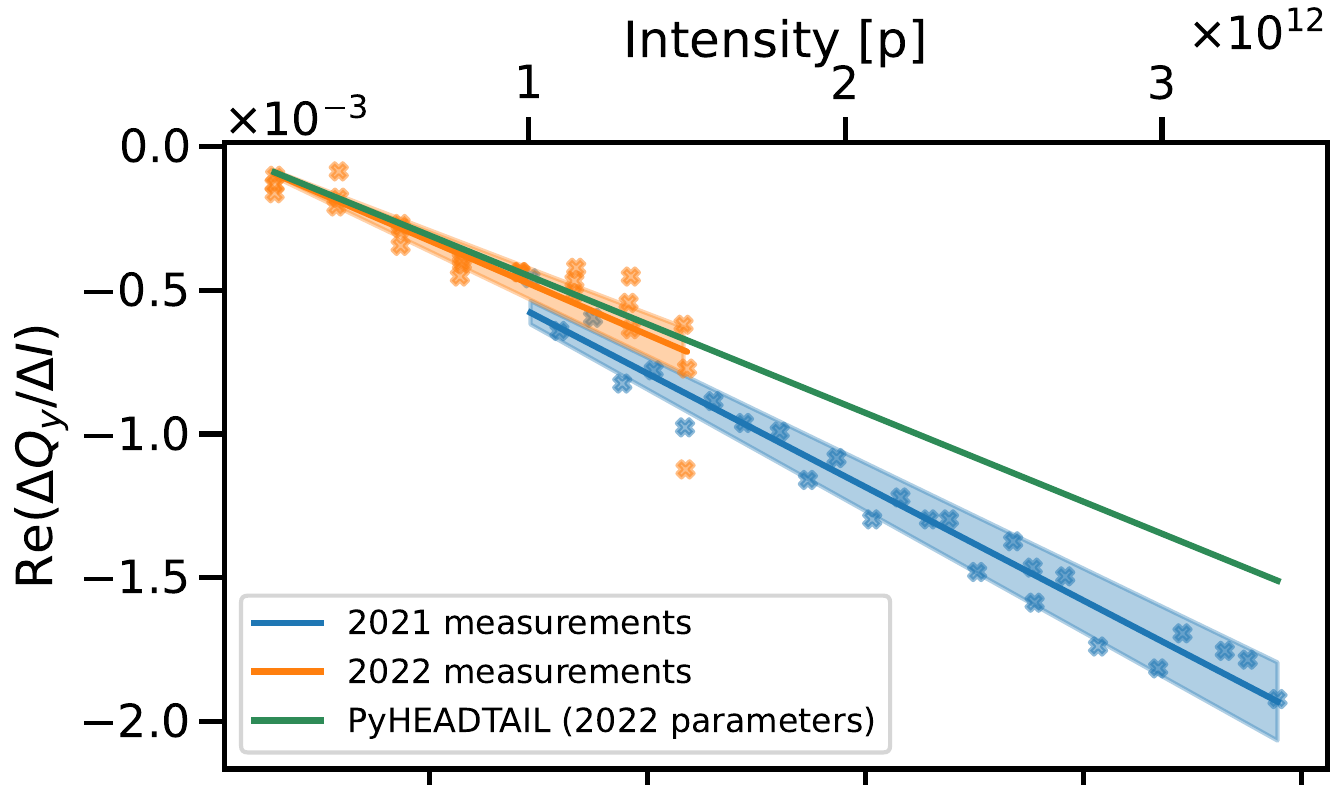}
\caption{Comparison between 2021 and 2022 measurements with PyHEADTAIL simulations, at extraction energy, in horizontal (top) and vertical (bottom).}
\label{fig:dQ_validate_model_FT}
\end{figure}

$Q''$ similar to those observed during the 2021 measurements, could not be reproduced in 2022, hence, the measured tune shift slopes are expected to be different. The validity of the model has been assessed by comparing the 2022 measurements and the corresponding simulations. The measured and simulated tune shifts are presented in Figs.~\ref{fig:dQ_validate_model_FB} and~\ref{fig:dQ_validate_model_FT}, with their slopes and confidence intervals summarized in Table~\ref{tab:comp_dQ}.

\begin{table*}[!htbp]
\begin{ruledtabular}
    \centering
    \begin{tabular}{cccccc}
        \hline
        Scenario & Plane & 2021, $\Delta Q / \Delta I$ [A$^{-1}$] & 2022, $\Delta Q / \Delta I$  [A$^{-1}$] & PyHEADTAIL, $\Delta Q / \Delta I$ [A$^{-1}$] \\
        \hline
        \multirow{2}{*}{Injection} & x & ($-4.50 \pm 4.86$) $\times 10^{-3}$ & ($1.58 \pm 1.69$) $\times 10^{-3}$ & -2.97 $\times 10^{-3}$ \\
         & y & ($-6.25 \pm 0.23$) $\times 10^{-2}$ & ($-5.22 \pm 0.10$) $\times 10^{-2}$ & -5.38 $\times 10^{-2}$ \\
        \hline
        \multirow{2}{*}{Extraction} & x & ($3.83 \pm 2.35$) $\times 10^{-3}$ & ($1.89 \pm 0.09$) $\times 10^{-3}$ & 2.29 $\times 10^{-3}$ \\
         & y & ($-7.66 \pm 0.54$) $\times 10^{-3}$ & ($-6.34 \pm 0.72$) $\times 10^{-3}$ & -5.99 $\times 10^{-3}$ \\
        \hline
    \end{tabular}
    \caption{Tune shift slope comparison between 2021 and 2022 measurements, and with PyHEADTAIL simulations (2022 parameters), at injection and extraction energies.}
    \label{tab:comp_dQ}
\end{ruledtabular}
\end{table*}

Besides the different chromaticities between the 2021 and 2022 measurements, the procedure has been improved by refining the chirp signal used to excite the beam~\cite{Joly:2024}, leading to significantly reduced uncertainties. At injection energy, while space charge effects lead to larger uncertainties in the horizontal plane, the vertical tune shift measurements achieve excellent precision with a confidence interval an order of magnitude smaller than the measured slope. 

The tune shifts at extraction energy are smaller than at injection energy due to the reduced indirect space charge contribution. However, the shorter bunch length at this energy extends the bunch spectrum to higher frequencies, providing validation of the impedance model over a wider frequency range. Notably, the simulated tune shifts closely match the 2022 measurements, falling within the confidence intervals in both planes.

At both energies, the variation in tune shift slope between 2021 and 2022 measurements is qualitatively consistent with the change in $Q''$ given in Table~\ref{tab:params_validate_model}, further validating our understanding of the mechanisms at work.

The PS impedance model successfully reproduces the measurements when including all the relevant mechanisms, particularly in the vertical plane, where simulated and measured tune shift slopes closely match. In addition, the 2022 measurements achieved unprecedented precision, allowing the observation of the horizontal main mode tune shift and the determination of its slope with a relative confidence interval of~\SI{5}{\percent}, as it can be seen in Appendix~\ref{app:C}.

\section{Instability growth rates}

On top of causing an intensity-dependent tune shift, the transverse impedance is also responsible, in some cases, for transverse instabilities. Instabilities exhibit complex behaviors with $Q'$~\cite{Blewett:118362}, $Q''$~\cite{Schenk:2018pae, Joly:2024}, and the Landau damping introduced by space charge and lattice non-linearities~\cite{Macridin:2015vua, Kornilov:2010zz, Kornilov:2015xta}. However, we focus here on the interplay between space charge and the slow-headtail instability, and not the fast-headtail instability as studied in detail in~\cite{Balbekov:2016qda, Zolkin:2018, Chin:2016lzk, Burov:2008be, Burov:2019, Burov:318826, Buffat:2022} for hadron accelerators and recently in~\cite{Antipov:2024ppx} for 4th generation light sources.

Transverse instabilities can also be used to validate the impedance model. More specifically, reproducing measured growth rates for various chromaticities and intensities in simulations is an adequate way to benchmark the real part of an impedance model. To complement the tune shift measurements in the horizontal plane, where some discrepancies remain, we focus on the horizontal growth rates at injection energy.


For the intensities considered in the following measurements, the PS is in the strong space charge regime with a space charge parameter $q_\mathrm{SC} = \Delta Q_\mathrm{SC,x} / 2 Q_\mathrm{s}$ ranging from 12 ($I_b = \SI{5e11}{p}$) to 75 ($I_b = \SI{3e12}{p}$).

Similarly to the tune shift simulations, octupolar order non-linearities, first and second-order chromaticity and non-linear synchrotron motion, are all included in PyHEADTAIL, as well as direct space charge, using the numerical parameters given in Table~\ref{tab:PIC_parameters2}. Note that the longitudinal resolution of the mesh grid has been increased with respect to the previous simulations, to resolve the transverse oscillations caused by instabilities along the bunch. In addition, the number of macroparticles has been doubled to populate sufficiently the bunch distribution tails. 
Finally, as the transverse emittance of a bunch increases while an instability is developing, the extent of the mesh grid has been increased to reach $\pm 8 \sigma_\mathrm{x,y}$ to delay the loss of particles as they leave the PIC grid. It translates into turn-by-turn data recorded over a larger number of turns with a preserved beam intensity. For octupolar order non-linearities, the same anharmonicity coefficients and intensity dependence on the transverse emittance, as in the tune shift simulations, were used.

\begin{table}[!htbp]
\begin{ruledtabular}
    \caption{Parameters for the PyHEADTAIL instability growth rate simulations with space charge}
    \begin{tabular}{ccc}
         \hline
         $n_\mathrm{macroparticles}$ & $n^\mathrm{cell}_\perp$ & Transverse grid span  \\
         \hline
         \SI{6e6}{} & 128 & $\pm 8 \sigma_\mathrm{x,y}$ \\
         \hline
         $n_\mathrm{segments}$ & $n^\mathrm{cell}_\mathrm{\parallel}$ & Longitudinal grid span \\
         \hline
         240 & 128 & $\pm 4 \sigma_\mathrm{z}$ \\
         \hline
    \end{tabular}
    \label{tab:PIC_parameters2}
\end{ruledtabular}
\end{table}

\begin{figure}[!htbp]
\centering
  \includegraphics[width=\linewidth]{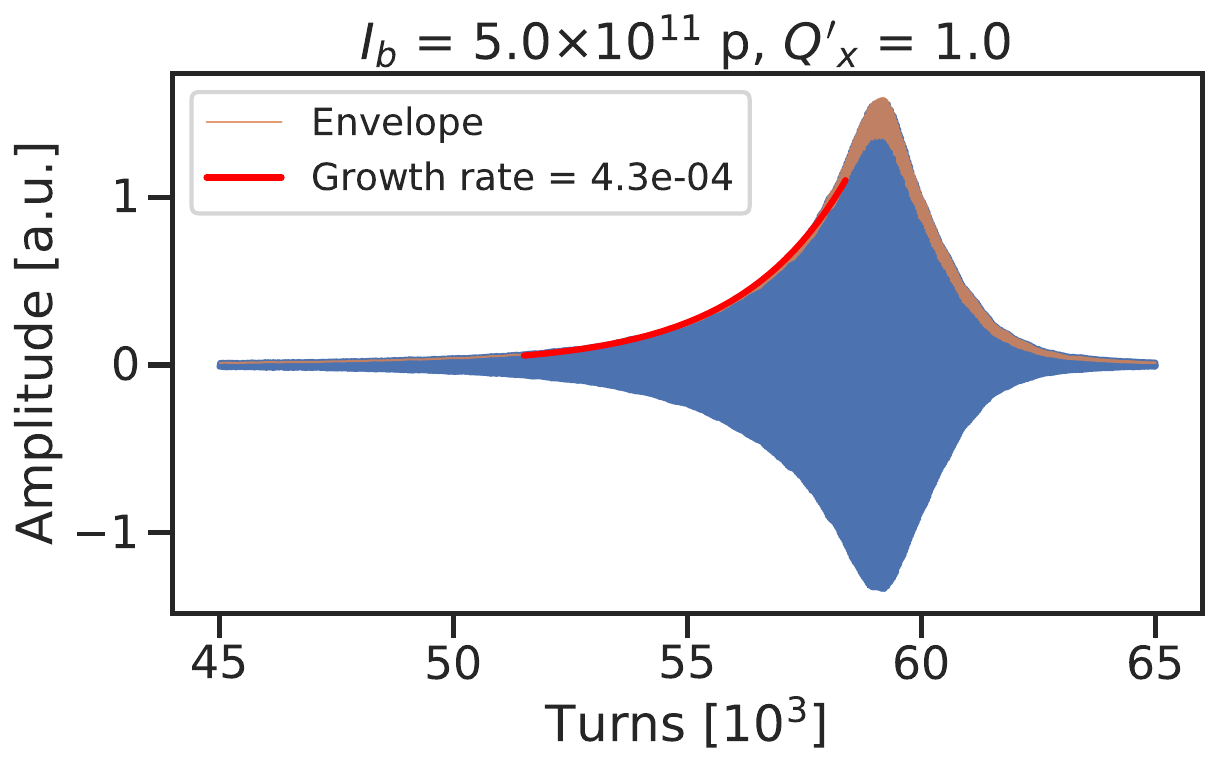}
  \includegraphics[width=\linewidth]{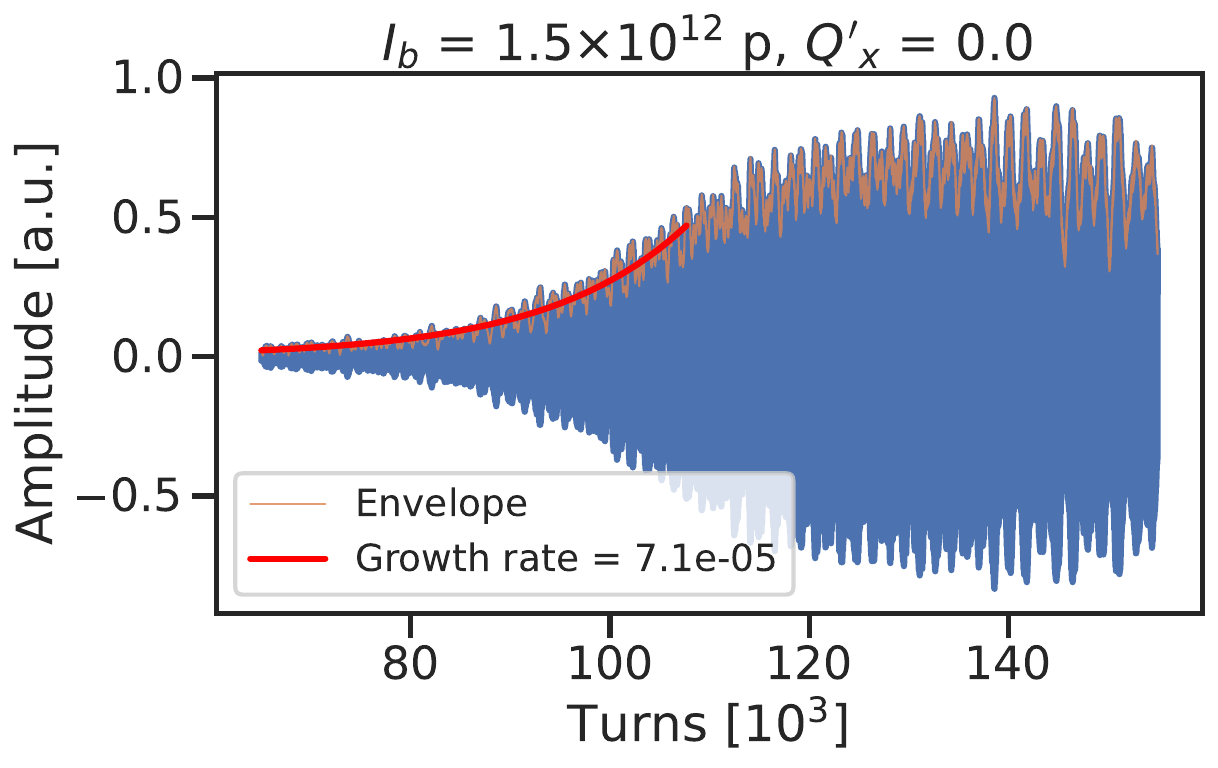}
  \includegraphics[width=\linewidth]{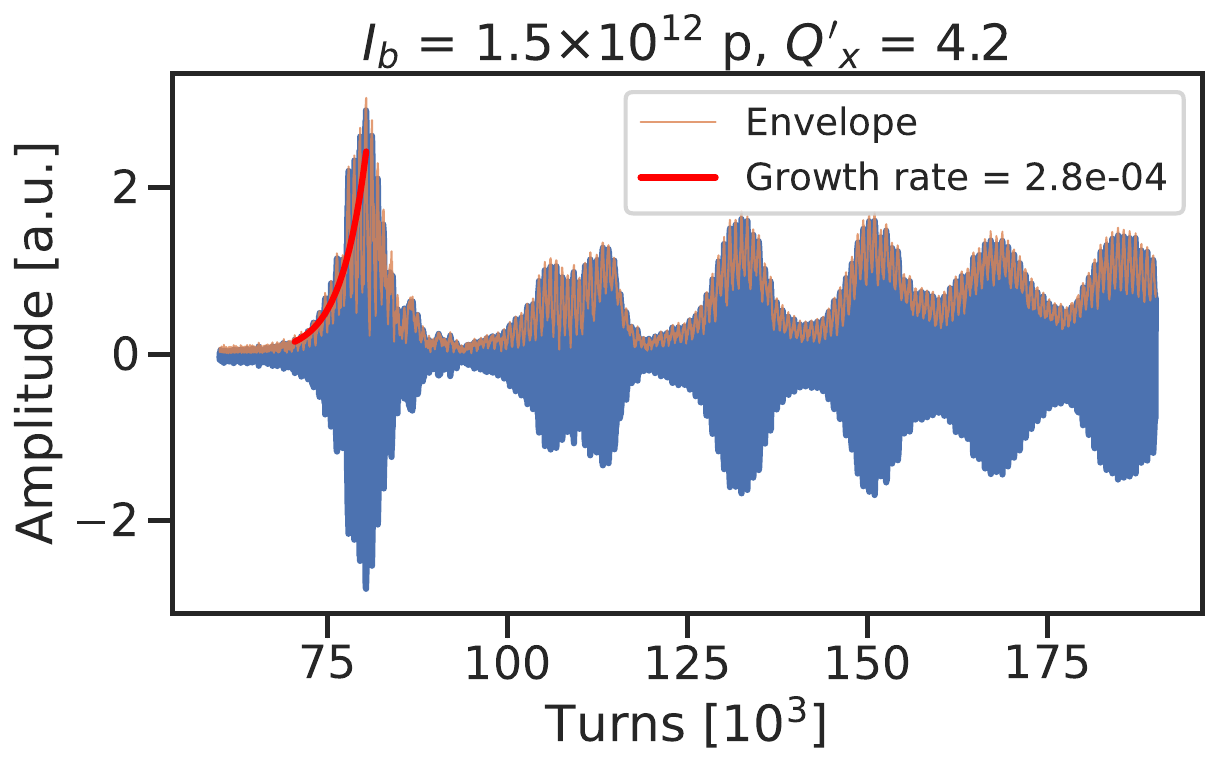}
\caption{Examples of BPM data recorded during the measurements shown in Fig.~\ref{fig:grw_SC}, with their respective growth rate fittings.}
\label{fig:inst_SC}
\end{figure}

In the presence of space charge, instabilities can exhibit multiple behaviors, as illustrated in Fig.~\ref{fig:inst_SC}. These range from a "textbook" exponential growth to a gradual increase in the center-of-mass motion that eventually saturates, or even to decoherence patterns characterized by successive bursts and decays of the center-of-mass motion. Without an intra-bunch motion measurement, such a signal cannot be distinguished from an instability mixed with a decoherence pattern~\cite{Karpov:2015hsa,karpov_early_2016} to a convective instability's signature~\cite{Burov:2018rmx}.
Therefore, an accurate estimation of the growth rate is challenging, as it was already observed in~\cite{Joly:2024acd}.

\begin{figure}[!htbp]
\centering
  \includegraphics[width=\linewidth]{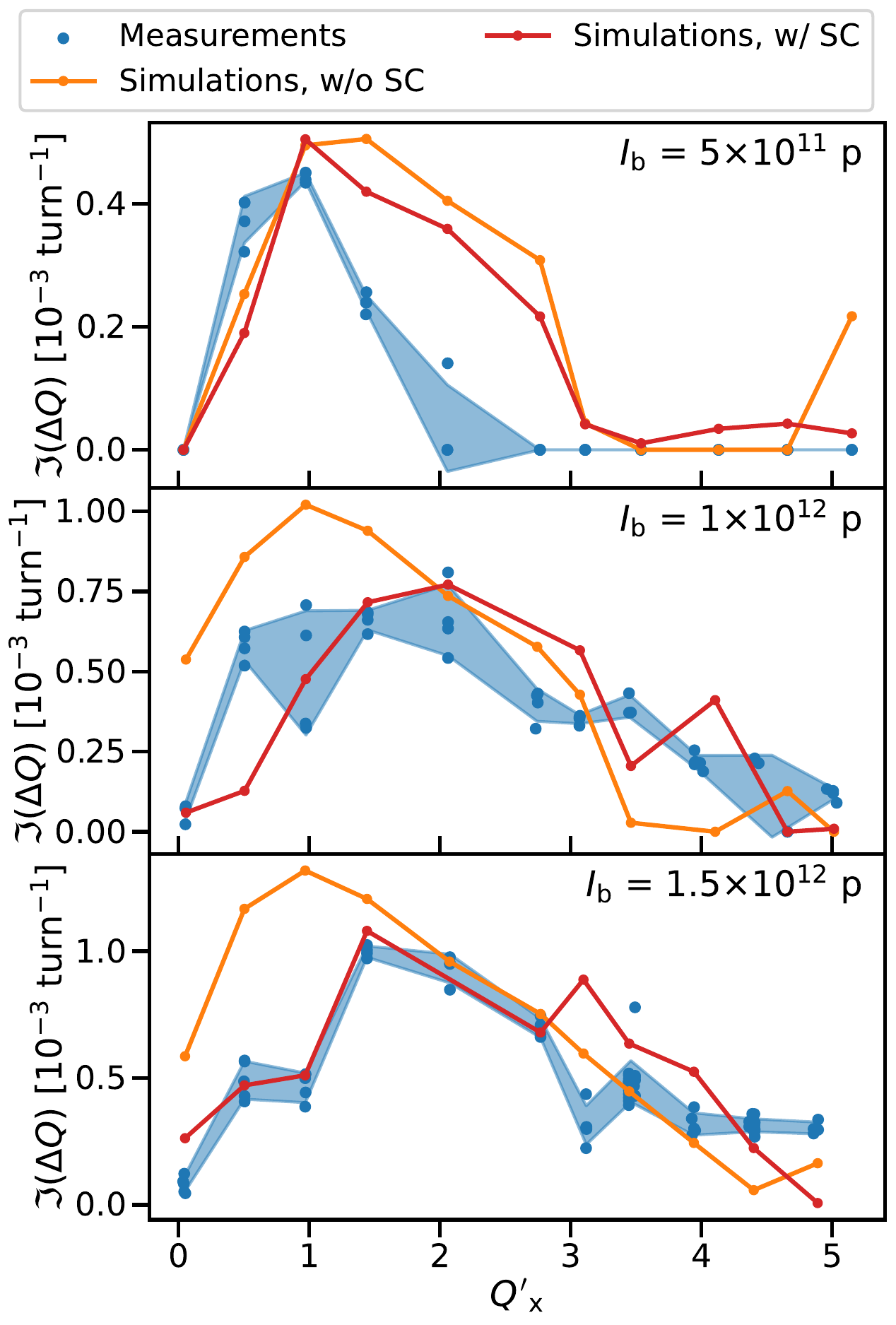}
\caption{Comparison between simulated horizontal instability growth
rates at injection, and those measured during the 2022 campaign, at injection, as a function of first-order chromaticity and for various beam intensities, with and without direct space charge.}
\label{fig:grw_SC}
\end{figure}

The measured growth rates were acquired during the 2022 measurement campaign for a range of $Q'$ and various intensities, as shown in Fig.~\ref{fig:grw_SC}. The growth rate fits were performed on the initial rise of the center-of-mass motion, stopping before the signal reached its maximum amplitude.

The shaded envelope around the growth rate values indicates the statistical confidence interval, calculated as the standard deviation of the growth rates obtained for each chromaticity value.

Regardless of the beam intensity, the growth rate pattern with $Q'$ follows qualitatively a hump shape. As the intensity increases, both the chromaticity value at which the maximum growth rate occurs and the width of the hump tend to increase. Simulations without space charge reproduce the measured growth rate fairly well, except for low $Q'$, where they overestimate the growth rate value. This discrepancy is resolved by including the effect of space charge in simulations, which then reproduces the measured growth rate envelope in most cases. 

For the lowest intensity, i.e.~$\SI{5e11}{p}$, the space charge force influence on the growth rate is negligible, and a discrepancy between measurements and simulations remains. A possible explanation could be the anharmonicity coefficients used, which may not realistically reflect the real ones.
Most importantly, the simulated growth rate values with and without space charge remain similar even for scenarios exhibiting a stronger space charge force. The presence of space charge tends to yield only a slight stabilization of the instability. In the strong space charge regime, the stabilization introduced by space charge tends to saturate when $\Delta Q_\mathrm{SC} \gg Q_\mathrm{s}$ as it was observed in~\cite{Kornilov:2010zz, Kornilov:2015xta}.

Similarly to the tune shift measurements, the PS impedance model successfully reproduces the instability growth rates across different beam intensities and a wide range of chromaticities. These measurements also allowed the horizontal components of the impedance to be investigated, completing the validation of the impedance model.

Additional growth rate measurements performed in 2021 at~$\SI{3e12}{p}$ can be found in Appendix~\ref{app:D}.

\section{CONCLUSION}

This paper presented the updated transverse impedance model of the CERN Proton Synchrotron following the Long Shutdown 2. It was shown that the most significant change that took place during LIU, was the increase of the injection energy, which led to a reduction of the imaginary part of the impedance while leaving its real part largely unchanged.
The impedance model includes all major impedance sources, namely the vacuum chamber, kicker magnets, vacuum equipment, septa, beam dumps, RF cavities, and beam instrumentation. 

To validate this impedance model, two complementary beam observables were used: impedance-induced tune shift and instability growth rates.
Their interplay with the first and second-order chromaticity, and space charge force was thoroughly investigated in simulations with PyHEADTAIL. The main observations were that $Q''$ strongly impacts the magnitude of the tune shift and the space charge introduces Landau damping, which is necessary to be accounted for to accurately reproduce the measured tune shifts and growth rates.

The tune shift measurements, particularly in the vertical plane, provided precise validation of the imaginary part of the impedance. On the other hand, the growth rate measurements complemented this by validating the real part of the impedance, especially in the horizontal plane, where tune shift measurements showed some discrepancies. The PS impedance model could reproduce the vertical tune shifts down to the level of the measurements' confidence interval and the instability growth rates at injection energy ($q_\mathrm{SC}$ ranging from 12 to 75) with a remarkable accuracy.

The PS impedance model has been successfully validated across a wide range of beam parameters. The model accurately reproduces the impedance-induced tune shifts and instability growth rates when all relevant mechanisms are included in the simulations. 
Hence, the impedance model is deemed valid and can reliably predict the beam stability in the PS, including scenarios considered for the Physics Beyond Colliders project.

The close agreement between measurements and simulations across different beam parameters and operating conditions demonstrates the reliability of our updated impedance model. These results have important implications for future PS operations and potential machine upgrades.

\section{ACKNOWLEDGEMENTS}
The author would like to thank the PS Operation team, M. Delrieux, and G. Imesch for their support and input during the beam-based measurements.
Moreover, the author would like to thank A. Huschauer for his explanations of the chromaticity control in the PS. 
Finally, the author would like to thank A. Oeftiger for invaluable help with the space-charge module of PyHEADTAIL and the insightful discussions on the interplay between the space-charge force and the transverse impedance.

\appendix

\section{PS transverse impedance model at injection}\label{app:A}

\begin{figure}[!htbp]
\centering
   \includegraphics[width=\linewidth]{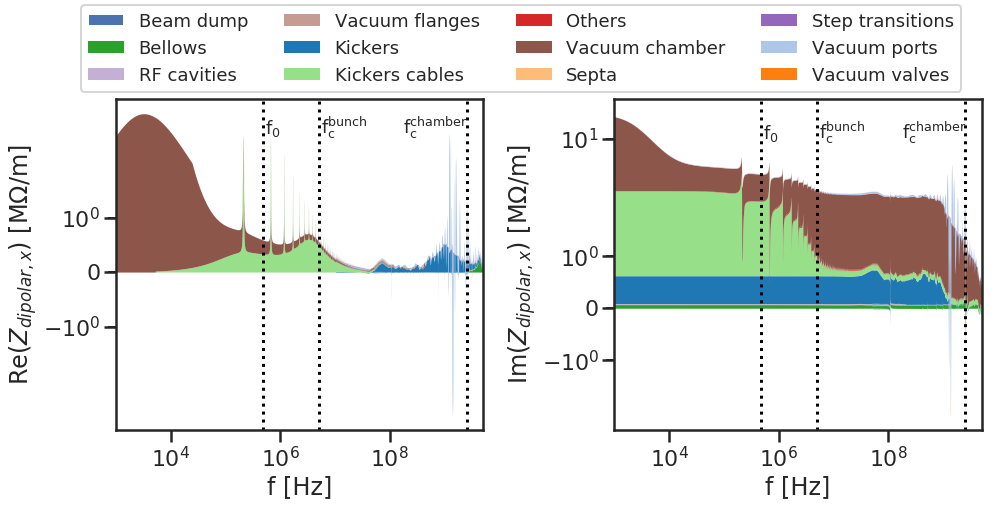}
\caption{Representation of the stacked real and imaginary parts of the horizontal dipolar impedance contributions accounted for in the Post LIU PS impedance model, at injection energy, in logarithmic scale. The revolution frequency $f_0$, the frequency corresponding to the bunch length at injection $f_\mathrm{c}^\mathrm{bunch}$, and the vacuum chamber cutoff frequency $f_\mathrm{c}^\mathrm{chamber}$ are displayed as dashed black lines.}
\label{fig:zxdip_liu}
\end{figure}

\begin{figure}[!htbp]
\centering
   \includegraphics[width=\linewidth]{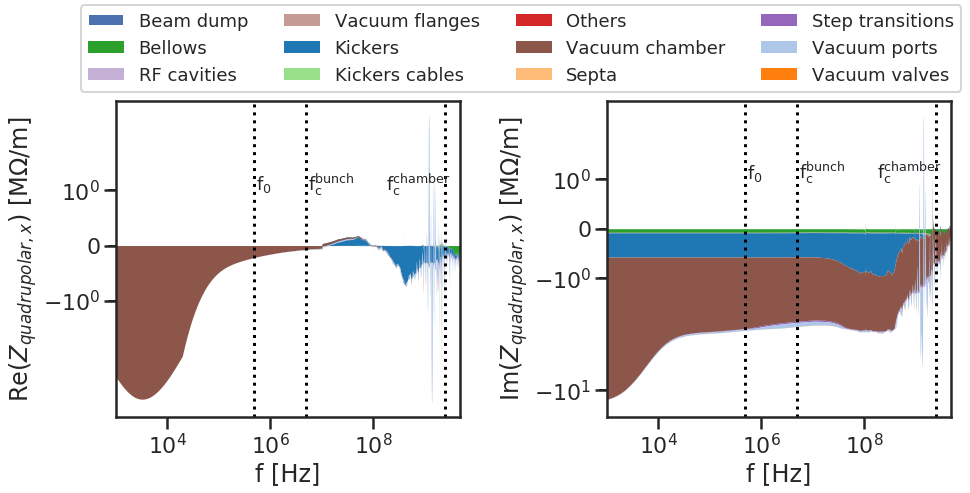}
\caption{Representation of the stacked real and imaginary parts of the horizontal quadrupolar impedance contributions accounted for in the Post LIU PS impedance model, at injection energy, in logarithmic scale. The revolution frequency $f_0$, the frequency corresponding to the bunch length at injection $f_\mathrm{c}^\mathrm{bunch}$, and the vacuum chamber cutoff frequency $f_\mathrm{c}^\mathrm{chamber}$ are displayed as dashed black lines.}
\label{fig:zxquad_liu}
\end{figure}

\begin{figure}[!htbp]
\centering
   \includegraphics[width=\linewidth]{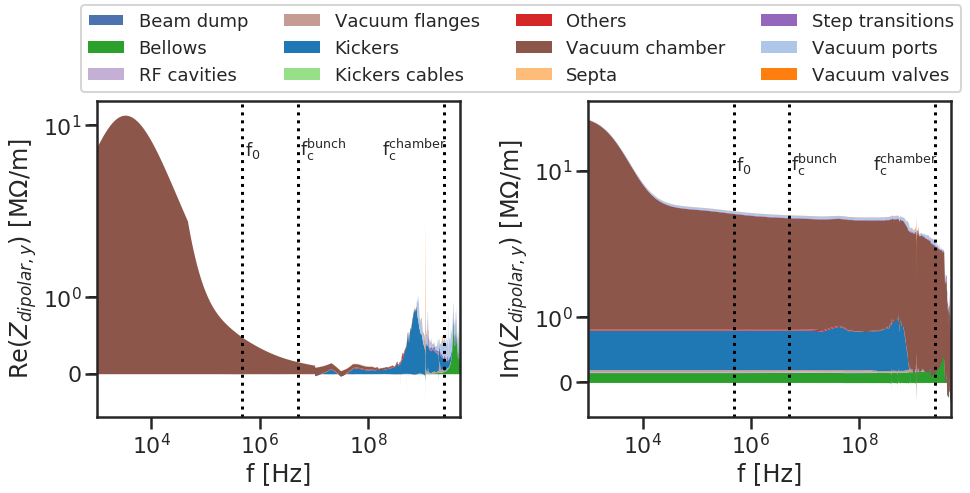}
\caption{Representation of the stacked real and imaginary parts of the vertical dipolar impedance contributions accounted for in the Post LIU PS impedance model, at injection energy, in logarithmic scale. The revolution frequency $f_0$, the frequency corresponding to the bunch length at injection $f_\mathrm{c}^\mathrm{bunch}$, and the vacuum chamber cutoff frequency $f_\mathrm{c}^\mathrm{chamber}$ are displayed as dashed black lines.}
\label{fig:zydip_liu}
\end{figure}

\begin{figure}[!htbp]
\centering
   \includegraphics[width=\linewidth]{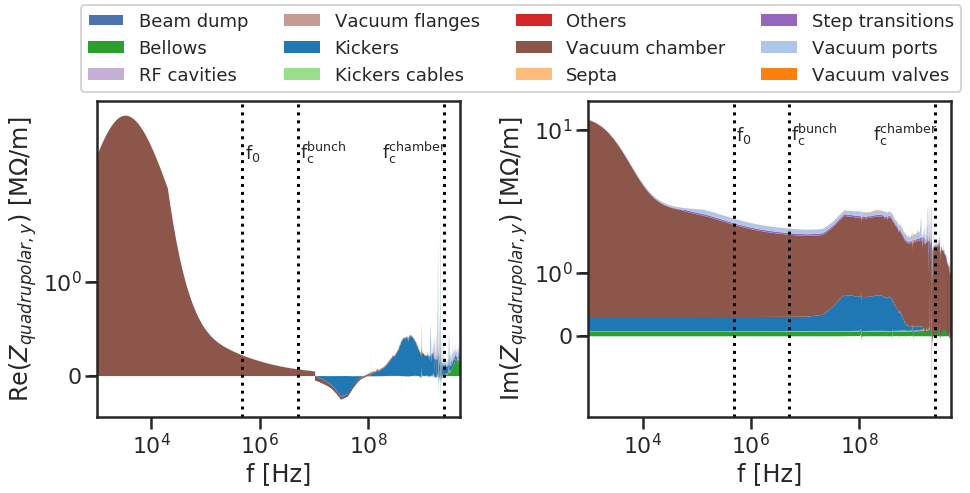}
\caption{Representation of the stacked real and imaginary parts of the vertical quadrupolar impedance contributions accounted for in the Post LIU PS impedance model, at injection energy, in logarithmic scale. The revolution frequency $f_0$, the frequency corresponding to the bunch length at injection $f_\mathrm{c}^\mathrm{bunch}$, and the vacuum chamber cutoff frequency $f_\mathrm{c}^\mathrm{chamber}$ are displayed as dashed black lines.}
\label{fig:zyquad_liu}
\end{figure}

The Post LIU PS transverse impedance model for each component is summarized in Figs.~\ref{fig:zxdip_liu} to~\ref{fig:zyquad_liu}.

\clearpage

\section{PS parameters and PyHEADTAIL settings}\label{app:B}

\begin{figure}[!htbp]
\centering
   \includegraphics[width=\linewidth]{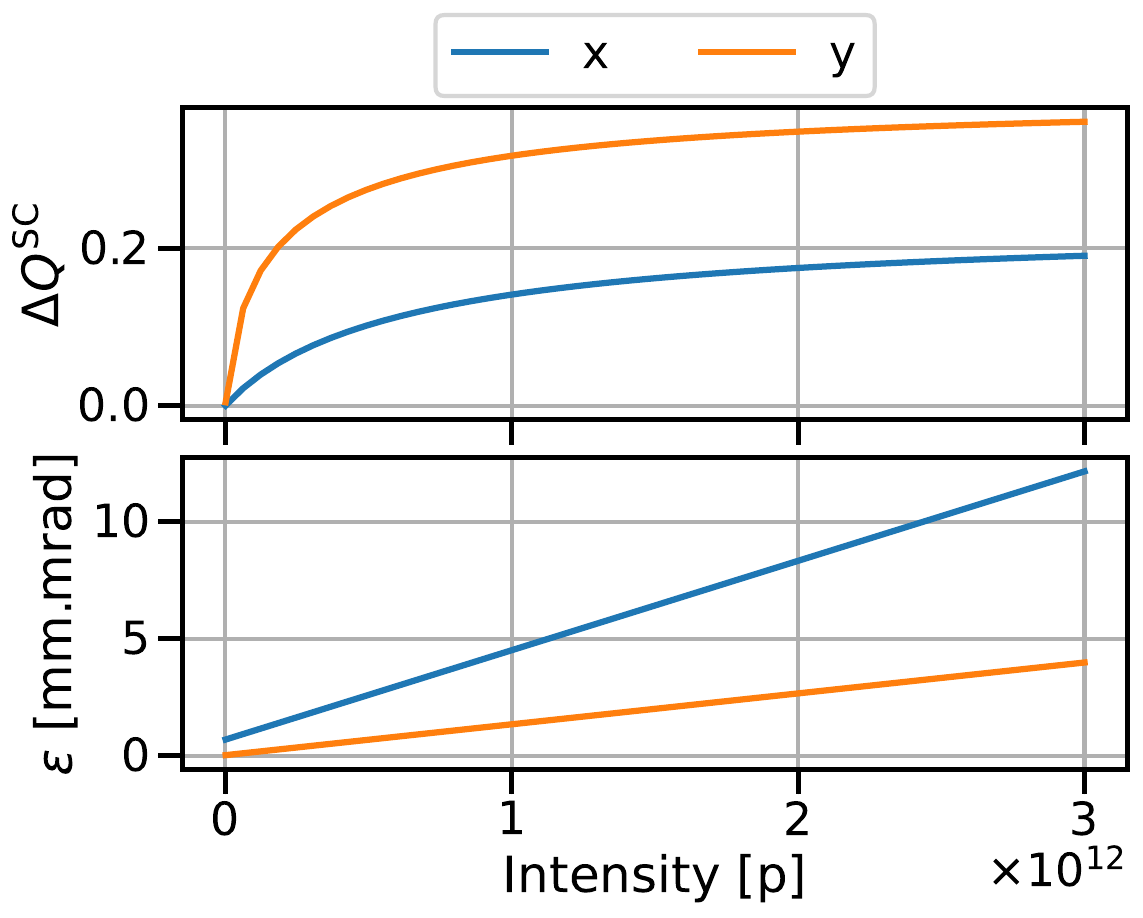}
\caption{Space charge tune shift (top) and transverse emittance (bottom) versus intensity for the custom nTOF cycle.}
\label{fig:dQ_SC_eps_TOF}
\end{figure}

\begin{figure}[!htbp]
\centering
   \includegraphics[width=\linewidth]{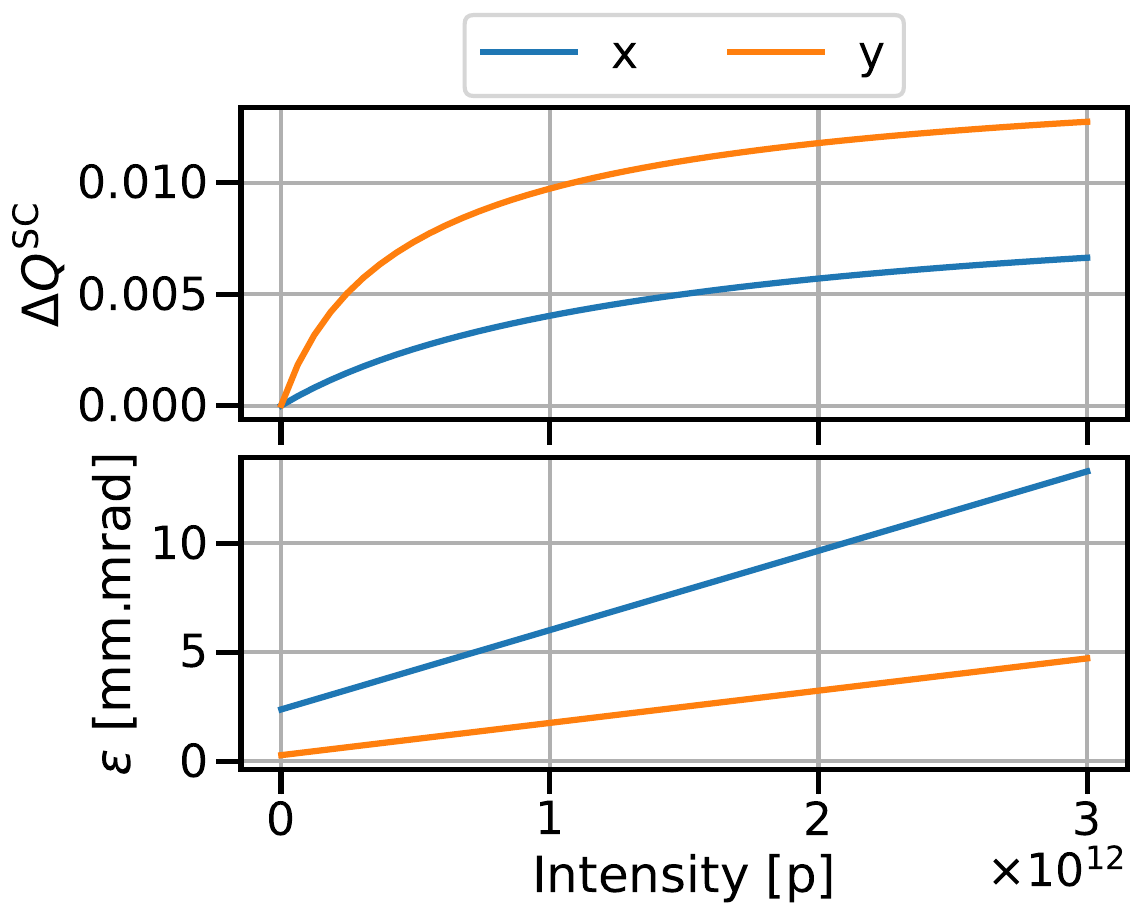}
\caption{Space charge tune shift (top) and transverse emittance (bottom) versus intensity for the custom EAST cycle.}
\label{fig:dQ_SC_eps_EAST}
\end{figure}

All the parameters listed in the following tables were experimentally measured and then used during the PyHEADTAIL simulations. Measurements and simulations are specific to proton beams. Two different custom cycles inspired by operational ones are presented.
The custom nTOF cycle was used during the impedance-induced tune shift and instability growth rate measurements at injection energy. Its intensity-dependent space charge tune shift and transverse emittances can be found in Fig.~\ref{fig:dQ_SC_eps_TOF} and its parameters in Tab.~\ref{tab:nTOF}.
The custom EAST cycle was used during the impedance-induced tune shift measurements at extraction energy. Its intensity-dependent space charge tune shift and transverse emittances can be found in Fig.~\ref{fig:dQ_SC_eps_EAST} and its parameters in Tab.~\ref{tab:east}.

\begin{table*}
\begin{ruledtabular}
    \begin{tabular}{ccc}
        \hline
        Parameter & Value & Unit \\
        \hline
        Circumference $C$ & 628.3185 & m \\
        \hline
        Kinetic energy $E_\mathrm{kinetic}$ & 2.0 & GeV \\
        \hline
        Lorentz factor $\gamma$ & 3.13 & \\
        \hline
        Revolution frequency $f_0$ & 452,153 & Hz \\
        \hline
        Horizontal and vertical tune $Q_\mathrm{x}/Q_\mathrm{y}$ & 6.07/6.31 & \\
        \hline
        Horizontal and vertical average beta functions $\beta_\mathrm{x}/\beta_\mathrm{y}$ & 16.5/15.8 & m$^{-1}$ \\
        \hline
        Anharmonicity coefficients $a_\mathrm{xx}/a_\mathrm{xy}/a_\mathrm{yy}$ & 6.11/-53.7/32.6 & m$^{-1}$ \\
        \hline
        Horizontal emittance $\epsilon_\mathrm{x}$ & $3.82 \times 10^{-18} * N_\mathrm{particles} + 6.90 \times 10^{-7}$ & mm.mrad \\
        \hline
        Vertical emittance $\epsilon_\mathrm{y}$ & $1.32 \times 10^{-18} * N_\mathrm{particles} + 2.50 \times 10^{-8}$ & mm.mrad \\
        \hline
        Horizontal and vertical bare first-order chromaticity $Q'_\mathrm{x}/Q'_\mathrm{y}$ & -5.0/-7.0 & \\
        \hline
        Horizontal and vertical bare second-order chromaticity $Q''_\mathrm{x}/Q''_\mathrm{y}$ & 89/-41 & \\
        \hline
        Momentum compaction factor $\alpha_\mathrm{p}$ & 0.025 & \\
        \hline
        Bunch length $\tau_\mathrm{b}$ & 131 & ns \\
        \hline
        Longitudinal emittance $\epsilon_\mathrm{l}$ & 1.55 & eV.s \\
        \hline
        Synchrotron tune $Q_\mathrm{s}$ & $1.36 \times 10^{-3}$ & \\
        \hline
        Harmonic number $h$ & 8 & \\
        \hline
        RF cavity voltage $V_\mathrm{RF}$ & 50 & kV \\
        \hline
        Number of macroparticles & 700,000 & \\
        \hline
        Number of slices & 1,000 & \\
        \hline
        Bunch slicing span & $\pm4$ & $\sigma_\mathrm{z}$ \\ 
        \hline
        Number of accumulated wake turns & 10 & turn \\
        \hline
        Initial transverse kick & 1 & mm \\
        \hline
        Smooth optics approximation & Yes & \\
        \hline
        Synchrotron motion & Non-linear & \\
        \hline
        \end{tabular}
    \caption{Machine, bunch, and simulation parameters for the nTOF cycle (injection energy).\label{tab:nTOF}}
\end{ruledtabular}
\end{table*}


\begin{table*}
\begin{ruledtabular}
    \begin{tabular}{ccc}
        \hline
        Parameter & Value & Unit \\
        \hline
        Circumference $C$ & 628.3185 & m \\
        \hline
        Kinetic energy $E_\mathrm{kinetic}$ & 25.4 & GeV \\
        \hline
        Lorentz factor $\gamma$ & 28.07 & \\
        \hline
        Revolution frequency $f_0$ & 476,832 & Hz \\
        \hline
        Horizontal and vertical tune $Q_\mathrm{x}/Q_\mathrm{y}$ & 6.23/6.27 & \\
        \hline
        Horizontal and vertical average beta functions $\beta_\mathrm{x}/\beta_\mathrm{y}$ & 16.0/16.0 & m$^{-1}$ \\
        \hline
        Anharmonicity coefficients $a_\mathrm{xx}/a_\mathrm{xy}/a_\mathrm{yy}$ & -110/35.6/53.3 & m$^{-1}$ \\
        \hline
        Horizontal emittance $\epsilon_\mathrm{x}$ & $3.64 \times 10^{-18} * N_\mathrm{particles} + 2.38 \times 10^{-6}$ & mm.mrad \\
        \hline
        Vertical emittance $\epsilon_\mathrm{y}$ & $1.48 \times 10^{-18} * N_\mathrm{particles} + 2.83 \times 10^{-7}$ & mm.mrad \\
        \hline
        Horizontal and vertical bare first-order chromaticity $Q'_\mathrm{x}/Q'_\mathrm{y}$ & 0.5/1.4 & \\
        \hline
        Horizontal and vertical bare second-order chromaticity $Q''_\mathrm{x}/Q''_\mathrm{y}$ & -46/-212 & \\
        \hline
        Momentum compaction factor $\alpha_\mathrm{p}$ & 0.027 & \\
        \hline
        Bunch length $\tau_\mathrm{b}$ & 42 & ns \\
        \hline
        Longitudinal emittance $\epsilon_\mathrm{l}$ & 2.09 & eV.s \\
        \hline
        Synchrotron tune $Q_\mathrm{s}$ & $4.98 \times 10^{-4}$ & \\
        \hline
        Harmonic number $h$ & 8 & \\
        \hline
        RF cavity voltage $V_\mathrm{RF}$ & 200 & kV \\
        \hline
        Number of macroparticles & 700,000 & \\
        \hline
        Number of slices & 1,000 & \\
        \hline
        Bunch slicing span & $\pm4$ & $\sigma_\mathrm{z}$ \\ 
        \hline
        Number of accumulated wake turns & 10 & turn \\
        \hline
        Initial transverse kick & 1 & mm \\
        \hline
        Smooth optics approximation & Yes & \\
        \hline
        Synchrotron motion & Non-linear & \\
        \hline
        \end{tabular}
    \caption{Machine, bunch, and simulation parameters for the EAST cycle (extraction energy).\label{tab:east}}
\end{ruledtabular}
\end{table*}

\clearpage

\section{Horizontal tune shift measurements at extraction energy}\label{app:C}

\begin{figure}[!htbp]
\centering
  \includegraphics[width=\linewidth]{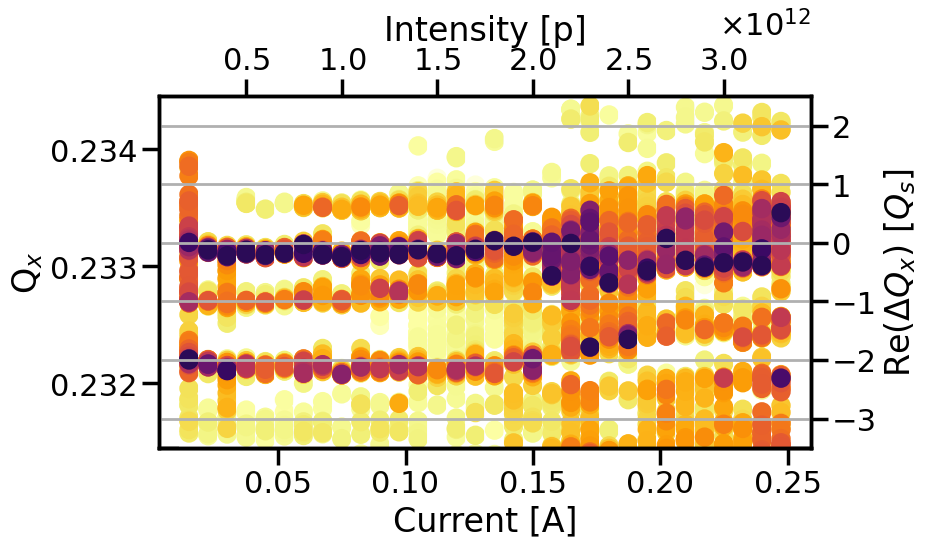}
\caption{Measured horizontal tune shift against beam intensity at extraction energy. The color scale represents the tune magnitude, from high (purple) to low (yellow).}
\label{fig:dQx}
\end{figure}

Measuring the tune shift in the horizontal plane at injection energy has proven to be challenging due to the magnitude of the tune shift slope with beam intensity (approximately $10^{-3}\,\mathrm{A}^{-1}$) and the large tune spread caused by direct space charge, surpassing the impedance-induced tune shift by orders of magnitude. 
At extraction energy, the tune shift slope magnitude is further reduced due to the reduced indirect space charge contribution. However, the direct space charge tune spread is reduced to a larger extent, thus enabling the observation not only the main mode but also additional azimuthal modes, as shown in Fig.~\ref{fig:dQx}.
, so does the direct space charge tune spread to a larger extent. Thus, It enables the observe not only the main mode but also additional azimuthal modes, as shown in Fig.~\ref{fig:dQx}. This acquisition was performed in 2022, with chromaticity measured as $Q'_\mathrm{x} = 0.8$ and $Q''_\mathrm{x} = -181$.

Thanks to the achieved precision, azimuthal modes from –2 to 1 can be clearly tracked across the entire intensity range. Moreover, a coupling between modes –1 and 0 is observed after \SI{0.15}{\A}. To our knowledge, this represents the first experimental observation of horizontal mode coupling in a hadron accelerator. A comparable level of precision has previously been attained only in a light source~\cite{Carver:2023xyy}.

\section{2021 instability measurements at injection energy}\label{app:D}

\begin{figure}[!htbp]
\centering
  \includegraphics[width=\linewidth]{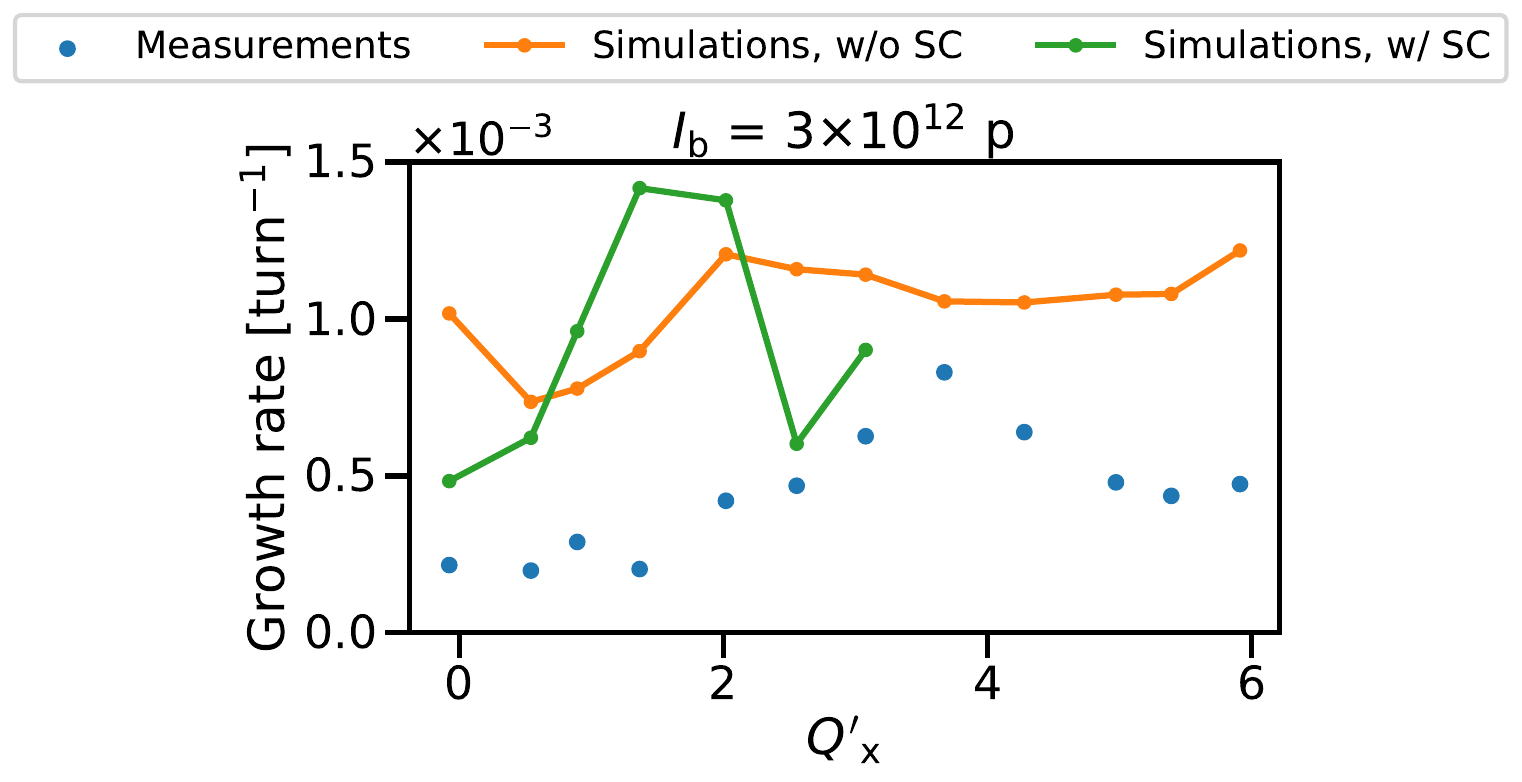}
\caption{Comparison between simulated horizontal instability growth
rates at injection, and those measured during the 2021 campaign, at injection, against first-order chromaticity for various beam intensities, with and without direct space.}
\label{fig:grw_SC_3e12}
\end{figure}

Prior to the measurements performed in 2022, another set of measurements was carried out in 2021 with an intensity of \SI{3e12}{p}. Contrary to the 2022 measurements, only one acquisition per chromaticity value was done, thus preventing the calculation of a statistical uncertainty for the growth rate values. Moreover, the chromaticity was only controlled through the PFW in 2021, introducing unknown $Q''$ values significantly larger than in 2022. The resulting growth rates can be found in Fig.~\ref{fig:grw_SC_3e12}.

The previously observed bump pattern with respect to chromaticity, is reproduced with a maximum further pushed to a larger $Q'$ value. However, the simulations with and without space charge fail at reproducing the same growth rates, overestimating them by approximately a factor of two. Moreover, it can be seen that some values are missing in the simulations with space charge. It is due to the presence of strong decoherence patterns that could not be distinguished from quickly damped instabilities, hence the instability simulations were discarded in such case. Besides, the presence of decoherence also hindered the growth rate fitting leading to noticeable growth rate variations from one chromaticity to the other.
At this stage, the unknown $Q''$ is the most plausible cause for the discrepancy between the 2021 measurements and the corresponding simulations.

\bibliographystyle{apsrev4-2}
\bibliography{apssamp}

\begin{thebibliography}{78}%
\makeatletter
\providecommand \@ifxundefined [1]{%
 \@ifx{#1\undefined}
}%
\providecommand \@ifnum [1]{%
 \ifnum #1\expandafter \@firstoftwo
 \else \expandafter \@secondoftwo
 \fi
}%
\providecommand \@ifx [1]{%
 \ifx #1\expandafter \@firstoftwo
 \else \expandafter \@secondoftwo
 \fi
}%
\providecommand \natexlab [1]{#1}%
\providecommand \enquote  [1]{``#1''}%
\providecommand \bibnamefont  [1]{#1}%
\providecommand \bibfnamefont [1]{#1}%
\providecommand \citenamefont [1]{#1}%
\providecommand \href@noop [0]{\@secondoftwo}%
\providecommand \href [0]{\begingroup \@sanitize@url \@href}%
\providecommand \@href[1]{\@@startlink{#1}\@@href}%
\providecommand \@@href[1]{\endgroup#1\@@endlink}%
\providecommand \@sanitize@url [0]{\catcode `\\12\catcode `\$12\catcode `\&12\catcode `\#12\catcode `\^12\catcode `\_12\catcode `\%12\relax}%
\providecommand \@@startlink[1]{}%
\providecommand \@@endlink[0]{}%
\providecommand \url  [0]{\begingroup\@sanitize@url \@url }%
\providecommand \@url [1]{\endgroup\@href {#1}{\urlprefix }}%
\providecommand \urlprefix  [0]{URL }%
\providecommand \Eprint [0]{\href }%
\providecommand \doibase [0]{https://doi.org/}%
\providecommand \selectlanguage [0]{\@gobble}%
\providecommand \bibinfo  [0]{\@secondoftwo}%
\providecommand \bibfield  [0]{\@secondoftwo}%
\providecommand \translation [1]{[#1]}%
\providecommand \BibitemOpen [0]{}%
\providecommand \bibitemStop [0]{}%
\providecommand \bibitemNoStop [0]{.\EOS\space}%
\providecommand \EOS [0]{\spacefactor3000\relax}%
\providecommand \BibitemShut  [1]{\csname bibitem#1\endcsname}%
\let\auto@bib@innerbib\@empty
\bibitem [{\citenamefont {Oeftiger}\ \emph {et~al.}(2022)\citenamefont {Oeftiger}, \citenamefont {Boine-Frankenheim}, \citenamefont {Chetvertkova}, \citenamefont {Kornilov}, \citenamefont {Rabusov},\ and\ \citenamefont {Sorge}}]{PhysRevAccelBeams.25.054402}%
  \BibitemOpen
  \bibfield  {author} {\bibinfo {author} {\bibfnamefont {A.}~\bibnamefont {Oeftiger}}, \bibinfo {author} {\bibfnamefont {O.}~\bibnamefont {Boine-Frankenheim}}, \bibinfo {author} {\bibfnamefont {V.}~\bibnamefont {Chetvertkova}}, \bibinfo {author} {\bibfnamefont {V.}~\bibnamefont {Kornilov}}, \bibinfo {author} {\bibfnamefont {D.}~\bibnamefont {Rabusov}},\ and\ \bibinfo {author} {\bibfnamefont {S.}~\bibnamefont {Sorge}},\ }\href {https://doi.org/10.1103/PhysRevAccelBeams.25.054402} {\bibfield  {journal} {\bibinfo  {journal} {Phys. Rev. Accel. Beams}\ }\textbf {\bibinfo {volume} {25}},\ \bibinfo {pages} {054402} (\bibinfo {year} {2022})}\BibitemShut {NoStop}%
\bibitem [{\citenamefont {Mohsen}\ \emph {et~al.}(2024)\citenamefont {Mohsen}, \citenamefont {Ainsworth},\ and\ \citenamefont {Burov}}]{mohsen:hb2023-thafp04}%
  \BibitemOpen
  \bibfield  {author} {\bibinfo {author} {\bibfnamefont {O.}~\bibnamefont {Mohsen}}, \bibinfo {author} {\bibfnamefont {R.}~\bibnamefont {Ainsworth}},\ and\ \bibinfo {author} {\bibfnamefont {A.}~\bibnamefont {Burov}},\ }in\ \href {https://doi.org/10.18429/JACoW-HB2023-THAFP04} {\emph {\bibinfo {booktitle} {Proc. HB'23}}},\ \bibinfo {series and number} {\bibinfo {number} {68}}\ (\bibinfo  {publisher} {JACoW Publishing, Geneva, Switzerland},\ \bibinfo {year} {2024})\ pp.\ \bibinfo {pages} {403--406}\BibitemShut {NoStop}%
\bibitem [{\citenamefont {Williamson}\ \emph {et~al.}(2024)\citenamefont {Williamson}, \citenamefont {Adams}, \citenamefont {Cavanagh}, \citenamefont {Kyle}, \citenamefont {de~Boer}, \citenamefont {Rafique},\ and\ \citenamefont {Warsop}}]{williamson:hb2023-tha1i2}%
  \BibitemOpen
  \bibfield  {author} {\bibinfo {author} {\bibfnamefont {R.}~\bibnamefont {Williamson}}, \bibinfo {author} {\bibfnamefont {D.}~\bibnamefont {Adams}}, \bibinfo {author} {\bibfnamefont {H.}~\bibnamefont {Cavanagh}}, \bibinfo {author} {\bibfnamefont {B.}~\bibnamefont {Kyle}}, \bibinfo {author} {\bibfnamefont {D.~P.}\ \bibnamefont {de~Boer}}, \bibinfo {author} {\bibfnamefont {H.}~\bibnamefont {Rafique}},\ and\ \bibinfo {author} {\bibfnamefont {C.}~\bibnamefont {Warsop}},\ }in\ \href {https://doi.org/10.18429/JACoW-HB2023-THA1I2} {\emph {\bibinfo {booktitle} {Proc. HB'23}}},\ \bibinfo {series and number} {\bibinfo {number} {68}}\ (\bibinfo  {publisher} {JACoW Publishing, Geneva, Switzerland},\ \bibinfo {year} {2024})\ pp.\ \bibinfo {pages} {331--337}\BibitemShut {NoStop}%
\bibitem [{\citenamefont {Williamson}(2023)}]{williamson2023a}%
  \BibitemOpen
  \bibfield  {author} {\bibinfo {author} {\bibfnamefont {R.}~\bibnamefont {Williamson}},\ }\emph {\bibinfo {title} {Studies of beam instabilities and associated beam-loss on a high-intensity proton synchroton}},\ \href {https://ora.ox.ac.uk/objects/uuid:fdedfe67-9e52-479e-b965-ad12fcac7149} {Ph.D. thesis},\ \bibinfo  {school} {University of Oxford} (\bibinfo {year} {2023})\BibitemShut {NoStop}%
\bibitem [{\citenamefont {Saha}\ \emph {et~al.}(2018)\citenamefont {Saha}, \citenamefont {Shobuda}, \citenamefont {Hotchi}, \citenamefont {Harada}, \citenamefont {Hayashi}, \citenamefont {Kinsho}, \citenamefont {Tamura}, \citenamefont {Tani}, \citenamefont {Yamamoto}, \citenamefont {Watanabe}, \citenamefont {Chin},\ and\ \citenamefont {Holmes}}]{PhysRevAccelBeams.21.024203}%
  \BibitemOpen
  \bibfield  {author} {\bibinfo {author} {\bibfnamefont {P.~K.}\ \bibnamefont {Saha}}, \bibinfo {author} {\bibfnamefont {Y.}~\bibnamefont {Shobuda}}, \bibinfo {author} {\bibfnamefont {H.}~\bibnamefont {Hotchi}}, \bibinfo {author} {\bibfnamefont {H.}~\bibnamefont {Harada}}, \bibinfo {author} {\bibfnamefont {N.}~\bibnamefont {Hayashi}}, \bibinfo {author} {\bibfnamefont {M.}~\bibnamefont {Kinsho}}, \bibinfo {author} {\bibfnamefont {F.}~\bibnamefont {Tamura}}, \bibinfo {author} {\bibfnamefont {N.}~\bibnamefont {Tani}}, \bibinfo {author} {\bibfnamefont {M.}~\bibnamefont {Yamamoto}}, \bibinfo {author} {\bibfnamefont {Y.}~\bibnamefont {Watanabe}}, \bibinfo {author} {\bibfnamefont {Y.~H.}\ \bibnamefont {Chin}},\ and\ \bibinfo {author} {\bibfnamefont {J.~A.}\ \bibnamefont {Holmes}},\ }\href {https://doi.org/10.1103/PhysRevAccelBeams.21.024203} {\bibfield  {journal} {\bibinfo  {journal} {Phys. Rev. Accel. Beams}\ }\textbf {\bibinfo {volume} {21}},\ \bibinfo {pages} {024203} (\bibinfo {year} {2018})}\BibitemShut
  {NoStop}%
\bibitem [{\citenamefont {Shobuda}\ \emph {et~al.}(2017)\citenamefont {Shobuda}, \citenamefont {Chin}, \citenamefont {Saha}, \citenamefont {Hotchi}, \citenamefont {Harada}, \citenamefont {Irie}, \citenamefont {Tamura}, \citenamefont {Tani}, \citenamefont {Toyama}, \citenamefont {Watanabe},\ and\ \citenamefont {Yamamoto}}]{10.1093/ptep/ptw169}%
  \BibitemOpen
  \bibfield  {author} {\bibinfo {author} {\bibfnamefont {Y.}~\bibnamefont {Shobuda}}, \bibinfo {author} {\bibfnamefont {Y.~H.}\ \bibnamefont {Chin}}, \bibinfo {author} {\bibfnamefont {P.~K.}\ \bibnamefont {Saha}}, \bibinfo {author} {\bibfnamefont {H.}~\bibnamefont {Hotchi}}, \bibinfo {author} {\bibfnamefont {H.}~\bibnamefont {Harada}}, \bibinfo {author} {\bibfnamefont {Y.}~\bibnamefont {Irie}}, \bibinfo {author} {\bibfnamefont {F.}~\bibnamefont {Tamura}}, \bibinfo {author} {\bibfnamefont {N.}~\bibnamefont {Tani}}, \bibinfo {author} {\bibfnamefont {T.}~\bibnamefont {Toyama}}, \bibinfo {author} {\bibfnamefont {Y.}~\bibnamefont {Watanabe}},\ and\ \bibinfo {author} {\bibfnamefont {M.}~\bibnamefont {Yamamoto}},\ }\href {https://doi.org/10.1093/ptep/ptw169} {\bibfield  {journal} {\bibinfo  {journal} {Progress of Theoretical and Experimental Physics}\ }\textbf {\bibinfo {volume} {2017}},\ \bibinfo {pages} {013G01} (\bibinfo {year} {2017})}\BibitemShut {NoStop}%
\bibitem [{\citenamefont {Antipov}\ \emph {et~al.}(2025)\citenamefont {Antipov}, \citenamefont {Gubaidulin}, \citenamefont {Agapov}, \citenamefont {Cortés~García},\ and\ \citenamefont {Gamelin}}]{Antipov:2024ppx}%
  \BibitemOpen
  \bibfield  {author} {\bibinfo {author} {\bibfnamefont {S.~A.}\ \bibnamefont {Antipov}}, \bibinfo {author} {\bibfnamefont {V.}~\bibnamefont {Gubaidulin}}, \bibinfo {author} {\bibfnamefont {I.}~\bibnamefont {Agapov}}, \bibinfo {author} {\bibfnamefont {E.~C.}\ \bibnamefont {Cortés~García}},\ and\ \bibinfo {author} {\bibfnamefont {A.}~\bibnamefont {Gamelin}},\ }\href {https://doi.org/10.1103/PhysRevAccelBeams.28.024401} {\bibfield  {journal} {\bibinfo  {journal} {Physical Review Accelerators and Beams}\ }\textbf {\bibinfo {volume} {28}},\ \bibinfo {pages} {024401} (\bibinfo {year} {2025})}\BibitemShut {NoStop}%
\bibitem [{\citenamefont {Bejar~Alonso}\ and\ \citenamefont {Rossi}(2015)}]{BejarAlonso:2069130}%
  \BibitemOpen
  \bibfield  {author} {\bibinfo {author} {\bibfnamefont {I.}~\bibnamefont {Bejar~Alonso}}\ and\ \bibinfo {author} {\bibfnamefont {L.}~\bibnamefont {Rossi}},\ }\href {https://cds.cern.ch/record/2069130} {\emph {\bibinfo {title} {{HiLumi} {LHC} Technical Design Report: Deliverable: D1.10}}},\ \bibinfo {type} {Tech. Rep.}\ \bibinfo {number} {{CERN}-{ACC}-2015-0140}\ (\bibinfo  {institution} {CERN},\ \bibinfo {year} {2015})\BibitemShut {NoStop}%
\bibitem [{\citenamefont {Damerau}\ \emph {et~al.}(2014)\citenamefont {Damerau}, \citenamefont {Funken}, \citenamefont {Garoby}, \citenamefont {Gilardoni}, \citenamefont {Goddard}, \citenamefont {Hanke}, \citenamefont {Lombardi}, \citenamefont {Manglunki}, \citenamefont {Meddahi}, \citenamefont {Mikulec}, \citenamefont {Rumolo}, \citenamefont {Shaposhnikova}, \citenamefont {Vretenar},\ and\ \citenamefont {Coupard}}]{Damerau:1976692}%
  \BibitemOpen
  \bibfield  {author} {\bibinfo {author} {\bibfnamefont {H.}~\bibnamefont {Damerau}}, \bibinfo {author} {\bibfnamefont {A.}~\bibnamefont {Funken}}, \bibinfo {author} {\bibfnamefont {R.}~\bibnamefont {Garoby}}, \bibinfo {author} {\bibfnamefont {S.}~\bibnamefont {Gilardoni}}, \bibinfo {author} {\bibfnamefont {B.}~\bibnamefont {Goddard}}, \bibinfo {author} {\bibfnamefont {K.}~\bibnamefont {Hanke}}, \bibinfo {author} {\bibfnamefont {A.}~\bibnamefont {Lombardi}}, \bibinfo {author} {\bibfnamefont {D.}~\bibnamefont {Manglunki}}, \bibinfo {author} {\bibfnamefont {M.}~\bibnamefont {Meddahi}}, \bibinfo {author} {\bibfnamefont {B.}~\bibnamefont {Mikulec}}, \bibinfo {author} {\bibfnamefont {G.}~\bibnamefont {Rumolo}}, \bibinfo {author} {\bibfnamefont {E.}~\bibnamefont {Shaposhnikova}}, \bibinfo {author} {\bibfnamefont {M.}~\bibnamefont {Vretenar}},\ and\ \bibinfo {author} {\bibfnamefont {J.}~\bibnamefont {Coupard}},\ }\href {https://doi.org/10.17181/CERN.7NHR.6HGC} {\emph {\bibinfo {title} {{LHC} Injectors Upgrade,
  Technical Design Report}}},\ \bibinfo {type} {Tech. Rep.}\ (\bibinfo  {institution} {CERN},\ \bibinfo {year} {2014})\BibitemShut {NoStop}%
\bibitem [{\citenamefont {Gareyte}\ and\ \citenamefont {Sacherer}(1975)}]{Gareyte:1974vf}%
  \BibitemOpen
  \bibfield  {author} {\bibinfo {author} {\bibfnamefont {J.}~\bibnamefont {Gareyte}}\ and\ \bibinfo {author} {\bibfnamefont {F.~J.}\ \bibnamefont {Sacherer}},\ }\href {https://cds.cern.ch/record/322647} {\emph {\bibinfo {title} {Head-tail type instabilities in the {CERN} {PS} and Booster, Pt 1; On the {PS} by J Gareyte; Pt 2, On the Booster by J Gareyte and F Sacherer}}},\ \bibinfo {type} {Tech. Rep.}\ \bibinfo {number} {{CERN}-{MPS}-{DL}-74-4, {MPS}-Int-{BR}-74-7}\ (\bibinfo  {institution} {CERN},\ \bibinfo {year} {1975})\BibitemShut {NoStop}%
\bibitem [{\citenamefont {Metral}\ \emph {et~al.}(2020)\citenamefont {Metral}, \citenamefont {Rumolo},\ and\ \citenamefont {Herr}}]{Metral:2020gny}%
  \BibitemOpen
  \bibfield  {author} {\bibinfo {author} {\bibfnamefont {E.}~\bibnamefont {Metral}}, \bibinfo {author} {\bibfnamefont {G.}~\bibnamefont {Rumolo}},\ and\ \bibinfo {author} {\bibfnamefont {W.}~\bibnamefont {Herr}},\ }in\ \href {https://doi.org/10.1007/978-3-030-34245-6_4} {\emph {\bibinfo {booktitle} {Particle Physics Reference Library}}},\ \bibinfo {editor} {edited by\ \bibinfo {editor} {\bibfnamefont {S.}~\bibnamefont {Myers}}\ and\ \bibinfo {editor} {\bibfnamefont {H.}~\bibnamefont {Schopper}}}\ (\bibinfo  {publisher} {Springer International Publishing},\ \bibinfo {year} {2020})\ pp.\ \bibinfo {pages} {105--181}\BibitemShut {NoStop}%
\bibitem [{\citenamefont {Migliorati}\ \emph {et~al.}({\natexlab{a}})\citenamefont {Migliorati}, \citenamefont {Aumon}, \citenamefont {Koukovini-Platia}, \citenamefont {Huschauer}, \citenamefont {Métral}, \citenamefont {Sterbini},\ and\ \citenamefont {Wang}}]{Migliorati:2018}%
  \BibitemOpen
  \bibfield  {author} {\bibinfo {author} {\bibfnamefont {M.}~\bibnamefont {Migliorati}}, \bibinfo {author} {\bibfnamefont {S.}~\bibnamefont {Aumon}}, \bibinfo {author} {\bibfnamefont {E.}~\bibnamefont {Koukovini-Platia}}, \bibinfo {author} {\bibfnamefont {A.}~\bibnamefont {Huschauer}}, \bibinfo {author} {\bibfnamefont {E.}~\bibnamefont {Métral}}, \bibinfo {author} {\bibfnamefont {G.}~\bibnamefont {Sterbini}},\ and\ \bibinfo {author} {\bibfnamefont {N.}~\bibnamefont {Wang}},\ }\href {https://doi.org/10.1103/PhysRevAccelBeams.21.120101} {\bibfield  {journal} {\bibinfo  {journal} {Physical Review Accelerators and Beams}\ }\textbf {\bibinfo {volume} {21}},\ \bibinfo {pages} {120101} ({\natexlab{a}})}\BibitemShut {NoStop}%
\bibitem [{\citenamefont {et~al.}(2023)}]{Joly:IPAC23-WEPL148}%
  \BibitemOpen
  \bibfield  {author} {\bibinfo {author} {\bibfnamefont {S.~J.}\ \bibnamefont {et~al.}},\ }in\ \href {https://doi.org/10.18429/JACoW-IPAC2023-WEPL148} {\emph {\bibinfo {booktitle} {Proc. IPAC'23}}},\ \bibinfo {series and number} {\bibinfo {number} {14}}\ (\bibinfo  {publisher} {JACoW Publishing, Geneva, Switzerland},\ \bibinfo {year} {2023})\ pp.\ \bibinfo {pages} {3450--3453}\BibitemShut {NoStop}%
\bibitem [{\citenamefont {Joly}\ \emph {et~al.}(2024)\citenamefont {Joly}, \citenamefont {Oeftiger}, \citenamefont {Iadarola}, \citenamefont {Zannini}, \citenamefont {Migliorati}, \citenamefont {Mounet},\ and\ \citenamefont {Salvant}}]{Joly:2024acd}%
  \BibitemOpen
  \bibfield  {author} {\bibinfo {author} {\bibfnamefont {S.}~\bibnamefont {Joly}}, \bibinfo {author} {\bibfnamefont {A.}~\bibnamefont {Oeftiger}}, \bibinfo {author} {\bibfnamefont {G.}~\bibnamefont {Iadarola}}, \bibinfo {author} {\bibfnamefont {C.}~\bibnamefont {Zannini}}, \bibinfo {author} {\bibfnamefont {M.}~\bibnamefont {Migliorati}}, \bibinfo {author} {\bibfnamefont {N.}~\bibnamefont {Mounet}},\ and\ \bibinfo {author} {\bibfnamefont {B.}~\bibnamefont {Salvant}},\ }\href {https://doi.org/10.1088/1748-0221/19/04/P04018} {\bibfield  {journal} {\bibinfo  {journal} {Journal of Instrumentation}\ }\textbf {\bibinfo {volume} {19}}\bibinfo  {number} { (4)}}\BibitemShut {NoStop}%
\bibitem [{\citenamefont {Beacham}\ \emph {et~al.}(2020)\citenamefont {Beacham}, \citenamefont {Burrage}, \citenamefont {Curtin}, \citenamefont {De~Roeck}, \citenamefont {Evans}, \citenamefont {Feng}, \citenamefont {Gatto}, \citenamefont {Gninenko}, \citenamefont {Hartin}, \citenamefont {Irastorza}, \citenamefont {Jaeckel}, \citenamefont {Jungmann}, \citenamefont {Kirch}, \citenamefont {Kling}, \citenamefont {Knapen}, \citenamefont {Lamont}, \citenamefont {Lanfranchi}, \citenamefont {Lazzeroni}, \citenamefont {Lindner}, \citenamefont {Martinez-Vidal}, \citenamefont {Moulson}, \citenamefont {Neri}, \citenamefont {Papucci}, \citenamefont {Pedraza}, \citenamefont {Petridis}, \citenamefont {Pospelov}, \citenamefont {Rozanov}, \citenamefont {Ruoso}, \citenamefont {Schuster}, \citenamefont {Semertzidis}, \citenamefont {Spadaro}, \citenamefont {Vallée},\ and\ \citenamefont {Wilkinson}}]{Beacham:2019nyx}%
  \BibitemOpen
\bibfield  {number} {  }\bibfield  {author} {\bibinfo {author} {\bibfnamefont {J.}~\bibnamefont {Beacham}}, \bibinfo {author} {\bibfnamefont {C.}~\bibnamefont {Burrage}}, \bibinfo {author} {\bibfnamefont {D.}~\bibnamefont {Curtin}}, \bibinfo {author} {\bibfnamefont {A.}~\bibnamefont {De~Roeck}}, \bibinfo {author} {\bibfnamefont {J.}~\bibnamefont {Evans}}, \bibinfo {author} {\bibfnamefont {J.~L.}\ \bibnamefont {Feng}}, \bibinfo {author} {\bibfnamefont {C.}~\bibnamefont {Gatto}}, \bibinfo {author} {\bibfnamefont {S.}~\bibnamefont {Gninenko}}, \bibinfo {author} {\bibfnamefont {A.}~\bibnamefont {Hartin}}, \bibinfo {author} {\bibfnamefont {I.}~\bibnamefont {Irastorza}}, \bibinfo {author} {\bibfnamefont {J.}~\bibnamefont {Jaeckel}}, \bibinfo {author} {\bibfnamefont {K.}~\bibnamefont {Jungmann}}, \bibinfo {author} {\bibfnamefont {K.}~\bibnamefont {Kirch}}, \bibinfo {author} {\bibfnamefont {F.}~\bibnamefont {Kling}}, \bibinfo {author} {\bibfnamefont {S.}~\bibnamefont {Knapen}}, \bibinfo {author} {\bibfnamefont
  {M.}~\bibnamefont {Lamont}}, \bibinfo {author} {\bibfnamefont {G.}~\bibnamefont {Lanfranchi}}, \bibinfo {author} {\bibfnamefont {C.}~\bibnamefont {Lazzeroni}}, \bibinfo {author} {\bibfnamefont {A.}~\bibnamefont {Lindner}}, \bibinfo {author} {\bibfnamefont {F.}~\bibnamefont {Martinez-Vidal}}, \bibinfo {author} {\bibfnamefont {M.}~\bibnamefont {Moulson}}, \bibinfo {author} {\bibfnamefont {N.}~\bibnamefont {Neri}}, \bibinfo {author} {\bibfnamefont {M.}~\bibnamefont {Papucci}}, \bibinfo {author} {\bibfnamefont {I.}~\bibnamefont {Pedraza}}, \bibinfo {author} {\bibfnamefont {K.}~\bibnamefont {Petridis}}, \bibinfo {author} {\bibfnamefont {M.}~\bibnamefont {Pospelov}}, \bibinfo {author} {\bibfnamefont {A.}~\bibnamefont {Rozanov}}, \bibinfo {author} {\bibfnamefont {G.}~\bibnamefont {Ruoso}}, \bibinfo {author} {\bibfnamefont {P.}~\bibnamefont {Schuster}}, \bibinfo {author} {\bibfnamefont {Y.}~\bibnamefont {Semertzidis}}, \bibinfo {author} {\bibfnamefont {T.}~\bibnamefont {Spadaro}}, \bibinfo {author} {\bibfnamefont
  {C.}~\bibnamefont {Vallée}},\ and\ \bibinfo {author} {\bibfnamefont {G.}~\bibnamefont {Wilkinson}},\ }\bibfield  {journal} {\bibinfo  {journal} {Journal of Physics G: Nuclear and Particle Physics}\ }\textbf {\bibinfo {volume} {47}},\ \href {https://doi.org/10.1088/1361-6471/ab4cd2} {10.1088/1361-6471/ab4cd2} (\bibinfo {year} {2020})\BibitemShut {NoStop}%
\bibitem [{\citenamefont {Rumolo}\ and\ \citenamefont {Zimmermann}(2002)}]{Rumolo:702717}%
  \BibitemOpen
  \bibfield  {author} {\bibinfo {author} {\bibfnamefont {G.}~\bibnamefont {Rumolo}}\ and\ \bibinfo {author} {\bibfnamefont {F.}~\bibnamefont {Zimmermann}},\ }\href {http://cds.cern.ch/record/702717} {\emph {\bibinfo {title} {Practical user guide for {HEADTAIL}}}},\ \bibinfo {type} {Tech. Rep.}\ \bibinfo {number} {{SL}-Note-2002-036-{AP}}\ (\bibinfo  {institution} {CERN},\ \bibinfo {year} {2002})\BibitemShut {NoStop}%
\bibitem [{\citenamefont {Oeftiger}(2019)}]{Oeftiger:2672381}%
  \BibitemOpen
  \bibfield  {author} {\bibinfo {author} {\bibfnamefont {A.}~\bibnamefont {Oeftiger}},\ }\href {https://doi.org/10.17181/CERN-ACC-NOTE-2019-0013} {\emph {\bibinfo {title} {An Overview of {PyHEADTAIL}}}},\ \bibinfo {type} {Tech. Rep.}\ (\bibinfo  {institution} {CERN},\ \bibinfo {year} {2019})\BibitemShut {NoStop}%
\bibitem [{\citenamefont {collaboration}(2021)}]{pyheadtail}%
  \BibitemOpen
  \bibfield  {author} {\bibinfo {author} {\bibfnamefont {P.}~\bibnamefont {collaboration}},\ }\href@noop {} {\bibinfo {title} {{PyHEADTAIL} code}},\ \bibinfo {howpublished} {\url{https://github.com/PyCOMPLETE/PyHEADTAIL}} (\bibinfo {year} {2021})\BibitemShut {NoStop}%
\bibitem [{\citenamefont {Salvant}\ \emph {et~al.}(2019)\citenamefont {Salvant} \emph {et~al.}}]{Salvant:IPAC19}%
  \BibitemOpen
  \bibfield  {author} {\bibinfo {author} {\bibfnamefont {B.}~\bibnamefont {Salvant}} \emph {et~al.},\ }in\ \href {https://doi.org/doi:10.18429/JACoW-IPAC2019-WEYPLS1} {\emph {\bibinfo {booktitle} {Proc. 10th International Particle Accelerator Conference (IPAC'19), Melbourne, Australia, 19-24 May 2019}}},\ \bibinfo {series and number} {\bibinfo {number} {10}}\ (\bibinfo  {publisher} {JACoW Publishing},\ \bibinfo {address} {Geneva, Switzerland},\ \bibinfo {year} {2019})\ pp.\ \bibinfo {pages} {2249--2254}\BibitemShut {NoStop}%
\bibitem [{\citenamefont {Ishibashi}\ \emph {et~al.}()\citenamefont {Ishibashi}, \citenamefont {Migliorati}, \citenamefont {Zhou}, \citenamefont {Shibata}, \citenamefont {Abe}, \citenamefont {Tobiyama}, \citenamefont {Suetsugu},\ and\ \citenamefont {Terui}}]{Ishibashi:2024}%
  \BibitemOpen
  \bibfield  {author} {\bibinfo {author} {\bibfnamefont {T.}~\bibnamefont {Ishibashi}}, \bibinfo {author} {\bibfnamefont {M.}~\bibnamefont {Migliorati}}, \bibinfo {author} {\bibfnamefont {D.}~\bibnamefont {Zhou}}, \bibinfo {author} {\bibfnamefont {K.}~\bibnamefont {Shibata}}, \bibinfo {author} {\bibfnamefont {T.}~\bibnamefont {Abe}}, \bibinfo {author} {\bibfnamefont {M.}~\bibnamefont {Tobiyama}}, \bibinfo {author} {\bibfnamefont {Y.}~\bibnamefont {Suetsugu}},\ and\ \bibinfo {author} {\bibfnamefont {S.}~\bibnamefont {Terui}},\ }\href {https://doi.org/10.1088/1748-0221/19/02/P02013} {\bibfield  {journal} {\bibinfo  {journal} {Journal of Instrumentation}\ }\textbf {\bibinfo {volume} {19}}\bibinfo  {number} { (2)}}\BibitemShut {NoStop}%
\bibitem [{\citenamefont {Mounet}(2012)}]{Mounet:2012}%
  \BibitemOpen
\bibfield  {number} {  }\bibfield  {author} {\bibinfo {author} {\bibfnamefont {N.}~\bibnamefont {Mounet}},\ }\emph {\bibinfo {title} {The {LHC} Transverse Coupled-Bunch Instability}},\ \href {https://doi.org/10.5075/EPFL-THESIS-5305} {\bibinfo {type} {phdthesis}},\ \bibinfo  {school} {Lausanne, {EPFL}} (\bibinfo {year} {2012})\BibitemShut {NoStop}%
\bibitem [{\citenamefont {Blednykh}\ \emph {et~al.}(2021)\citenamefont {Blednykh}, \citenamefont {Bassi}, \citenamefont {Smaluk},\ and\ \citenamefont {Lindberg}}]{PhysRevAccelBeams.24.104801}%
  \BibitemOpen
  \bibfield  {author} {\bibinfo {author} {\bibfnamefont {A.}~\bibnamefont {Blednykh}}, \bibinfo {author} {\bibfnamefont {G.}~\bibnamefont {Bassi}}, \bibinfo {author} {\bibfnamefont {V.}~\bibnamefont {Smaluk}},\ and\ \bibinfo {author} {\bibfnamefont {R.}~\bibnamefont {Lindberg}},\ }\href {https://doi.org/10.1103/PhysRevAccelBeams.24.104801} {\bibfield  {journal} {\bibinfo  {journal} {Physical Review Accelerators and Beams}\ }\textbf {\bibinfo {volume} {24}},\ \bibinfo {pages} {104801} (\bibinfo {year} {2021})}\BibitemShut {NoStop}%
\bibitem [{\citenamefont {Wang}\ \emph {et~al.}(2022)\citenamefont {Wang}, \citenamefont {Bane}, \citenamefont {Li}, \citenamefont {Luo}, \citenamefont {Omolayo}, \citenamefont {Penn}, \citenamefont {De~Santis}, \citenamefont {Steier},\ and\ \citenamefont {Venturini}}]{Wang:2022tnt}%
  \BibitemOpen
  \bibfield  {author} {\bibinfo {author} {\bibfnamefont {D.}~\bibnamefont {Wang}}, \bibinfo {author} {\bibfnamefont {K.}~\bibnamefont {Bane}}, \bibinfo {author} {\bibfnamefont {D.}~\bibnamefont {Li}}, \bibinfo {author} {\bibfnamefont {T.}~\bibnamefont {Luo}}, \bibinfo {author} {\bibfnamefont {O.}~\bibnamefont {Omolayo}}, \bibinfo {author} {\bibfnamefont {G.}~\bibnamefont {Penn}}, \bibinfo {author} {\bibfnamefont {S.}~\bibnamefont {De~Santis}}, \bibinfo {author} {\bibfnamefont {C.}~\bibnamefont {Steier}},\ and\ \bibinfo {author} {\bibfnamefont {M.}~\bibnamefont {Venturini}},\ }\href {https://doi.org/10.1016/j.nima.2022.166524} {\bibfield  {journal} {\bibinfo  {journal} {Nuclear Instruments and Methods in Physics Research Section A: Accelerators, Spectrometers, Detectors and Associated Equipment}\ }\textbf {\bibinfo {volume} {1031}},\ \bibinfo {pages} {166524} (\bibinfo {year} {2022})}\BibitemShut {NoStop}%
\bibitem [{\citenamefont {Smaluk}\ \emph {et~al.}({\natexlab{a}})\citenamefont {Smaluk}, \citenamefont {Martin}, \citenamefont {Fielder},\ and\ \citenamefont {Bartolini}}]{Smaluk:2022tnt}%
  \BibitemOpen
  \bibfield  {author} {\bibinfo {author} {\bibfnamefont {V.}~\bibnamefont {Smaluk}}, \bibinfo {author} {\bibfnamefont {I.}~\bibnamefont {Martin}}, \bibinfo {author} {\bibfnamefont {R.}~\bibnamefont {Fielder}},\ and\ \bibinfo {author} {\bibfnamefont {R.}~\bibnamefont {Bartolini}},\ }\href {https://doi.org/10.1103/PhysRevSTAB.18.064401} {\bibfield  {journal} {\bibinfo  {journal} {Physical Review Special Topics - Accelerators and Beams}\ }\textbf {\bibinfo {volume} {18}},\ \bibinfo {pages} {064401} ({\natexlab{a}})}\BibitemShut {NoStop}%
\bibitem [{\citenamefont {Carver}\ \emph {et~al.}(2023)\citenamefont {Carver}, \citenamefont {Brochard}, \citenamefont {Buratin}, \citenamefont {Carmignani}, \citenamefont {Ewald}, \citenamefont {Hoummi}, \citenamefont {Liuzzo}, \citenamefont {Perron}, \citenamefont {Roche},\ and\ \citenamefont {White}}]{Carver:2023xyy}%
  \BibitemOpen
  \bibfield  {author} {\bibinfo {author} {\bibfnamefont {L.}~\bibnamefont {Carver}}, \bibinfo {author} {\bibfnamefont {T.}~\bibnamefont {Brochard}}, \bibinfo {author} {\bibfnamefont {E.}~\bibnamefont {Buratin}}, \bibinfo {author} {\bibfnamefont {N.}~\bibnamefont {Carmignani}}, \bibinfo {author} {\bibfnamefont {F.}~\bibnamefont {Ewald}}, \bibinfo {author} {\bibfnamefont {L.}~\bibnamefont {Hoummi}}, \bibinfo {author} {\bibfnamefont {S.}~\bibnamefont {Liuzzo}}, \bibinfo {author} {\bibfnamefont {T.}~\bibnamefont {Perron}}, \bibinfo {author} {\bibfnamefont {B.}~\bibnamefont {Roche}},\ and\ \bibinfo {author} {\bibfnamefont {S.}~\bibnamefont {White}},\ }\href {https://doi.org/10.1103/PhysRevAccelBeams.26.044402} {\bibfield  {journal} {\bibinfo  {journal} {Physical Review Accelerators and Beams}\ }\textbf {\bibinfo {volume} {26}},\ \bibinfo {pages} {044402} (\bibinfo {year} {2023})}\BibitemShut {NoStop}%
\bibitem [{\citenamefont {Iadarola}\ \emph {et~al.}(2024)\citenamefont {Iadarola}, \citenamefont {Abramov}, \citenamefont {Belanger}, \citenamefont {Buffat}, \citenamefont {Maria}, \citenamefont {Demetriadou}, \citenamefont {Deniau}, \citenamefont {Croce}, \citenamefont {Hermes}, \citenamefont {Kicsiny}, \citenamefont {Kruyt}, \citenamefont {Latina}, \citenamefont {Mether}, \citenamefont {Niedermayer}, \citenamefont {Paraschou}, \citenamefont {Pieloni}, \citenamefont {Seidel}, \citenamefont {Sterbini}, \citenamefont {der Veken}, \citenamefont {van Riesen-Haupt},\ and\ \citenamefont {Łopaciuk}}]{xsuite}%
  \BibitemOpen
  \bibfield  {author} {\bibinfo {author} {\bibfnamefont {G.}~\bibnamefont {Iadarola}}, \bibinfo {author} {\bibfnamefont {A.}~\bibnamefont {Abramov}}, \bibinfo {author} {\bibfnamefont {P.}~\bibnamefont {Belanger}}, \bibinfo {author} {\bibfnamefont {X.}~\bibnamefont {Buffat}}, \bibinfo {author} {\bibfnamefont {R.~D.}\ \bibnamefont {Maria}}, \bibinfo {author} {\bibfnamefont {D.}~\bibnamefont {Demetriadou}}, \bibinfo {author} {\bibfnamefont {L.}~\bibnamefont {Deniau}}, \bibinfo {author} {\bibfnamefont {D.~D.}\ \bibnamefont {Croce}}, \bibinfo {author} {\bibfnamefont {P.}~\bibnamefont {Hermes}}, \bibinfo {author} {\bibfnamefont {P.}~\bibnamefont {Kicsiny}}, \bibinfo {author} {\bibfnamefont {P.}~\bibnamefont {Kruyt}}, \bibinfo {author} {\bibfnamefont {A.}~\bibnamefont {Latina}}, \bibinfo {author} {\bibfnamefont {L.}~\bibnamefont {Mether}}, \bibinfo {author} {\bibfnamefont {P.}~\bibnamefont {Niedermayer}}, \bibinfo {author} {\bibfnamefont {K.}~\bibnamefont {Paraschou}}, \bibinfo {author} {\bibfnamefont
  {T.}~\bibnamefont {Pieloni}}, \bibinfo {author} {\bibfnamefont {M.}~\bibnamefont {Seidel}}, \bibinfo {author} {\bibfnamefont {G.}~\bibnamefont {Sterbini}}, \bibinfo {author} {\bibfnamefont {F.~V.}\ \bibnamefont {der Veken}}, \bibinfo {author} {\bibfnamefont {L.}~\bibnamefont {van Riesen-Haupt}},\ and\ \bibinfo {author} {\bibfnamefont {S.}~\bibnamefont {Łopaciuk}},\ }in\ \href {https://doi.org/10.18429/JACoW-HB2023-TUA2I1} {\emph {\bibinfo {booktitle} {Proc. HB'23}}},\ \bibinfo {series and number} {\bibinfo {number} {68}}\ (\bibinfo  {publisher} {JACoW Publishing, Geneva, Switzerland},\ \bibinfo {year} {2024})\ pp.\ \bibinfo {pages} {73--80}\BibitemShut {NoStop}%
\bibitem [{\citenamefont {Borland}(2000)}]{elegant}%
  \BibitemOpen
  \bibfield  {author} {\bibinfo {author} {\bibfnamefont {M.}~\bibnamefont {Borland}},\ }\href {https://doi.org/10.2172/761286} {\emph {\bibinfo {title} {{ELEGANT}: A flexible {SDDS}-compliant code for accelerator simulation}}},\ \bibinfo {type} {Tech. Rep.}\ \bibinfo {number} {{LS}-287, 761286}\ (\bibinfo  {institution} {Advanced Photon Source},\ \bibinfo {year} {2000})\BibitemShut {NoStop}%
\bibitem [{\citenamefont {Gamelin}\ \emph {et~al.}(2021)\citenamefont {Gamelin}, \citenamefont {Foosang},\ and\ \citenamefont {Nagaoka}}]{Gamelin:IPAC21-MOPAB070}%
  \BibitemOpen
  \bibfield  {author} {\bibinfo {author} {\bibfnamefont {A.}~\bibnamefont {Gamelin}}, \bibinfo {author} {\bibfnamefont {W.}~\bibnamefont {Foosang}},\ and\ \bibinfo {author} {\bibfnamefont {R.}~\bibnamefont {Nagaoka}},\ }in\ \href {https://doi.org/10.18429/JACoW-IPAC2021-MOPAB070} {\emph {\bibinfo {booktitle} {Proc. IPAC'21}}},\ \bibinfo {series and number} {\bibinfo {number} {12}}\ (\bibinfo  {publisher} {JACoW Publishing, Geneva, Switzerland},\ \bibinfo {year} {2021})\ pp.\ \bibinfo {pages} {282--285}\BibitemShut {NoStop}%
\bibitem [{\citenamefont {Joly}\ \emph {et~al.}(2023)\citenamefont {Joly}, \citenamefont {Salvant}, \citenamefont {Imesch}, \citenamefont {Delrieux}, \citenamefont {Migliorati},\ and\ \citenamefont {Mounet}}]{Joly:IPAC23-WEPL149}%
  \BibitemOpen
  \bibfield  {author} {\bibinfo {author} {\bibfnamefont {S.}~\bibnamefont {Joly}}, \bibinfo {author} {\bibfnamefont {B.}~\bibnamefont {Salvant}}, \bibinfo {author} {\bibfnamefont {G.}~\bibnamefont {Imesch}}, \bibinfo {author} {\bibfnamefont {M.}~\bibnamefont {Delrieux}}, \bibinfo {author} {\bibfnamefont {M.}~\bibnamefont {Migliorati}},\ and\ \bibinfo {author} {\bibfnamefont {N.}~\bibnamefont {Mounet}},\ }in\ \href {https://doi.org/10.18429/JACoW-IPAC2023-WEPL149} {\emph {\bibinfo {booktitle} {Proc. IPAC'23}}},\ \bibinfo {series and number} {\bibinfo {number} {14}}\ (\bibinfo  {publisher} {JACoW Publishing, Geneva, Switzerland},\ \bibinfo {year} {2023})\ pp.\ \bibinfo {pages} {3454--3457}\BibitemShut {NoStop}%
\bibitem [{\citenamefont {Carver}\ \emph {et~al.}(2021)\citenamefont {Carver} \emph {et~al.}}]{Carver:IPAC21-MOPAB117}%
  \BibitemOpen
  \bibfield  {author} {\bibinfo {author} {\bibfnamefont {L.~R.}\ \bibnamefont {Carver}} \emph {et~al.},\ }in\ \href {https://doi.org/10.18429/JACoW-IPAC2021-MOPAB117} {\emph {\bibinfo {booktitle} {Proc. IPAC'21}}}\ (\bibinfo  {publisher} {JACoW Publishing, Geneva, Switzerland},\ \bibinfo {year} {2021})\ pp.\ \bibinfo {pages} {425--428}\BibitemShut {NoStop}%
\bibitem [{\citenamefont {Smaluk}\ \emph {et~al.}({\natexlab{b}})\citenamefont {Smaluk}, \citenamefont {Yang}, \citenamefont {Blednykh}, \citenamefont {Tian},\ and\ \citenamefont {Ha}}]{Smaluk:2017}%
  \BibitemOpen
  \bibfield  {author} {\bibinfo {author} {\bibfnamefont {V.}~\bibnamefont {Smaluk}}, \bibinfo {author} {\bibfnamefont {X.}~\bibnamefont {Yang}}, \bibinfo {author} {\bibfnamefont {A.}~\bibnamefont {Blednykh}}, \bibinfo {author} {\bibfnamefont {Y.}~\bibnamefont {Tian}},\ and\ \bibinfo {author} {\bibfnamefont {K.}~\bibnamefont {Ha}},\ }\href {https://doi.org/10.1016/j.nima.2017.07.067} {\bibfield  {journal} {\bibinfo  {journal} {Nuclear Instruments and Methods in Physics Research Section A: Accelerators, Spectrometers, Detectors and Associated Equipment}\ }\textbf {\bibinfo {volume} {871}},\ \bibinfo {pages} {59} ({\natexlab{b}})}\BibitemShut {NoStop}%
\bibitem [{\citenamefont {Mart?¡}\ \emph {et~al.}(2019)\citenamefont {Mart?¡}, \citenamefont {Benedetti}, \citenamefont {G??nzel},\ and\ \citenamefont {Iriso}}]{Marti:IPAC19-MOPGW066}%
  \BibitemOpen
  \bibfield  {author} {\bibinfo {author} {\bibfnamefont {Z.}~\bibnamefont {Mart?¡}}, \bibinfo {author} {\bibfnamefont {G.}~\bibnamefont {Benedetti}}, \bibinfo {author} {\bibfnamefont {T.~F.~G.}\ \bibnamefont {G??nzel}},\ and\ \bibinfo {author} {\bibfnamefont {U.}~\bibnamefont {Iriso}},\ }in\ \href {https://doi.org/10.18429/JACoW-IPAC2019-MOPGW066} {\emph {\bibinfo {booktitle} {Proc. IPAC'19}}}\ (\bibinfo  {publisher} {JACoW Publishing, Geneva, Switzerland},\ \bibinfo {year} {2019})\ pp.\ \bibinfo {pages} {240--242}\BibitemShut {NoStop}%
\bibitem [{\citenamefont {Zannini}\ \emph {et~al.}(2015)\citenamefont {Zannini}, \citenamefont {Bartosik}, \citenamefont {Iadarola}, \citenamefont {Rumolo},\ and\ \citenamefont {Salvant}}]{Zannini:IPAC15-MOPJE049}%
  \BibitemOpen
  \bibfield  {author} {\bibinfo {author} {\bibfnamefont {C.}~\bibnamefont {Zannini}}, \bibinfo {author} {\bibfnamefont {H.}~\bibnamefont {Bartosik}}, \bibinfo {author} {\bibfnamefont {G.}~\bibnamefont {Iadarola}}, \bibinfo {author} {\bibfnamefont {G.}~\bibnamefont {Rumolo}},\ and\ \bibinfo {author} {\bibfnamefont {B.}~\bibnamefont {Salvant}},\ }in\ \href {https://doi.org/10.18429/JACoW-IPAC2015-MOPJE049} {\emph {\bibinfo {booktitle} {Proc. IPAC'15}}}\ (\bibinfo  {publisher} {JACoW Publishing, Geneva, Switzerland},\ \bibinfo {year} {2015})\ pp.\ \bibinfo {pages} {402--405}\BibitemShut {NoStop}%
\bibitem [{\citenamefont {de~la Fuente}\ \emph {et~al.}(2024)\citenamefont {de~la Fuente}, \citenamefont {Bartosik}, \citenamefont {Solé}, \citenamefont {Rumolo},\ and\ \citenamefont {Zannini}}]{delaFuente:2024oro}%
  \BibitemOpen
  \bibfield  {author} {\bibinfo {author} {\bibfnamefont {E.}~\bibnamefont {de~la Fuente}}, \bibinfo {author} {\bibfnamefont {H.}~\bibnamefont {Bartosik}}, \bibinfo {author} {\bibfnamefont {I.~M.}\ \bibnamefont {Solé}}, \bibinfo {author} {\bibfnamefont {G.}~\bibnamefont {Rumolo}},\ and\ \bibinfo {author} {\bibfnamefont {C.}~\bibnamefont {Zannini}},\ }in\ \href {https://doi.org/10.18429/JACoW-HB2023-THAFP01} {\emph {\bibinfo {booktitle} {Proc. HB'23}}},\ \bibinfo {series and number} {\bibinfo {number} {68}}\ (\bibinfo  {publisher} {JACoW Publishing, Geneva, Switzerland},\ \bibinfo {year} {2024})\ pp.\ \bibinfo {pages} {393--396}\BibitemShut {NoStop}%
\bibitem [{\citenamefont {Collaboration}(2019)}]{madx}%
  \BibitemOpen
  \bibfield  {author} {\bibinfo {author} {\bibfnamefont {M.}~\bibnamefont {Collaboration}},\ }\href@noop {} {\bibinfo {title} {{MAD} - methodical accelerator design}},\ \bibinfo {howpublished} {\url{https://github.com/MethodicalAcceleratorDesign/MAD-X}} (\bibinfo {year} {2019})\BibitemShut {NoStop}%
\bibitem [{\citenamefont {Collaboration}(2024)}]{pywit}%
  \BibitemOpen
  \bibfield  {author} {\bibinfo {author} {\bibfnamefont {X.}~\bibnamefont {Collaboration}},\ }\href@noop {} {\bibinfo {title} {{XWakes} tool}},\ \bibinfo {howpublished} {\url{https://github.com/xsuite/xwakes}} (\bibinfo {year} {2024})\BibitemShut {NoStop}%
\bibitem [{\citenamefont {SIMULIA}(2023)}]{cst}%
  \BibitemOpen
  \bibfield  {author} {\bibinfo {author} {\bibnamefont {SIMULIA}},\ }\href@noop {} {\bibinfo {title} {{CST} studio suite}},\ \bibinfo {howpublished} {\url{https://www.3ds.com/products-services/simulia/products/cst-studio-suite/?utm_source=cst.com&utm_medium=301&utm_campaign=cst}} (\bibinfo {year} {2023})\BibitemShut {NoStop}%
\bibitem [{\citenamefont {Mounet}\ and\ \citenamefont {Eskil}(2023)}]{neffint}%
  \BibitemOpen
  \bibfield  {author} {\bibinfo {author} {\bibfnamefont {N.}~\bibnamefont {Mounet}}\ and\ \bibinfo {author} {\bibfnamefont {V.}~\bibnamefont {Eskil}},\ }\href@noop {} {\bibinfo {title} {Non-equidistant filon fourier {INTegration} ({NEFFFINT}) algorithm.}},\ \bibinfo {howpublished} {\url{https://neffint.readthedocs.io/en/latest/}} (\bibinfo {year} {2023})\BibitemShut {NoStop}%
\bibitem [{\citenamefont {Ventura}(2013)}]{Ventura:1625120}%
  \BibitemOpen
  \bibfield  {author} {\bibinfo {author} {\bibfnamefont {L.}~\bibnamefont {Ventura}},\ }\emph {\bibinfo {title} {Study of longitudinal multibunch instabilities for {LHC}-type beams at the {CERN} Proton Synchrotron}},\ \href {https://cds.cern.ch/record/1625120} {\bibinfo {type} {phdthesis}},\ \bibinfo  {school} {Rome U.} (\bibinfo {year} {2013})\BibitemShut {NoStop}%
\bibitem [{\citenamefont {Persichelli}(2015)}]{Persichelli:2015}%
  \BibitemOpen
  \bibfield  {author} {\bibinfo {author} {\bibfnamefont {S.}~\bibnamefont {Persichelli}},\ }\emph {\bibinfo {title} {The beam coupling impedance model of {CERN} Proton Synchrotron}},\ \href {https://cds.cern.ch/record/2027523} {\bibinfo {type} {phdthesis}},\ \bibinfo  {school} {Rome U.} (\bibinfo {year} {2015})\BibitemShut {NoStop}%
\bibitem [{\citenamefont {Sébastien}(2021)}]{Impedance_model}%
  \BibitemOpen
  \bibfield  {author} {\bibinfo {author} {\bibfnamefont {J.}~\bibnamefont {Sébastien}},\ }\href@noop {} {\bibinfo {title} {Pre-{LIU} {PS} transverse impedance model}},\ \bibinfo {howpublished} {\url{https://gitlab.cern.ch/IRIS/PS_IW_model}} (\bibinfo {year} {2021})\BibitemShut {NoStop}%
\bibitem [{\citenamefont {Popovic}(2020)}]{popovic}%
  \BibitemOpen
  \bibfield  {author} {\bibinfo {author} {\bibfnamefont {B.}~\bibnamefont {Popovic}},\ }\href {https://indico.cern.ch/event/879306/contributions/3725675/attachments/1978397/3293536/IWG37_Imped_Model_Update_300120_FINAL.pdf} {\bibinfo {title} {{PS} longitudinal model}}\ (\bibinfo {year} {2020})\ 01,\ 2020\BibitemShut {NoStop}%
\bibitem [{\citenamefont {Joly}(2024)}]{Joly:2024}%
  \BibitemOpen
  \bibfield  {author} {\bibinfo {author} {\bibfnamefont {S.}~\bibnamefont {Joly}},\ }\emph {\bibinfo {title} {{PS} impedance model and related beam stability}},\ \href {https://hdl.handle.net/11573/1734571} {\bibinfo {type} {phdthesis}},\ \bibinfo  {school} {Rome U.} (\bibinfo {year} {2024})\BibitemShut {NoStop}%
\bibitem [{\citenamefont {Sébastien}\ and\ \citenamefont {Benoît}(2020)}]{Impedance_website}%
  \BibitemOpen
  \bibfield  {author} {\bibinfo {author} {\bibfnamefont {J.}~\bibnamefont {Sébastien}}\ and\ \bibinfo {author} {\bibfnamefont {S.}~\bibnamefont {Benoît}},\ }\href@noop {} {\bibinfo {title} {{CERN} impedance website}},\ \bibinfo {howpublished} {\url{https://impedance.web.cern.ch/}} (\bibinfo {year} {2020})\BibitemShut {NoStop}%
\bibitem [{\citenamefont {Sacherer}(1977)}]{Blewett:118362}%
  \BibitemOpen
  \bibfield  {author} {\bibinfo {author} {\bibfnamefont {F.~J.}\ \bibnamefont {Sacherer}},\ }in\ \href {https://doi.org/10.5170/CERN-1977-013} {\emph {\bibinfo {booktitle} {Theoretical aspects of the behaviour of beams in accelerators and storage rings}}}\ (\bibinfo {year} {1977})\ pp.\ \bibinfo {pages} {198--218}\BibitemShut {NoStop}%
\bibitem [{\citenamefont {Wasef}\ \emph {et~al.}(2013)\citenamefont {Wasef} \emph {et~al.}}]{Wasef:IPAC13-WEPEA070}%
  \BibitemOpen
  \bibfield  {author} {\bibinfo {author} {\bibfnamefont {R.}~\bibnamefont {Wasef}} \emph {et~al.},\ }in\ \href {https://jacow.org/IPAC2013/papers/WEPEA070.pdf} {\emph {\bibinfo {booktitle} {Proc. IPAC'13}}}\ (\bibinfo  {publisher} {JACoW Publishing, Geneva, Switzerland},\ \bibinfo {year} {2013})\ pp.\ \bibinfo {pages} {2669--2671}\BibitemShut {NoStop}%
\bibitem [{\citenamefont {Asvesta}\ \emph {et~al.}(2020)\citenamefont {Asvesta}, \citenamefont {Bartosik}, \citenamefont {Gilardoni}, \citenamefont {Huschauer}, \citenamefont {Machida}, \citenamefont {Papaphilippou},\ and\ \citenamefont {Wasef}}]{Asvesta:2020qtd}%
  \BibitemOpen
  \bibfield  {author} {\bibinfo {author} {\bibfnamefont {F.}~\bibnamefont {Asvesta}}, \bibinfo {author} {\bibfnamefont {H.}~\bibnamefont {Bartosik}}, \bibinfo {author} {\bibfnamefont {S.}~\bibnamefont {Gilardoni}}, \bibinfo {author} {\bibfnamefont {A.}~\bibnamefont {Huschauer}}, \bibinfo {author} {\bibfnamefont {S.}~\bibnamefont {Machida}}, \bibinfo {author} {\bibfnamefont {Y.}~\bibnamefont {Papaphilippou}},\ and\ \bibinfo {author} {\bibfnamefont {R.}~\bibnamefont {Wasef}},\ }\href {https://doi.org/10.1103/PhysRevAccelBeams.23.091001} {\bibfield  {journal} {\bibinfo  {journal} {Physical Review Accelerators and Beams}\ }\textbf {\bibinfo {volume} {23}},\ \bibinfo {pages} {091001} (\bibinfo {year} {2020})}\BibitemShut {NoStop}%
\bibitem [{\citenamefont {Lee}(2004)}]{Lee}%
  \BibitemOpen
  \bibfield  {author} {\bibinfo {author} {\bibfnamefont {S.~Y.}\ \bibnamefont {Lee}},\ }\href {https://doi.org/10.1142/5761} {\emph {\bibinfo {title} {Accelerator Physics}}},\ \bibinfo {edition} {2nd}\ ed.\ (\bibinfo  {publisher} {{WORLD} {SCIENTIFIC}},\ \bibinfo {year} {2004})\BibitemShut {NoStop}%
\bibitem [{\citenamefont {Salvant}(2010)}]{Salvant:2010dda}%
  \BibitemOpen
  \bibfield  {author} {\bibinfo {author} {\bibfnamefont {B.}~\bibnamefont {Salvant}},\ }\emph {\bibinfo {title} {Impedance model of the {CERN} {SPS} and aspects of {LHC} single-bunch stability}},\ \href {https://doi.org/10.5075/EPFL-THESIS-4585} {\bibinfo {type} {phdthesis}},\ \bibinfo  {school} {Lausanne, {EPFL}} (\bibinfo {year} {2010})\BibitemShut {NoStop}%
\bibitem [{\citenamefont {Joly}\ \emph {et~al.}(2021)\citenamefont {Joly}, \citenamefont {Migliorati}, \citenamefont {Mounet},\ and\ \citenamefont {Salvant}}]{Joly:IPAC21-WEPAB224}%
  \BibitemOpen
  \bibfield  {author} {\bibinfo {author} {\bibfnamefont {S.}~\bibnamefont {Joly}}, \bibinfo {author} {\bibfnamefont {M.}~\bibnamefont {Migliorati}}, \bibinfo {author} {\bibfnamefont {N.}~\bibnamefont {Mounet}},\ and\ \bibinfo {author} {\bibfnamefont {B.}~\bibnamefont {Salvant}},\ }in\ \href {https://doi.org/10.18429/JACoW-IPAC2021-WEPAB224} {\emph {\bibinfo {booktitle} {Proc. IPAC'21}}},\ \bibinfo {series and number} {\bibinfo {number} {12}}\ (\bibinfo  {publisher} {JACoW Publishing, Geneva, Switzerland},\ \bibinfo {year} {2021})\ pp.\ \bibinfo {pages} {3149--3152}\BibitemShut {NoStop}%
\bibitem [{\citenamefont {Yokoya}(1993)}]{yokoya}%
  \BibitemOpen
  \bibfield  {author} {\bibinfo {author} {\bibfnamefont {K.}~\bibnamefont {Yokoya}},\ }\href {https://cds.cern.ch/record/248630} {\emph {\bibinfo {title} {Resistive wall impedance of beam pipes of general cross section}}},\ \bibinfo {type} {Tech. Rep.}\ (\bibinfo  {institution} {KEK},\ \bibinfo {year} {1993})\BibitemShut {NoStop}%
\bibitem [{\citenamefont {Migliorati}\ \emph {et~al.}({\natexlab{b}})\citenamefont {Migliorati}, \citenamefont {Palumbo}, \citenamefont {Zannini}, \citenamefont {Biancacci},\ and\ \citenamefont {Vaccaro}}]{Migliorati:2705426}%
  \BibitemOpen
  \bibfield  {author} {\bibinfo {author} {\bibfnamefont {M.}~\bibnamefont {Migliorati}}, \bibinfo {author} {\bibfnamefont {L.}~\bibnamefont {Palumbo}}, \bibinfo {author} {\bibfnamefont {C.}~\bibnamefont {Zannini}}, \bibinfo {author} {\bibfnamefont {N.}~\bibnamefont {Biancacci}},\ and\ \bibinfo {author} {\bibfnamefont {V.}~\bibnamefont {Vaccaro}},\ }\href {https://doi.org/10.1103/PhysRevAccelBeams.22.121001} {\bibfield  {journal} {\bibinfo  {journal} {Physical Review Accelerators and Beams}\ }\textbf {\bibinfo {volume} {22}},\ \bibinfo {pages} {121001} ({\natexlab{b}})}\BibitemShut {NoStop}%
\bibitem [{\citenamefont {Koukovini-Platia}\ \emph {et~al.}(2018)\citenamefont {Koukovini-Platia}, \citenamefont {Bartosik}, \citenamefont {Migliorati},\ and\ \citenamefont {Rumolo}}]{Koukovini-Platia:2640483}%
  \BibitemOpen
  \bibfield  {author} {\bibinfo {author} {\bibfnamefont {E.}~\bibnamefont {Koukovini-Platia}}, \bibinfo {author} {\bibfnamefont {H.}~\bibnamefont {Bartosik}}, \bibinfo {author} {\bibfnamefont {M.}~\bibnamefont {Migliorati}},\ and\ \bibinfo {author} {\bibfnamefont {G.}~\bibnamefont {Rumolo}},\ }\bibfield  {journal} {\bibinfo  {journal} {Proceedings of the 61{\textbackslash}textsuperscript\{st\} {ICFA} {ABDW} on High-Intensity and High-Brightness Hadron Beams}\ }\textbf {\bibinfo {volume} {{HB}2018}},\ \href {https://doi.org/10.18429/JACOW-HB2018-WEA2WA01} {10.18429/JACOW-HB2018-WEA2WA01} (\bibinfo {year} {2018})\BibitemShut {NoStop}%
\bibitem [{\citenamefont {Laskar}(1990)}]{naff}%
  \BibitemOpen
  \bibfield  {author} {\bibinfo {author} {\bibfnamefont {J.}~\bibnamefont {Laskar}},\ }\href {https://doi.org/10.1016/0019-1035(90)90084-M} {\bibfield  {journal} {\bibinfo  {journal} {Icarus}\ }\textbf {\bibinfo {volume} {88}},\ \bibinfo {pages} {266} (\bibinfo {year} {1990})}\BibitemShut {NoStop}%
\bibitem [{\citenamefont {Foteini}\ \emph {et~al.}(2023)\citenamefont {Foteini}, \citenamefont {Nikos},\ and\ \citenamefont {Panagiotis}}]{pynaff}%
  \BibitemOpen
  \bibfield  {author} {\bibinfo {author} {\bibfnamefont {A.}~\bibnamefont {Foteini}}, \bibinfo {author} {\bibfnamefont {K.}~\bibnamefont {Nikos}},\ and\ \bibinfo {author} {\bibfnamefont {Z.}~\bibnamefont {Panagiotis}},\ }\href@noop {} {\bibinfo {title} {{PyNAFF} python module}},\ \bibinfo {howpublished} {\url{https://pypi.org/project/PyNAFF/}} (\bibinfo {year} {2023})\BibitemShut {NoStop}%
\bibitem [{\citenamefont {Weisstein}(2023)}]{LS_exp}%
  \BibitemOpen
  \bibfield  {author} {\bibinfo {author} {\bibfnamefont {E.~W.}\ \bibnamefont {Weisstein}},\ }\href@noop {} {\bibinfo {title} {Least squares fitting-exponential}},\ \bibinfo {howpublished} {\url{https://mathworld.wolfram.com/LeastSquaresFittingExponential.html}} (\bibinfo {year} {2023}),\ \bibinfo {note} {from {MathWorld} -- A Wolfram Web Resource}\BibitemShut {NoStop}%
\bibitem [{\citenamefont {Sands}(1969)}]{sands1969head}%
  \BibitemOpen
  \bibfield  {author} {\bibinfo {author} {\bibfnamefont {M.}~\bibnamefont {Sands}},\ }\href {https://inspirehep.net/files/397c085909f38a15f8e047f5de33975e} {\emph {\bibinfo {title} {{THE} {HEAD} - {TAIL} {EFFECT}: {AN} {INSTABILITY} {MECHANISM} {IN} {STORAGE} {RINGS}}}},\ \bibinfo {type} {Tech. Rep.}\ \bibinfo {number} {{SLAC}-{TN}-69-008}\ (\bibinfo  {institution} {{SLAC}},\ \bibinfo {year} {1969})\BibitemShut {NoStop}%
\bibitem [{\citenamefont {Cocq}\ \emph {et~al.}(1998)\citenamefont {Cocq}, \citenamefont {Jones},\ and\ \citenamefont {Schmickler}}]{Cocq:370134}%
  \BibitemOpen
  \bibfield  {author} {\bibinfo {author} {\bibfnamefont {D.}~\bibnamefont {Cocq}}, \bibinfo {author} {\bibfnamefont {O.~R.}\ \bibnamefont {Jones}},\ and\ \bibinfo {author} {\bibfnamefont {H.}~\bibnamefont {Schmickler}}\ }(\bibinfo {year} {1998})\ pp.\ \bibinfo {pages} {281--288}\BibitemShut {NoStop}%
\bibitem [{\citenamefont {Huschauer}(2016)}]{Huschauer:2194332}%
  \BibitemOpen
  \bibfield  {author} {\bibinfo {author} {\bibfnamefont {A.}~\bibnamefont {Huschauer}},\ }\emph {\bibinfo {title} {Beam Dynamics Studies for High-Intensity Beams in the {CERN} Proton Synchrotron}},\ \href {https://cds.cern.ch/record/2194332} {\bibinfo {type} {phdthesis}},\ \bibinfo  {school} {Vienna, Tech. U} (\bibinfo {year} {2016})\BibitemShut {NoStop}%
\bibitem [{\citenamefont {Huschauer}(2012)}]{Huschauer_WP}%
  \BibitemOpen
  \bibfield  {author} {\bibinfo {author} {\bibfnamefont {A.}~\bibnamefont {Huschauer}},\ }\emph {\bibinfo {title} {Working point and resonance studies at the {CERN} Proton Synchrotron}},\ \href {https://cds.cern.ch/record/1501943} {\bibinfo {type} {mastersthesis}},\ \bibinfo  {school} {Vienna, Tech. U.} (\bibinfo {year} {2012})\BibitemShut {NoStop}%
\bibitem [{\citenamefont {Chao}\ and\ \citenamefont {Chao}(1993)}]{chao1993physics}%
  \BibitemOpen
  \bibfield  {author} {\bibinfo {author} {\bibfnamefont {A.~W.}\ \bibnamefont {Chao}}\ and\ \bibinfo {author} {\bibfnamefont {A.~W.}\ \bibnamefont {Chao}},\ }\href@noop {} {\emph {\bibinfo {title} {Physics of collective beam instabilities in high energy accelerators}}},\ Wiley series in beam physics and accelerator technology\ (\bibinfo  {publisher} {Wiley},\ \bibinfo {year} {1993})\BibitemShut {NoStop}%
\bibitem [{\citenamefont {Zannini}\ and\ \citenamefont {et}(2019)}]{abp_injectors_day}%
  \BibitemOpen
  \bibfield  {author} {\bibinfo {author} {\bibfnamefont {C.}~\bibnamefont {Zannini}}\ and\ \bibinfo {author} {\bibfnamefont {a.}~\bibnamefont {et}},\ }\href {https://indico.cern.ch/event/799216/#3-impedance-and-z-database-151} {\bibinfo {title} {Impedance and z database}}\ (\bibinfo {organization} {CERN},\ \bibinfo {year} {2019})\ 04,\ 2019.\EOS\ ({ABP} injectors day)\BibitemShut {NoStop}%
\bibitem [{\citenamefont {Schenk}\ \emph {et~al.}(2018)\citenamefont {Schenk}, \citenamefont {Buffat}, \citenamefont {Li},\ and\ \citenamefont {Maillard}}]{Schenk:2018pae}%
  \BibitemOpen
  \bibfield  {author} {\bibinfo {author} {\bibfnamefont {M.}~\bibnamefont {Schenk}}, \bibinfo {author} {\bibfnamefont {X.}~\bibnamefont {Buffat}}, \bibinfo {author} {\bibfnamefont {K.}~\bibnamefont {Li}},\ and\ \bibinfo {author} {\bibfnamefont {A.}~\bibnamefont {Maillard}},\ }\href {https://doi.org/10.1103/PhysRevAccelBeams.21.084402} {\bibfield  {journal} {\bibinfo  {journal} {Physical Review Accelerators and Beams}\ }\textbf {\bibinfo {volume} {21}},\ \bibinfo {pages} {084402} (\bibinfo {year} {2018})}\BibitemShut {NoStop}%
\bibitem [{\citenamefont {Berg}\ and\ \citenamefont {Ruggiero}(1996)}]{Berg:318826}%
  \BibitemOpen
  \bibfield  {author} {\bibinfo {author} {\bibfnamefont {J.~S.}\ \bibnamefont {Berg}}\ and\ \bibinfo {author} {\bibfnamefont {F.}~\bibnamefont {Ruggiero}},\ }\href {https://cds.cern.ch/record/318826} {\emph {\bibinfo {title} {Landau damping with two-dimensional betatron tune spread}}},\ \bibinfo {type} {Tech. Rep.}\ \bibinfo {number} {{CERN}-{SL}-96-071-{AP}}\ (\bibinfo  {institution} {{CERN}},\ \bibinfo {year} {1996})\BibitemShut {NoStop}%
\bibitem [{\citenamefont {Oeftiger}(2016)}]{Oeftiger:318826}%
  \BibitemOpen
  \bibfield  {author} {\bibinfo {author} {\bibfnamefont {A.}~\bibnamefont {Oeftiger}},\ }\emph {\bibinfo {title} {Space Charge Effects and Advanced Modelling for {CERN} Low Energy Machines}},\ \href {https://cds.cern.ch/record/2233212} {\bibinfo {type} {phdthesis}},\ \bibinfo  {school} {Ecole Polytechnique, Lausanne, {LPAP}} (\bibinfo {year} {2016})\BibitemShut {NoStop}%
\bibitem [{\citenamefont {Burov}(2022)}]{Burov:318826}%
  \BibitemOpen
  \bibfield  {author} {\bibinfo {author} {\bibfnamefont {A.}~\bibnamefont {Burov}},\ }\href {https://doi.org/10.1140/epjp/s13360-022-02614-w} {\bibfield  {journal} {\bibinfo  {journal} {The European Physical Journal Plus}\ }\textbf {\bibinfo {volume} {137}},\ \bibinfo {pages} {624} (\bibinfo {year} {2022})}\BibitemShut {NoStop}%
\bibitem [{\citenamefont {Kornilov}\ and\ \citenamefont {Boine-Frankenheim}(2010)}]{Kornilov:2010zz}%
  \BibitemOpen
  \bibfield  {author} {\bibinfo {author} {\bibfnamefont {V.}~\bibnamefont {Kornilov}}\ and\ \bibinfo {author} {\bibfnamefont {O.}~\bibnamefont {Boine-Frankenheim}},\ }\href {https://doi.org/10.1103/PhysRevSTAB.13.114201} {\bibfield  {journal} {\bibinfo  {journal} {Physical Review Special Topics - Accelerators and Beams}\ }\textbf {\bibinfo {volume} {13}},\ \bibinfo {pages} {114201} (\bibinfo {year} {2010})}\BibitemShut {NoStop}%
\bibitem [{\citenamefont {Macridin}\ \emph {et~al.}(2015)\citenamefont {Macridin}, \citenamefont {Burov}, \citenamefont {Stern}, \citenamefont {Amundson},\ and\ \citenamefont {Spentzouris}}]{Macridin:2015vua}%
  \BibitemOpen
  \bibfield  {author} {\bibinfo {author} {\bibfnamefont {A.}~\bibnamefont {Macridin}}, \bibinfo {author} {\bibfnamefont {A.}~\bibnamefont {Burov}}, \bibinfo {author} {\bibfnamefont {E.}~\bibnamefont {Stern}}, \bibinfo {author} {\bibfnamefont {J.}~\bibnamefont {Amundson}},\ and\ \bibinfo {author} {\bibfnamefont {P.}~\bibnamefont {Spentzouris}},\ }\href {https://doi.org/10.1103/PhysRevSTAB.18.074401} {\bibfield  {journal} {\bibinfo  {journal} {Physical Review Special Topics - Accelerators and Beams}\ }\textbf {\bibinfo {volume} {18}},\ \bibinfo {pages} {074401} (\bibinfo {year} {2015})}\BibitemShut {NoStop}%
\bibitem [{\citenamefont {Kornilov}\ \emph {et~al.}(2015)\citenamefont {Kornilov}, \citenamefont {Boine-Frankenheim}, \citenamefont {Adams}, \citenamefont {Jones}, \citenamefont {Pine}, \citenamefont {Warsop},\ and\ \citenamefont {Williamson}}]{Kornilov:2015xta}%
  \BibitemOpen
  \bibfield  {author} {\bibinfo {author} {\bibfnamefont {V.}~\bibnamefont {Kornilov}}, \bibinfo {author} {\bibfnamefont {O.}~\bibnamefont {Boine-Frankenheim}}, \bibinfo {author} {\bibfnamefont {D.}~\bibnamefont {Adams}}, \bibinfo {author} {\bibfnamefont {B.}~\bibnamefont {Jones}}, \bibinfo {author} {\bibfnamefont {B.}~\bibnamefont {Pine}}, \bibinfo {author} {\bibfnamefont {C.}~\bibnamefont {Warsop}},\ and\ \bibinfo {author} {\bibfnamefont {R.}~\bibnamefont {Williamson}},\ }in\ \href {https://accelconf.web.cern.ch/HB2014/papers/weo1lr02.pdf} {\emph {\bibinfo {booktitle} {54th {ICFA} Advanced Beam Dynamics Workshop on High-Intensity and High-Brightness Hadron Beams}}}\ (\bibinfo {year} {2015})\ pp.\ \bibinfo {pages} {240--244},\ \bibinfo {note} {{WEO}1LR02}\BibitemShut {NoStop}%
\bibitem [{\citenamefont {Balbekov}(2017)}]{Balbekov:2016qda}%
  \BibitemOpen
  \bibfield  {author} {\bibinfo {author} {\bibfnamefont {V.}~\bibnamefont {Balbekov}},\ }\href {https://doi.org/10.1103/PhysRevAccelBeams.20.034401} {\bibfield  {journal} {\bibinfo  {journal} {Physical Review Accelerators and Beams}\ }\textbf {\bibinfo {volume} {20}},\ \bibinfo {pages} {034401} (\bibinfo {year} {2017})}\BibitemShut {NoStop}%
\bibitem [{\citenamefont {Zolkin}\ \emph {et~al.}(2018)\citenamefont {Zolkin}, \citenamefont {Burov},\ and\ \citenamefont {Pandey}}]{Zolkin:2018}%
  \BibitemOpen
  \bibfield  {author} {\bibinfo {author} {\bibfnamefont {T.}~\bibnamefont {Zolkin}}, \bibinfo {author} {\bibfnamefont {A.}~\bibnamefont {Burov}},\ and\ \bibinfo {author} {\bibfnamefont {B.}~\bibnamefont {Pandey}},\ }\href {https://doi.org/10.1103/PhysRevAccelBeams.21.104201} {\bibfield  {journal} {\bibinfo  {journal} {Physical Review Accelerators and Beams}\ }\textbf {\bibinfo {volume} {21}},\ \bibinfo {pages} {104201} (\bibinfo {year} {2018})}\BibitemShut {NoStop}%
\bibitem [{\citenamefont {Chin}\ \emph {et~al.}(2016)\citenamefont {Chin}, \citenamefont {Chao},\ and\ \citenamefont {Blaskiewicz}}]{Chin:2016lzk}%
  \BibitemOpen
  \bibfield  {author} {\bibinfo {author} {\bibfnamefont {Y.~H.}\ \bibnamefont {Chin}}, \bibinfo {author} {\bibfnamefont {A.~W.}\ \bibnamefont {Chao}},\ and\ \bibinfo {author} {\bibfnamefont {M.~M.}\ \bibnamefont {Blaskiewicz}},\ }\href {https://doi.org/10.1103/PhysRevAccelBeams.19.014201} {\bibfield  {journal} {\bibinfo  {journal} {Physical Review Accelerators and Beams}\ }\textbf {\bibinfo {volume} {19}},\ \bibinfo {pages} {014201} (\bibinfo {year} {2016})}\BibitemShut {NoStop}%
\bibitem [{\citenamefont {Burov}(2009)}]{Burov:2008be}%
  \BibitemOpen
  \bibfield  {author} {\bibinfo {author} {\bibfnamefont {A.}~\bibnamefont {Burov}},\ }\href {https://doi.org/10.1103/PhysRevSTAB.12.044202} {\bibfield  {journal} {\bibinfo  {journal} {Physical Review Special Topics - Accelerators and Beams}\ }\textbf {\bibinfo {volume} {12}},\ \bibinfo {pages} {044202} (\bibinfo {year} {2009})}\BibitemShut {NoStop}%
\bibitem [{\citenamefont {Burov}(2019)}]{Burov:2019}%
  \BibitemOpen
  \bibfield  {author} {\bibinfo {author} {\bibfnamefont {A.}~\bibnamefont {Burov}},\ }\href {https://doi.org/10.1103/PhysRevAccelBeams.22.034202} {\bibfield  {journal} {\bibinfo  {journal} {Physical Review Accelerators and Beams}\ }\textbf {\bibinfo {volume} {22}},\ \bibinfo {pages} {034202} (\bibinfo {year} {2019})}\BibitemShut {NoStop}%
\bibitem [{\citenamefont {Buffat}\ and\ \citenamefont {Bartosik}(2022)}]{Buffat:2022}%
  \BibitemOpen
  \bibfield  {author} {\bibinfo {author} {\bibfnamefont {X.}~\bibnamefont {Buffat}}\ and\ \bibinfo {author} {\bibfnamefont {H.}~\bibnamefont {Bartosik}},\ }in\ \href {https://doi.org/10.18429/JACoW-IPAC2022-WEPOTK058} {\emph {\bibinfo {booktitle} {Proc. IPAC'22}}},\ \bibinfo {series and number} {\bibinfo {number} {13}}\ (\bibinfo  {publisher} {JACoW Publishing, Geneva, Switzerland},\ \bibinfo {year} {2022})\ pp.\ \bibinfo {pages} {2193--2196}\BibitemShut {NoStop}%
\bibitem [{\citenamefont {Karpov}\ \emph {et~al.}(2015)\citenamefont {Karpov}, \citenamefont {Boine-Frankenheim},\ and\ \citenamefont {Kornilov}}]{Karpov:2015hsa}%
  \BibitemOpen
  \bibfield  {author} {\bibinfo {author} {\bibfnamefont {I.}~\bibnamefont {Karpov}}, \bibinfo {author} {\bibfnamefont {O.}~\bibnamefont {Boine-Frankenheim}},\ and\ \bibinfo {author} {\bibfnamefont {V.}~\bibnamefont {Kornilov}},\ }in\ \href {https://inspirehep.net/files/790c7a1c1540298b181e609d315dd95e} {\emph {\bibinfo {booktitle} {54th {ICFA} Advanced Beam Dynamics Workshop on High-Intensity and High-Brightness Hadron Beams}}}\ (\bibinfo {year} {2015})\BibitemShut {NoStop}%
\bibitem [{\citenamefont {Karpov}\ \emph {et~al.}(2016)\citenamefont {Karpov}, \citenamefont {Kornilov},\ and\ \citenamefont {Boine-Frankenheim}}]{karpov_early_2016}%
  \BibitemOpen
  \bibfield  {author} {\bibinfo {author} {\bibfnamefont {I.}~\bibnamefont {Karpov}}, \bibinfo {author} {\bibfnamefont {V.}~\bibnamefont {Kornilov}},\ and\ \bibinfo {author} {\bibfnamefont {O.}~\bibnamefont {Boine-Frankenheim}},\ }\href {https://doi.org/10.1103/PhysRevAccelBeams.19.124201} {\bibfield  {journal} {\bibinfo  {journal} {Physical Review Accelerators and Beams}\ }\textbf {\bibinfo {volume} {19}},\ \bibinfo {pages} {124201} (\bibinfo {year} {2016})}\BibitemShut {NoStop}%
\bibitem [{\citenamefont {Burov}(2018)}]{Burov:2018rmx}%
  \BibitemOpen
  \bibfield  {author} {\bibinfo {author} {\bibfnamefont {A.}~\bibnamefont {Burov}},\ }\href {https://doi.org/10.48550/arXiv.1808.08498} {\bibinfo {title} {Core-halo collective instabilities}} (\bibinfo {year} {2018}),\ \Eprint {https://arxiv.org/abs/1808.08498 [physics]} {1808.08498 [physics]} \BibitemShut {NoStop}%
\end{thebibliography}%

\end{document}